\documentclass[prd,aps,nofootinbib,showpacs]{revtex4}

\usepackage{amsmath,amsfonts,amssymb,amsthm}           
\usepackage{dsfont}                                    
\usepackage{graphicx, epsfig}                          
\usepackage{multirow}

\usepackage{grffile}                                  
\usepackage{epstopdf}                                 

\usepackage{color}
\usepackage{verbatim}
\usepackage{simplewick}
\usepackage{hyperref}

\newcommand{\idMatrix}{\mathds{1}}
\newcommand{\pd}{\partial}
\newcommand{\ii}{\mathrm{i}}

\newcommand{\diff}{\mathrm{d}}

\newcommand{\Lagr}{\mathcal{L}}
\newcommand{\bigO}{\mathcal{O}}           
\DeclareMathOperator{\tr}{tr}            
\DeclareMathOperator{\IIm}{\mathrm{Im}}  
\newcommand{\hc}{^\dagger}               
\newcommand{\cc}{^\ast}               
\newcommand{\avg}[1]{\langle{#1}\rangle} 
\newcommand{\degree}{^\circ}             
\newcommand{\cket}[1]{\ensuremath{\left|{#1}\right>}}  
\newcommand{\brac}[1]{\ensuremath{\left<{#1}\right|}}  
\newcommand{\NProd}[1]{{\,:\!{#1}\!:\,}}                 
\newcommand{\hv}[1]{\hat{\bvec{#1}}}                   
\newcommand{\bvec}[1]{\boldsymbol{#1}}                 
\newcommand{\twoCol}[2]{\begin{pmatrix} #1 \\ #2 \end{pmatrix}}
\newcommand{\twoMatrix}[4]{ \begin{pmatrix}{#1} & {#2} \\ {#3} & {#4} \end{pmatrix} }

\newcommand{\vsp}{\vskip0.7em}

\DeclareMathOperator{\diag}{\mathrm{diag}}     

\newcommand{\diagPart}{\textrm{diag}}
\newcommand{\offdiagPart}{\textrm{offdiag}}

\newcommand{\contractAMB}[4][1ex]{%
   \contraction[#1]{}{#2}{#3}{#4}%
   {#2}{#3}{#4}%
}

\newcommand{\contractWithIndices}[5][1ex]{%
    \contraction[#1]{}{#2}{{}^{#3}}{#4}%
    {#2}{^#3}{#4}^{#5}%
}

\begin{document}

\title{Effects of nonstandard neutrino self-interactions and magnetic moment on collective Majorana neutrino oscillations}
\author{Oleg~G.~Kharlanov} \email{kharlanov@physics.msu.ru}
\affiliation{Faculty of Physics, Lomonosov Moscow State University, 1/2 Leninskie Gory, 119991 Moscow, Russia}

\author{Pavel~I.~Shustov}
\affiliation{Space Research Institute of the Russian Academy of Sciences, 84/32 Profsoyuznaya st., 117997 Moscow, Russia}

%

\pacs{14.60.Pq, 14.60.St, 97.60.Bw}

\begin{abstract}
    We derive the effective Hamiltonian describing collective oscillations of Majorana neutrinos with
    a transition magnetic moment, allowing for the presence of scalar and pseudoscalar nonstandard neutrino self-interactions (NSSIs).
    Using this Hamiltonian, we analyze new flavor instability channels of collective oscillations in a
    core-collapse supernova environment that open up in the presence of a small but nonzero neutrino magnetic moment.
    It turns out that, contrary to certain claims in the literature, within the minimally extended Standard Model (i.e., without NSSIs),
    no new instabilities arise within the linear order, nor do they produce any observable signatures
    in the neutrino flavor-energy spectra, at least for magnetic moments up to $10^{-15}~\mu_{\text{B}}$
    and quite realistic fields of the order of $10^{12}$~Gauss. On the other hand, in the presence of NSSIs,
    new fast and slow instabilities mixing neutrinos and antineutrinos appear,
    which show up in the spectra even for tiny magnetic moments of the order of $10^{-24}~\mu_{\text{B}}$,
    leading to considerable distortions of the spectra
    and nonstandard spectral splits. We study sensitivity of collective oscillations to these,
    NSSI-induced instabilities in detail and discuss the observability of the NSSI couplings triggering them.
\end{abstract}

    \maketitle

    \section{Introduction}
    \label{sec:Intro}

    During the most violent phase of a core-collapse supernova explosion, emitted neutrinos carry away the lion's share
    of the explosion energy and turn out to be the first signal from a newborn supernova coming from its innermost regions
    \cite{Mirizzi2016_SNnus,Giunti_book}. An observation of supernova neutrinos could thus yield priceless information both on the physical processes
    at work during the explosion and on the physics of the neutrino itself in such an extreme environment, possibly
    unveiling signatures of certain Beyond-Standard-Model (BSM) phenomena \cite{Raffelt_book_StarsAsLabs, Sung2019_BSM, Shalgar2019_BSM, deGouvea2012_2013, Esteban2007_NSI}.
    Indeed, thanks to the progress in experimental techniques of neutrino detection made since the SN~1987A event, modern neutrino observatories,
    as well as those planned to start operating in the coming years, should be able to yield a number of events
    quite sufficient to draw the neutrino energy spectra. For instance, for a 10~kpc-away explosion, the JUNO detector would yield as many
    as several thousand events in both the neutrino and the antineutrino channels, including the inverse-beta-decay detection channel
    with an unprecedented energy resolution about $1\%$ \cite{JUNO}. Inevitably though, the energy spectra thus observed would be
    a result of neutrino oscillations on the way from the supernova core (the protoneutron star) to the detector, so, in order to
    `decypher' the spectra, one should be able to `rewind' these oscillations back to the stellar interior.

    For a typical core-collapse supernova, the neutrino densities turn out to be so high next to their last scattering surface
    (also referred to as the neutrino sphere) that neutrino-neutrino forward scattering processes become important for the flavor evolution,
    ushering into the physics of collective neutrino oscillations \cite{Duan2010_review}. This self-induced, nonlinear process shapes the
    flavor-energy spectra in the region deep under the Mikheev--Smirnov--Wolfenstein (MSW) resonance surface, i.e.,
    where noncollective oscillations would be suppressed by a gigantic matter potential \cite{Wolfenstein1978, MikheevSmirnov1985}.
    One of the main drivers of collective flavor transformations is the instabilities, which are intrinsically present
    in the nonlinear evolution equations on collective oscillations
    \cite{Duan2010_review, Raffelt2013_AngularInstabilities, Duan2015_NLM, Johns2020_FastInstabilities, Glas2020_FastInstabilities, Capozzi2020_FastInstabilities, Duan2006_CollOsc}.
    It has been identified, indeed, that a hierarchy of these instabilities, both in the linearized and fully nonlinear regimes, can lead to specific
    neutrino spectral features, such as the so-called spectral swaps (splits) and the behavior resembling turbulence
    \cite{Duan2006_swaps, Duan2019_NLM, Duan2020_NRM}.

    New types of instabilities often arise when new degrees of freedom come into play. For example, fast instabilities show up beyond the
    so-called single-angle scheme \cite{Raffelt2013_AngularInstabilities, Johns2020_FastInstabilities, Glas2020_FastInstabilities, Capozzi2020_FastInstabilities, Mirizzi2015_transInv};
    lifting the assumption of translational invariance of the solution leads to turbulent flavor patterns breaking this invariance
    on a wide range of spatial scales \cite{Duan2019_NLM, Duan2020_NRM, Duan2019_DR}.
    In view of this, it seems appealing to analyze the effect of the neutrino anomalous magnetic moment \cite{Giunti2009_nuEMP}, which mixes the two
    helicity states and thus doubles the number of nontrivially interacting degrees of freedom, on the spectrum of instabilities.
    This also sounds natural because of superstrong magnetic fields typical
    for collapsing stars. This problem has been studied in a number of papers, including the derivation of the effective Hamiltonian
    \cite{deGouvea2012_2013, Dvornikov2012_Heff, Cirigliano2015_spinQK} and the impact on the flavor evolution within the single-angle
    \cite{deGouvea2012_2013, Kharlanov2019} and multiangle frameworks \cite{Abbar2020}. Notably, the effective Hamiltonians derived/used
    in Refs.~\cite{Dvornikov2012_Heff, Cirigliano2015_spinQK, Kharlanov2019, Abbar2020} and Ref.~\cite{deGouvea2012_2013} are different in
    the self-interaction term and lead to drastically different flavor evolutions: in the latter case, new types of
    instabilities arise that strongly deform the neutrino flavor-energy spectra even for tiny values of the magnetic moment
    \cite{deGouvea2012_2013}.

    Certainly, it seems interesting not only to settle the question of the correct Hamiltonian, but rather to ask a more general question:
    are there any other factors producing neutrino flavor signatures in an interplay with the effects of a (tiny) neutrino magnetic
    moment? At least one answer to this question is discussed in the present paper. Namely, we study the effect of the so-called
    NonStandard neutrino Self-Interactions (NSSIs, see Refs.~\cite{NSSI_Ohlsson2013, NSSI_Farzan2018, NSSI_Bhupal2019} for a review)
    on the collective flavor evolution of Majorana neutrinos triggered by the nonzero neutrino transition
    magnetic moment. Recently, NSSIs, i.e., four-fermion neutrino-neutrino interactions that are absent in the electroweak sector of
    the Standard Model, have attracted a lot of attention for being able to affect the observable neutrino spectra
    \cite{NSSI_Ge2019, NSSI_Dighe2018, NSSI_Das2017, NSSI_Yang2018_SPint, NSSI_Khlopov1988}. One can distinguish two classes of NSSIs,
    those with flavor-dependent V--A interactions and those with a non-V--A tensor structure of the four-fermion interaction.
    The NSSIs of the second class, specifically, those involving scalar $(\bar\nu\nu)^2$ and pseudoscalar $(\bar\nu\gamma_5\nu)^2$ terms, are especially
    interesting to us in our context, since, as we show below, their presence opens up a new instability channel in oscillations
    of Majorana neutrinos with a nonzero magnetic moment. It is worth admitting here that another instability channel due
    to scalar-pseudoscalar NSSIs has already been studied in Ref.~\cite{NSSI_Yang2018_SPint}; however, the corresponding unstable modes
    do not mix the neutrino helicities and the nonvanishing magnetic moment is not crucial for their development.

    According to our analysis which follows, it turns out that presence of a scalar-pseudoscalar NSSI violently deforms the
    neutrino spectra even for minuscule values of the magnetic moment, at least down to $10^{-24}\mu_{\text{B}}$ for non-exaggerated stellar fields
    $B \sim 10^{12}\text{ Gauss}$, while in the absence of NSSIs (i.e., for the neutrino interactions within the Standard Model),
    the effect of the magnetic moment is virtually unobservable even for much greater magnetic moments. Moreover, certain
    NSSI signatures originating in collective oscillations deep inside the supernova, such as the anomalous neutrino-to-antineutrino number ratio,
    can be safely transported to the surface and further to the detector. This means that supernova neutrino spectra can be used to probe the presence of (pseudo)scalar NSSIs,
    possibly stemming from exchange of (pseudo)scalar BSM particles.
    The steps we take within our analysis of the NSSI-induced instabilities are presented in the following sections.
    Namely, in Sec.~\ref{sec:EvolutionEquation}, we rederive the Hamiltonian for collective Majorana neutrino oscillations with
    NSSIs and a nonzero neutrino magnetic moment and compare the no-NSSI case with the results of previous derivations.
    Further, in Sec.~\ref{sec:Stability}, we carry out a linear stability analysis of our flavor evolution equations, also showing
    why in the absence of NSSIs, the effect of the magnetic moment should be small. Focusing then on the most interesting
    NSSI-induced unstable modes mixing the two neutrino helicities, we determine their growth rates, revealing fast and slow branches,
    as well as a specific intermediate, matter density dependent one. The linear analysis is accompanied by a full numerical
    simulation in Sec.~\ref{sec:Simulation}, to study what happens beyond the linear stability regime and to determine the
    sensitivities of the spectra to the neutrino magnetic moment and the NSSI couplings. The final section~\ref{sec:Discussion}
    summarizes the results obtained and discusses their possible generalizations. In the Appendix, several identities are
    proved or listed that we use in the derivation of the effective Hamiltonian in the presence of NSSIs.

    \section{Evolution equation for collective oscillations in the presence of NSSIs}
    \label{sec:EvolutionEquation}

    Let us now settle the question of the effective Hamiltonian for collective oscillations of Majorana neutrinos with nonzero magnetic moment
    we mentioned above and, more importantly, derive the terms in this Hamiltonian that describe the scalar and pseudoscalar
    NSSIs. For that, let us derive the evolution equations for the neutrino flavor density matrix accounting for the forward
    scattering processes, starting from field theory. We use relativistic units $\hbar = c = 1$ throughout the paper.

    We consider Majorana neutrinos in a small representative volume $V$, interacting with background electrons, protons, and neutrons,
    plus an external electromagnetic field $F_{\mu\nu}(x)$. After the electroweak symmetry is broken, the terms in the Lagrangian
    that contribute to neutrino forward scattering read \cite{PDG, Giunti_book, Giunti2009_nuEMP, NSSI_Yang2018_SPint}
    \begin{eqnarray}\label{Lagr}
       \Lagr_{\nu}   &=& \Lagr_{\text{vac}}^{(2)} + \Lagr_{\text{mat}}^{(2)} + \Lagr_{\text{AMM}}^{(2)} + \Lagr_{\text{VA}}^{(4)} + \Lagr_{\text{SP}}^{(4)}, \\
       \Lagr_{\text{vac}}^{(2)}  &=& \frac12\sum\limits_{a = 1}^{N_{\text{f}}} \bar\nu_a (\ii \gamma^\mu \pd_\mu - m_a) \nu_a,  \label{L0}\\
       \Lagr_{\text{mat}}^{(2)} &=& -G_{\text{F}}\sqrt2 \bar\nu_{f'}\gamma^\mu_{\text{L}} \nu_f
                                 \Biggl\{ -\delta_{f,f'} \Bigl[ \bar{p} (2s_{\text{W}}^2 \gamma_\mu - \gamma_{\mu\text{L}}) p
                                         + \bar{n}\gamma_{\mu\text{L}}n \Bigr] +
                                        \bar{e} \Bigl[\delta_{f,f'}(2s_{\text{W}}^2 \gamma_\mu - \gamma_{\mu\text{L}})
                                        + 2 \delta_{f,e} \delta_{f',e}\gamma_{\mu\text{L}}\Bigr] e \Biggr\}, \label{L_mat}\\
       \Lagr_{\text{AMM}}^{(2)} &=& -\frac{\ii}{4} \mathfrak{m}_{ab} \bar\nu_{a} \sigma_{\mu\nu} F^{\mu\nu} \nu_{b},  \label{L2}\\
       \Lagr_{\text{VA}}^{(4)} &=& -\frac{G_{\text{F}}}{\sqrt2} : (\bar\nu_a\gamma^\mu_{\text{L}} \nu_a)^2 :,    \label{L4}\\
       \Lagr_{\text{SP}}^{(4)} &=& -\frac{G_{\text{F}}}{4\sqrt2}
                                   \left\{g_{\text{S}} :(\bar\nu_a \nu_a)^2: \; + \;\; g_{\text{P}} :(\bar\nu_a \gamma_5 \nu_a)^2: \right\}. \label{L_NSSI}
    \end{eqnarray}
    Here $\nu_a(x), \, a=1,\ldots,N_{\text{f}}$, are the neutrino fields with Majorana masses $m_a$ and $e(x), p(x), n(x)$
    are the fields describing electrons, protons, and neutrons, respectively. By definition, the neutrino fields obey Majorana constraints
    $\bar\nu_a(x) = \nu_a^{\mathrm{T}}(x)C$, where $C = -\ii \gamma^2 \gamma^0$ is the charge conjugation matrix
    and the bar denotes a Dirac conjugate $\bar\psi \equiv \psi\hc\gamma^0$. The Dirac matrices $\gamma^\mu_{\text{L}} \equiv \gamma^\mu (1-\gamma_5)/2$,
    $\sigma_{\mu\nu} \equiv \frac{\ii}{2} [\gamma_\mu, \gamma_\nu]$; $s_{\text{W}} \equiv \sin\theta_{\text{W}}$
    is the sine of the weak mixing angle, and $G_{\text{F}}$ is the Fermi constant; colons in Eqs.~\eqref{L4}, \eqref{L_NSSI} denote
    normal ordering of field operators. The neutrino flavor states are defined via a PMNS matrix \cite{MNS, PDG} assumed to be unitary,
    \begin{equation}
        \nu_f(x) \equiv U_{fa} \nu_a(x), \quad \nu_a(x) \equiv U\cc_{fa} \nu_f(x),
    \end{equation}
    where the summation over the mass indices $a,b,\ldots$ and flavor indices $f,f',\ldots$ is assumed, the latter
    take on $e,\mu, \tau$ values in the three-flavor case and $e,x$ values in the two-flavor case. The electromagnetic interaction term~\eqref{L2}
    contains the neutrino magnetic moment matrix $\mathfrak{m}_{ab}$, which is real and antisymmetric in the Majorana case,
    so that only transition magnetic moments are allowed \cite{Giunti2009_nuEMP}. Finally, the quartic terms~\eqref{L4},
    $\eqref{L_NSSI}$ describe the electroweak (V--A) and the nonstandard (scalar/pseudoscalar, SP) neutrino interactions,
    and the latter is parameterized by two dimensionless NSSI couplings $g_{\text{S},\text{P}}$.

    Within the forward-scattering approximation, i.e., while momentum-changing processes of the form
    $\nu_{a}(\bvec{p}) \to \nu_{b}(\bvec{p}')$, $\nu_{a}(\bvec{p}) \to \bar\nu_{b}(\bvec{p}')$ with $\bvec{p}' \ne \bvec{p}$
    play a minor role, we can ignore coherent superpositions of different neutrino momentum states (while keeping track of
    coherent superpositions of \emph{mass} states due to oscillations). Accordingly, the neutrino state(s) $\cket{\Phi}$ we will
    consider below factorize into a infinite tensor product of state vectors $\cket{\Phi_{\bvec{p}}}$,
    each of them describing neutrinos with a fixed momentum $\bvec{p}$ and belonging to the corresponding Fock space,
    \begin{eqnarray}
       \cket{\Phi} &=& \bigotimes\limits_{\bvec{p}} \cket{\Phi_{\bvec{p}}},\\
       \label{Phi_p}
       \cket{\Phi_{\bvec{p}}} &=& \sum\limits_{A_1\ldots A_{N_{\bvec{p}}}}
                         C_{A_1\ldots A_{N_{\bvec{p}}}}(\bvec{p}) \; \hat{a}\hc_{A_1\bvec{p}}\ldots \hat{a}\hc_{A_{N_{\bvec{p}}}\bvec{p}}
                         \cket{0_{\bvec{p}}}.
    \end{eqnarray}
    Here, for the sake of the calculations which will follow, we adopt a notation $A = (a\alpha), B = (b\beta),
    \ldots$ for the creation operators $\hat{a}\hc_{A\bvec{p}}$, with the Latin indices $a,b,\ldots = 1,\ldots,N_{\text{f}}$ denoting the neutrino mass states and
    the Greek ones $\alpha,\beta,\ldots = \pm1$ standing for the neutrino helicities (multiplied by two).
    As mentioned above, the state vector $\cket{\Phi_{\bvec{p}}}$ describes neutrinos with momentum $\bvec{p}$,
    whose total number $N_{\bvec{p}} \in \{ 0, 1, \ldots, 2 N_{\text{f}} \}$ is limited due to the Pauli exclusion principle; moreover,
    in the forward scattering regime, these numbers are conserved for every individual $\bvec{p}$ and
    \begin{equation}\label{particleNumber_p}
        \sum\limits_A \hat{a}\hc_{A\bvec{p}} \hat{a}_{A\bvec{p}} \; \cdot \cket{\Phi} = N_{\bvec{p}} \cket{\Phi}.
    \end{equation}
    Thus, the time evolution of the neutrino state affects only the set of $c$-number-valued coefficients $C_{A_1\ldots A_{N_{\bvec{p}}}}(\bvec{p})$.
    It is much more convenient, however, to work with a $2 N_{\text{f}}\times 2 N_{\text{f}}$ neutrino flavor density matrix
    instead of these coefficients
    \begin{eqnarray}\label{rho_def}
        \rho_{AB}(\bvec{p}) &\equiv& \brac{\Phi} \hat{a}\hc_{B\bvec{p}} \hat{a}_{A\bvec{p}} \cket\Phi, \\
        \label{rho_Ehrenfest}
        \frac{\pd \rho_{AB}(\bvec{p})}{\pd t} &=& \ii \brac{\Phi} [\hat{H}, \hat{a}\hc_{B\bvec{p}} \hat{a}_{A\bvec{p}} ] \cket{\Phi},
    \end{eqnarray}
    where $\hat{H}$ is the Hamiltonian corresponding to the Lagrangian \eqref{Lagr}. Our task now is to transform Ehrenfest equation \eqref{rho_Ehrenfest}
    into an effective von Neumann equation for the density matrix
    \begin{equation}\label{Heisenberg_rho}
        \ii\frac{\pd \rho_{AB}(\bvec{p})}{\pd t} = h_{AC}(\bvec{p}) \rho_{CB}(\bvec{p}) - h_{CB}(\bvec{p}) \rho_{AC}(\bvec{p})
                                              \equiv [h(\bvec{p}), \rho(\bvec{p})]_{AB},
    \end{equation}
    where $h(\bvec{p})$ is the desired $2 N_{\text{f}}\times 2 N_{\text{f}}$ $c$-number matrix of the effective Hamiltonian, possibly depending
    on the density matrix $\rho$. In the derivation of the effective Hamiltonian, which follows, let us work in the Schroedinger picture and
    consider the contributions of the quadratic and quartic parts of $\hat{H}$ separately.

    In the ultrarelativistic approximation we are using, the Schroedinger-picture neutrino field operators $\nu_a(\bvec{x})$ read
    \begin{equation}\label{nu_operator}
        \nu_a(\bvec{x}) = \frac{1}{\sqrt{V}} \sum\limits_{\bvec{p}, \alpha}
                                        \{ u_\alpha(\bvec{p}) e^{\ii \bvec{p}\cdot \bvec{x}} \hat{a}_{a\alpha\bvec{p}} +
                                           u_{-\alpha}(\bvec{p}) e^{-\ii \bvec{p}\cdot \bvec{x}} \hat{a}\hc_{a\alpha\bvec{p}} \},
                                           \qquad a = 1,\ldots,N_{\text{f}},
    \end{equation}
    where the 3-momentum $\bvec{p} \in (2\pi / L)\mathds{Z}^3$ is quantized in the normalization volume $V \equiv L\times L \times L$.
    The neutrino annihilation operators $\hat{a}_{a\alpha\bvec{p}}$ enter together with the plane-wave solutions
    $u_{\alpha}(\bvec{p}) e^{\ii \bvec{p}\cdot \bvec{x}} / \sqrt{V}$ describing particles with helicity $\alpha/2 = \pm1/2$. In what
    follows, for brevity, we will refer to negative- and positive-helicity states as neutrino and antineutrino states,
    respectively, and also refer to $\alpha = \pm1$ as the helicity values, instead of the rigorous helicities $\alpha/2 = \pm1/2$.
    Some obvious expressions for the polarization bispinors $u_{\alpha}(\bvec{p})$, as well as for their bilinear combinations we
    will need below, are listed in Appendix~\ref{app:MajoranaIdentities}.

    The quadratic part of the Hamiltonian $\hat{H}$ in the Ehrenfest equation~\eqref{rho_Ehrenfest}, up to a insignificant
    multiple of the identity operator, can obviously be written as
    \begin{equation}\label{H2}
        \hat{H}^{(2)} = \sum\limits_{C,D,\bvec{q},\bvec{r}}
                        \lambda_{CD}(\bvec{q},\bvec{r}) \hat{a}\hc_{C\bvec{q}} \hat{a}_{D\bvec{r}}
                      + \hat{a}\hat{a} \text{ terms} + \hat{a}\hc\hat{a}\hc \text{ terms}.
    \end{equation}
    The second and third terms produce vanishing contributions to the Ehrenfest equation, since
    $\cket\Phi$ possesses definite particle numbers $N_{\bvec{p}}$ in all momentum modes.
    As for the particle number-conserving $\hat{a}\hc\hat{a}$ term, the corresponding contribution to the time derivative
    of the density matrix is evaluated straightforwardly
    \begin{eqnarray}
        \ii \frac{\pd\rho_{AB}(\bvec{p})}{\partial t} &\supset& -\brac{\Phi} [\hat{H}^{(2)}, \hat{a}\hc_{B\bvec{p}} \hat{a}_{A\bvec{p}} ] \cket{\Phi} =
        -\sum\limits_{C,D,\bvec{q},\bvec{r}}\lambda_{CD}(\bvec{q},\bvec{r})
        \brac\Phi
            \delta_{\bvec{r}\bvec{p}} \delta_{BD} \hat{a}\hc_{C\bvec{q}}\hat{a}_{A\bvec{p}}
            -\delta_{\bvec{q}\bvec{p}}\delta_{AC} \hat{a}\hc_{B\bvec{p}} \hat{a}_{D\bvec{r}}
        \cket\Phi \nonumber\\
        &=&
        [\lambda(\bvec{p},\bvec{p}), \rho(\bvec{p})]_{AB}
        \label{a_expValue}
    \end{eqnarray}
    and has the desired von Neumann structure~\eqref{Heisenberg_rho} (the $\supset$ sign means that quartic terms have been omitted on the r.h.s.).
    Forward scattering clearly manifests itself in only the diagonal ($\bvec{q} = \bvec{r}$) entries of the coefficient function $\lambda(\bvec{q}, \bvec{r})$
    affecting the evolution of the density matrix. Thus, to find the noncollective part of the desired effective Hamiltonian
    $h(\bvec{p})$ in Eq.~\eqref{Heisenberg_rho}, one should simply list the three $\hat{a}\hc\hat{a}$ terms in $\hat{H}^{(2)}$
    coming from the three Lagrangians $\Lagr^{(2)}_{\text{vac}}$, $\Lagr^{(2)}_{\text{mat}}$, and $\Lagr^{(2)}_{\text{AMM}}$
    describing the vacuum neutrino mixing and their interaction with matter and magnetic field.
    The first term gives nothing but a free Hamiltonian,
    \begin{eqnarray}\label{h_vac_indices}
        \hat{H}_{\text{vac}}^{(2)} &=& \sum\limits_{\bvec{q}, C} \sqrt{\bvec{q}^2 + m_c^2} \, \hat{a}\hc_{C\bvec{q}} \hat{a}_{C\bvec{q}}
        = \sum\limits_{\bvec{q}, C,D} \lambda_{CD}^{\text{vac}}(\bvec{q},\bvec{q}) \hat{a}\hc_{C\bvec{q}} \hat{a}_{D\bvec{q}}, \\
        \lambda^{\text{vac}}(\bvec{q},\bvec{q}) &=& \twoMatrix{\sqrt{\bvec{q}^2 + M^2}}{0}{0}{\sqrt{\bvec{q}^2 + M^2}}
            \approx |\bvec{q}|\twoMatrix{\idMatrix}{0}{0}{\idMatrix} + \twoMatrix{M^2/2|\bvec{q}|}{0}{0}{M^2/2|\bvec{q}|},
            \label{lambda_vac}
    \end{eqnarray}
    where $(M^2)_{cd} = m_c^2 \delta_{cd}$ is the neutrino mass-squared matrix (in the mass basis here) and, as usual, the term
    proportional to the identity matrix does not contribute to the commutator in Eq.~\eqref{a_expValue} and can be omitted.
    The matrix notation we have adopted in Eq.~\eqref{lambda_vac} and to be used hereinafter is as follows:
    the two block lines/columns of a $2 N_{\text{f}} \times 2 N_{\text{f}}$ matrix $\lambda^{\text{vac}}$ correspond to the two helicities
    $\gamma, \delta = -1, +1$, following a pattern
    \begin{equation}
        \rho \equiv \twoMatrix{\rho_{--}}{\rho_{-+}}{\rho_{+-}}{\rho_{++}}
             \equiv \twoMatrix{\rho_{\nu\nu}}{\rho_{\nu\bar\nu}}{\rho_{\bar\nu\nu}}{\rho_{\bar\nu\bar\nu}}.
    \end{equation}
    In other words, the first and the second diagonal blocks describe neutrinos and antineutrinos, respectively, while
    the two off-diagonal blocks describe coherent mixtures of neutrinos and antineutrinos.

    The matter (MSW) term in $\hat{H}^{(2)}$ can be retrieved by treating background matter and the corresponding $e, p, n$ operators
    in the mean-field fashion
    \begin{equation}\label{H_matter_general}
        \hat{H}_{\text{mat}}^{(2)} = -\int_V \Lagr^{(2)}_{\text{mat}} \diff^3x
            \to G_{\text{F}}\sqrt2 \, \avg{\mathds{J}^\mu_{ab}} \int_V \bar\nu_a \gamma_{\mu\text{L}} \nu_b \diff^3x.
    \end{equation}
    The  background matter current $\mathds{J}^\mu_{ab}$ above is directly extracted from Eq.~\eqref{L_mat} and after averaging
    over neutral nonmoving nonmagnetized matter gives
    \begin{eqnarray}
        \avg{\mathds{J}^\mu_{ab}} &=&
        -\delta_{ab} \Bigl\langle\bar{p} (2s_{\text{W}}^2 \gamma^\mu - \gamma^\mu_{\text{L}}) p -
                           \bar{e} (2s_{\text{W}}^2 \gamma^\mu - \gamma^\mu_{\text{L}}) e +
                           \bar{n}\gamma^\mu_{\text{L}}n
                     \Bigr\rangle
        + 2 (\mathds{P}_e)_{ab} \langle \bar{e} \gamma^\mu_{\text{L}}e \rangle
        \nonumber\\
        &=& \delta^\mu_0 \Bigl\{ n_e (\mathds{P}_e)_{ab} - \frac{n_n}{2} \delta_{ab} \Bigr\}, \label{J_matter_avg}
    \end{eqnarray}
    where the projector onto the electron neutrino state $(\mathds{P}_e)_{ab} \equiv U\cc_{fa} \cdot \delta_{f,e}\delta_{f',e}\cdot
    U_{f'b}$ and $n_{e,n}$ are the electron and neutron number densities, respectively.
    Indeed, expectations of all axial vectors vanish, while for polar vectors, $\avg{\bar\ell \gamma^\mu \ell} = \delta^\mu_0
    n_\ell$, $\ell = e, p, n$. It remains now to directly substitute the neutrino operators \eqref{nu_operator} into
    \eqref{H_matter_general} and extract the terms contributing to forward scattering
    (for the bilinear expressions of the form $\bar{u}(\ldots)u$, see Appendix~\ref{app:MajoranaIdentities})
    \begin{eqnarray}
        \hat{H}_{\text{mat}}^{(2)} &\supset& G_{\text{F}}\sqrt2 \sum_{\bvec{q},A,B}
        \avg{\mathds{J}^0_{ab}}
        \Bigl\{
            \bar{u}_\alpha(\bvec{q}) \gamma^0_{\text{L}} u_\beta(\bvec{q}) \hat{a}\hc_{A\bvec{q}} \hat{a}_{B\bvec{q}} -
            \bar{u}_{-\alpha}(\bvec{q}) \gamma^0_{\text{L}} u_{-\beta}(\bvec{q}) \hat{a}\hc_{B\bvec{q}} \hat{a}_{A\bvec{q}}
        \Bigr\} = \sum_{\bvec{q},A,B} \lambda_{AB}^{\text{mat}}(\bvec{q},\bvec{q}) \hat{a}\hc_{A\bvec{q}} \hat{a}_{B\bvec{q}},
        \\
        \lambda^{\text{mat}}(\bvec{q},\bvec{q}) &=&
            G_{\text{F}}\sqrt2 \twoMatrix{\avg{\mathds{J}^0}}{0}{0}{-\avg{\mathds{J}^0}^{\text{T}}}
            =G_{\text{F}}\sqrt2 \twoMatrix{n_e \mathds{P}_e - \frac{n_n}{2} \idMatrix}{0}{0}{-n_e \mathds{P}_e^{\text{T}} + \frac{n_n}{2}
            \idMatrix}. \label{lambda_mat}
    \end{eqnarray}

    Finally, the magnetic moment term is evaluated in the same way giving
    \begin{eqnarray}
        \hat{H}_{\text{AMM}}^{(2)} &=& -\frac{\ii}{2}\mathfrak{m}_{ab} \bvec{B}\cdot \int_V \bar\nu_a \bvec\Sigma \nu_b \diff^3x
        \supset
        \sum_{\bvec{q},A,B} \lambda_{AB}^{\text{AMM}}(\bvec{q},\bvec{q}) \hat{a}\hc_{A\bvec{q}} \hat{a}_{B\bvec{q}}, \\
        \lambda^{\text{AMM}}(\bvec{q},\bvec{q}) &=& -\ii\sqrt{2}
            \twoMatrix{0}{(\bvec\zeta_+(\bvec{p})\cdot\bvec{B})\mathfrak{m}}{(\bvec\zeta_-(\bvec{p})\cdot\bvec{B})\mathfrak{m}}{0},
            \label{lambda_AMM}
    \end{eqnarray}
    where complex vectors $\bvec\zeta_\pm(\bvec{p}) = \frac{1}{\sqrt2}\chi_\mp\hc(\bvec{p}) \bvec\sigma \chi_\pm(\bvec{p})$
    are matrix elements of the Pauli matrices between states with opposite helicities (see Appendix~\ref{app:MajoranaIdentities} for the definition details and
    properties of these vectors).

    \vsp
    Thus, we have found the noncollective part of the effective neutrino Hamiltonian to be
    $\lambda^{\text{vac}}(\bvec{p}, \bvec{p}) + \lambda^{\text{mat}}(\bvec{p}, \bvec{p}) + \lambda^{\text{AMM}}(\bvec{p}, \bvec{p})$, and we
    can now resort to the quartic terms in the neutrino Hamiltonian $\hat{H}$. These terms, after commutation with
    $\hat{a}\hc_{B\bvec{p}} \hat{a}_{A\bvec{p}}$ in the Ehrenfest equation \eqref{rho_Ehrenfest}, lead to quartic expectations of the form
    $\brac\Phi \hat{a}\hc \hat{a}\hc \hat{a} \hat{a} \cket\Phi$. In order to close our
    system of equations on the density matrix, namely, to obtain \eqref{Heisenberg_rho}, we will now apply a kind of Wick's theorem
    and express such quartic expectations in terms of quadratic ones
    $\brac\Phi \hat{a}\hc_{B\bvec{p}} \hat{a}_{A\bvec{p}}\cket\Phi = \rho_{AB}(\bvec{p})$.
    Let us start from the contribution of the electroweak Lagrangian \eqref{L4} having a V--A structure and
    insert the corresponding Hamiltonian into Eq.~\eqref{rho_Ehrenfest}
    \begin{equation}
        \ii\frac{\pd \rho_{AB}(\bvec{p})}{\pd t} \supset \ii\left(\frac{\pd \rho_{AB}(\bvec{p})}{\pd t}\right)_{\text{VA}} =
        -\frac{G_{\text{F}}}{\sqrt{2}} \int_V \diff^3x\,
        \brac\Phi
                \bigl[ \NProd{(\bar\nu_c(\bvec{x}) \gamma^\mu_{\text{L}} \nu_c(\bvec{x}))^2}, \; \hat{a}\hc_{B\bvec{p}} \hat{a}_{A\bvec{p}} \bigr]
        \cket\Phi.
    \end{equation}
    First of all, commutation with $\hat{a}\hc_{B\bvec{p}} \hat{a}_{A\bvec{p}}$ transforms
    creation and annihilation operators into creation and annihilation ones, respectively, i.e., it does not spoil the normal ordering.
    Thus, the commutator can safely be put inside the normal ordering, leading to an expectation of
    \begin{equation}\label{commutator1}
        \NProd{\bigl[(\bar\nu_c(\bvec{x}) \gamma^\mu_{\text{L}} \nu_c(\bvec{x}))^2, \; \hat{a}\hc_{B\bvec{p}} \hat{a}_{A\bvec{p}} \bigr]} =
        2 \NProd{\bar\nu_c(\bvec{x}) \gamma^\mu_{\text{L}} \nu_c(\bvec{x}) \;
        \bigl[\bar\nu_d(\bvec{x}) \gamma_{\mu\text{L}} \nu_d(\bvec{x}), \hat{a}\hc_{B\bvec{p}} \hat{a}_{A\bvec{p}} \bigr]}.
    \end{equation}
    Next, due to symmetries of bilinear expressions in Majorana fields (see, e.g., Ref.~\cite{Giunti_book})
    \begin{equation}\label{Majorana_bilinears_symmetry}
        \begin{array}{c}
            \NProd{\bar\psi\omega} = \NProd{\bar\omega\psi}, \quad
            \NProd{\bar\psi\gamma_5\omega} = \NProd{\bar\omega\gamma_5\psi}, \quad
            \NProd{\bar\psi\gamma^\mu\omega} = -\NProd{\bar\omega\gamma^\mu\psi}, \\
            \NProd{\bar\psi\gamma^\mu\gamma_5\omega} = \NProd{\bar\omega\gamma^\mu\gamma_5\psi}, \quad
            \NProd{\bar\psi\sigma_{\mu\nu}\omega} = -\NProd{\bar\omega\sigma_{\mu\nu}\psi},
        \end{array}
    \end{equation}
    the V--A currents can be replaced by axial vector ones,
    $\bar\nu_d(\bvec{x}) \gamma_{\mu\text{L}} \nu_d(\bvec{x}) = -\frac12 \bar\nu_d(\bvec{x}) \gamma_{\mu}\gamma_5 \nu_d(\bvec{x})$.
    Moreover, by virtue of the same symmetry properties, the latter one is a symmetric quadratic form in $\nu_d$, and its commutator with
    $\hat{a}\hc_{B\bvec{p}} \hat{a}_{A\bvec{p}}$ leads to two identical terms, yielding
    $-\bar\nu_d(\bvec{x}) \gamma_{\mu}\gamma_5 \bigl[\nu_d(\bvec{x}), \hat{a}\hc_{B\bvec{p}} \hat{a}_{A\bvec{p}} \bigr]$. As a result,
    the V--A contribution to the evolution equation for the density matrix takes the form
    \begin{eqnarray}
        \ii\left(\frac{\pd \rho_{AB}(\bvec{p})}{\pd t}\right)_{\text{VA}} &=&
        -\frac{G_{\text{F}}}{\sqrt{2}} \int_V \diff^3x\,
        \brac\Phi
                \NProd{
                        \bar\nu_c(\bvec{x}) \gamma^\mu\gamma_5 \nu_c(\bvec{x})
                        \cdot
                        \bar\nu_d(\bvec{x}) \gamma_{\mu}\gamma_5 \bigl[\nu_d(\bvec{x}), \hat{a}\hc_{B\bvec{p}} \hat{a}_{A\bvec{p}} \bigr]
                }
        \cket\Phi
        \nonumber\\
        &=&
            -\frac{G_{\text{F}}}{\sqrt{2V}} \int_V \diff^3x\,
            \brac\Phi
                    \NProd{
                            \bar\nu_c(\bvec{x}) \gamma^\mu\gamma_5 \nu_c(\bvec{x})
                            \cdot
                            \bar\nu_d(\bvec{x}) \gamma_{\mu}\gamma_5
                            \bigl\{ \delta_{bd} e^{\ii \bvec{p}\cdot\bvec{x}} u_\beta(\bvec{p}) \hat{a}_{A\bvec{p}} -
                                    \delta_{ad} e^{-\ii \bvec{p}\cdot\bvec{x}} u_{-\alpha}(\bvec{p}) \hat{a}\hc_{B\bvec{p}}
                            \bigr\}
                    }
            \cket\Phi \nonumber\\
        &\equiv& -\frac{G_{\text{F}}}{\sqrt{2V}} \int_V \diff^3x\,
            \brac\Phi
                    \NProd{
                            \bar\varphi \gamma^\mu\gamma_5 \chi
                            \cdot
                            \bar\psi \gamma_{\mu}\gamma_5 \omega
                    }
            \cket\Phi,
            \label{rhoDeriv_VA}
    \end{eqnarray}
    where we have symbolically denoted $\varphi = \chi = \nu_c(\bvec{x})$, $\psi = \nu_d(\bvec{x})$, and
    $\omega = \delta_{bd} e^{\ii \bvec{p}\cdot\bvec{x}} u_\beta(\bvec{p}) \hat{a}_{A\bvec{p}} - \delta_{ad} e^{-\ii \bvec{p}\cdot\bvec{x}} u_{-\alpha}(\bvec{p})
    \hat{a}\hc_{B\bvec{p}}$. A quartic expectation value we have encountered can now be (approximately) reduced to products
    of quadratic expectations, or contractions, by virtue of the Wick's theorem (see Appendix~\ref{app:WickTheorem} for
    details)
    \begin{equation}\label{Wick1}
        \brac\Phi
            \NProd{
                    \bar\varphi \gamma^\mu\gamma_5 \chi
                    \cdot
                    \bar\psi \gamma_{\mu}\gamma_5 \omega
            }
        \cket\Phi \approx
        \contraction[1.4ex]{}{\bar\varphi}{\gamma^\mu\gamma_5}{\chi}
        \contraction[0.8ex]{\bar\varphi \gamma^\mu\gamma_5 \chi \cdot}{\bar\psi}{\gamma_{\mu}\gamma_5}{\omega}{}
        \bar\varphi \gamma^\mu\gamma_5 \chi \cdot \bar\psi \gamma_{\mu}\gamma_5 \omega
        +
        \contraction[1.4ex]{}{\bar\varphi}{\gamma^\mu\gamma_5\chi \cdot}{\bar\psi}
        \contraction[1ex]{\bar\varphi \gamma^\mu\gamma_5}{\chi}{\cdot \bar\psi \gamma_{\mu}\gamma_5}{\omega}
        \bar\varphi \gamma^\mu\gamma_5 \chi \cdot \bar\psi \gamma_{\mu}\gamma_5 \omega
        +
        \contraction[1.5ex]{}{\bar\varphi}{\gamma^\mu\gamma_5 \chi \cdot \bar\psi \gamma_{\mu}\gamma_5}{\omega}
        \contraction[0.7ex]{\bar\varphi \gamma^\mu\gamma_5}{\psi}{\cdot}{\bar\psi}
        \bar\varphi \gamma^\mu\gamma_5 \chi \cdot \bar\psi \gamma_{\mu}\gamma_5 \omega,
    \end{equation}
    where a contraction of two spinor fields $\contractWithIndices{\varphi}{i}{\chi}{j} \equiv \brac\Phi \NProd{\varphi^i\chi^j} \cket\Phi$,
    with $i,j = 1,2,3,4$ numbering the components of a bispinor. To avoid using the index notation, however, we note that the second and the third
    terms in the above expression are equal due to~\eqref{Majorana_bilinears_symmetry} and apply the Fierz identity
    to them \cite{Fierz}, arriving at
    \begin{equation}\label{Fierz_VA}
        \brac\Phi
            \NProd{
                    \bar\varphi \gamma^\mu\gamma_5 \chi
                    \cdot
                    \bar\psi \gamma_{\mu}\gamma_5 \omega
            }
        \cket\Phi \approx
        \contraction[1.4ex]{}{\bar\varphi}{\gamma^\mu\gamma_5}{\chi}
        \contraction[0.8ex]{\bar\varphi \gamma^\mu\gamma_5 \chi \cdot}{\bar\psi}{\gamma_{\mu}\gamma_5}{\omega}{}
        \bar\varphi \gamma^\mu\gamma_5 \chi \cdot \bar\psi \gamma_{\mu}\gamma_5 \omega
        + 2\contractAMB{\bar\varphi}{}{\omega} \cdot \contractAMB{\bar\psi}{}{\chi}
        - 2\contractAMB{\bar\varphi}{\gamma_5}{\omega} \cdot \contractAMB{\bar\psi}{\gamma_5}{\chi}
        + \contractAMB{\bar\varphi}{\gamma^\mu}{\omega} \cdot \contractAMB{\bar\psi}{\gamma_\mu}{\chi}
        + \contractAMB{\bar\varphi}{\gamma^\mu \gamma_5}{\omega} \cdot \contractAMB{\bar\psi}{\gamma_\mu \gamma_5}{\chi}.
    \end{equation}
    Now, every of these contractions is a sum of expectation values of the form $\brac\Phi \hat{a}\hc \hat{a} \cket\Phi$,
    which are nothing but specific entries of the density matrix $\rho$, namely,
    \begin{eqnarray}
        \contractAMB{\bar\varphi}{\gamma^\mu\gamma_5}{\chi} &=& \frac{2}{V} \sum_{\bvec{q}, c, \gamma} \gamma (1,
        \hat{\bvec{q}}) \rho_{c\gamma, c\gamma}(\bvec{q}) \equiv -\frac{2}{V} \sum_{\bvec{q}} (1,\hat{\bvec{q}}) \tr\bigl(\rho(\bvec{q}) \mathcal{G}\bigr), \\
        \contractAMB{\bar\psi}{\gamma^\mu\gamma_5}{\omega} &=& \frac{(1,\hat{\bvec{p}})}{\sqrt{V}} (\beta - \alpha)
        \rho_{AB}(\bvec{p}) \equiv \frac{(1,\hat{\bvec{p}})}{\sqrt{V}} \bigl[\mathcal{G}, \rho(\bvec{p})\bigr]_{AB},\\
        \contractAMB{\bar\psi}{}{\chi} &=& \contractAMB{\bar\psi}{\gamma_5}{\chi} = 0, \label{contraction_SP} \\
        \contractAMB{\bar\psi}{\gamma^\mu}{\chi} &=& \frac{1}{V} \sum_{\bvec{q}, \gamma} (1, \hat{\bvec{q}}) \wp_{c\gamma, d\gamma}(\bvec{q}), \label{contractions4}\\
        \contractAMB{\bar\psi}{\gamma^\mu \gamma_5}{\chi} &=& \frac{1}{V} \sum_{\bvec{q}, \gamma} \gamma (1, \hat{\bvec{q}})
                                                              \wp_{c\gamma, d\gamma}(\bvec{q}), \label{contractions5}\\
        \contractAMB{\bar\varphi}{\gamma^\mu}{\omega} &=& \frac{(1,\hat{\bvec{p}})}{\sqrt{V}}
                                 \big\{\delta_{bd} \rho_{A,c\beta}(\bvec{p}) + \delta_{ad} \rho_{c\alpha, B}(\bvec{p})\big\}, \\
        \contractAMB{\bar\varphi}{\gamma^\mu\gamma_5}{\omega} &=& \frac{(1,\hat{\bvec{p}})}{\sqrt{V}}
                                 \big\{\beta \delta_{bd} \rho_{A,c\beta}(\bvec{p}) - \alpha \delta_{ad} \rho_{c\alpha,
                                 B}(\bvec{p})\big\}.
    \end{eqnarray}
    The above identities are obtained in a straightforward way from the neutrino operator \eqref{nu_operator} and the bilinears
    \eqref{u_bilinears_first}--\eqref{u_bilinears_last}. In the first two identities, we have introduced a matrix distinguishing neutrinos
    and antineutrinos
    \begin{equation}
        \mathcal{G} \equiv \twoMatrix{\idMatrix}{0}{0}{-\idMatrix}, \qquad \mathcal{G}_{c\gamma, d\delta} = -\gamma \delta_{cd}\delta_{\gamma\delta},
    \end{equation}
    while another matrix featuring in identities \eqref{contractions4} and \eqref{contractions5} is a difference between the
    density matrix and the charge conjugate of its transpose
    \begin{gather}
        \wp_{c\gamma, d\delta}(\bvec{q}) \equiv \rho_{c\gamma, d\delta}(\bvec{q}) - \rho_{d\, -\delta, c\, -\gamma}(\bvec{q}),
        \qquad
        \wp(\bvec{q}) = \rho(\bvec{q}) - \rho^{\text{cT}}(\bvec{q}), \\
        \rho^{\text{c}} \equiv \mathcal{C} \rho \mathcal{C}, \qquad
        \mathcal{C} \equiv \twoMatrix{0}{\idMatrix}{\idMatrix}{0}.
    \end{gather}
    The charge conjugation, as defined here, simply swaps the neutrino lines/columns with the antineutrino lines/columns of
    the density matrix.

    Finally, after substituting the contractions listed above into the Fierz transformed Wick's theorem
    \eqref{Fierz_VA} and using the latter expression in the Ehrenfest equation \eqref{rhoDeriv_VA},
    we obtain a contribution to the evolution of neutrino density matrix due to electroweak neutrino-neutrino interactions
    \begin{eqnarray}
        \ii\left(\frac{\pd \rho(\bvec{p})}{\pd t}\right)_{\text{VA}} &=& \Bigl[h_{\text{self}}^{\text{VA}}(\bvec{p}),
        \rho(\bvec{p})\Bigr], \\
        h_{\text{self}}^{\text{VA}}(\bvec{p}) &=& \frac{G_{\text{F}}\sqrt2}{V} \sum_{\bvec{q}} (1 - \hat{\bvec{p}} \cdot \hat{\bvec{q}})
        \; \Bigl\{ \tr\bigl(\rho(\bvec{q})\mathcal{G}\bigr) \mathcal{G} + \wp^\diagPart(\bvec{q}) \Bigr\}.
        \label{h_VA}
    \end{eqnarray}
    The $\wp^\diagPart(\bvec{q})$ matrix above is a block-diagonal part of $\wp(\bvec{q})$
    \begin{equation}\label{diagPart}
        \wp \equiv \twoMatrix{\wp_{\nu\nu}}{\wp_{\nu\bar\nu}}{\wp_{\bar\nu\nu}}{\wp_{\bar\nu\bar\nu}}, \qquad
        \wp^\diagPart \equiv \frac12(\wp + \mathcal{G} \wp \mathcal{G})
         = \twoMatrix{\wp_{\nu\nu}}{0}{0}{\wp_{\bar\nu\bar\nu}}.
    \end{equation}

    It is now time to find the term in the evolution equation coming from the Scalar (S) and Pseudoscalar (P) nonstandard neutrino interactions.
    This is done following exactly the same steps as in the V--A interaction case, but starting from Lagrangian~\eqref{L_NSSI}. Namely, the S/P contribution
    to the time derivative of the density matrix reads
    \begin{equation}
        \ii\left(\frac{\pd \rho_{AB}(\bvec{p})}{\pd t}\right)_{\text{SP}}
        = -\frac{G_{\text{F}}}{\sqrt{2V}} \int_V \diff^3x\,
            \brac\Phi
                    \NProd{
                            g_{\text{S}} \bar\varphi \chi \cdot \bar\psi \omega +
                            g_{\text{P}} \bar\varphi \gamma_5 \chi \cdot \bar\psi \gamma_5 \omega
                    }
            \cket\Phi,
            \label{rhoDeriv_SP}
    \end{equation}
    and after applying the Wick's theorem, we arrive at
    \begin{equation}\label{Wick2}
        \ii\left(\frac{\pd \rho_{AB}(\bvec{p})}{\pd t}\right)_{\text{SP}} =
        -\frac{G_{\text{F}}}{\sqrt{2V}} \int_V \diff^3x\,
            \left\{
                g_{\text{S}}\Bigl(
                                \contraction[1.4ex]{}{\bar\varphi}{}{\chi}
                                \contraction[0.8ex]{\bar\varphi\chi \cdot}{\bar\psi}{}{\omega}{}
                                \bar\varphi \chi \cdot \bar\psi \omega
                                +2
                                \contraction[1.5ex]{}{\bar\varphi}{\chi \cdot \bar\psi }{\omega}
                                \contraction[0.7ex]{\bar\varphi}{\psi}{\cdot}{\bar\psi}
                                \bar\varphi \chi \cdot \bar\psi \omega
                \Bigr)
                +
                g_{\text{P}}\Bigl(
                                \contraction[1.4ex]{}{\bar\varphi}{\gamma_5}{\chi}
                                \contraction[0.8ex]{\bar\varphi\gamma_5\chi \cdot}{\bar\psi}{\gamma_5}{\omega}{}
                                \bar\varphi \gamma_5 \chi \cdot \bar\psi \gamma_5 \omega
                                +2
                                \contraction[1.5ex]{}{\bar\varphi}{\gamma_5\chi \cdot \bar\psi \gamma_5}{\omega}
                                \contraction[0.7ex]{\bar\varphi\gamma_5}{\psi}{\cdot}{\bar\psi}
                                \bar\varphi \gamma_5 \chi \cdot \bar\psi \gamma_5 \omega
                \Bigr)
            \right\}.
    \end{equation}
    Note that the first contractions in each of the two parentheses vanish due to Eq.~\eqref{u_bilinears_SP}. For the second contractions,
    we use the Fierz identity \cite{Fierz}, omitting the vanishing scalar and pseudoscalar terms in it \eqref{contraction_SP}, and arrive at
    \begin{equation}\label{rhoDeriv_SP_2}
        \ii\left(\frac{\pd \rho_{AB}(\bvec{p})}{\pd t}\right)_{\text{SP}}
        =\frac{G_{\text{F}}}{\sqrt{2V}} \int_V \diff^3x\,
            \left\{
                  g_- \contractAMB{\bar\varphi}{\gamma^\mu}{\omega} \cdot \contractAMB{\bar\psi}{\gamma_\mu}{\chi}
                - g_- \contractAMB{\bar\varphi}{\gamma^\mu\gamma_5}{\omega} \cdot \contractAMB{\bar\psi}{\gamma_\mu\gamma_5}{\chi}
                + \frac{g_+}{2} \contractAMB{\bar\varphi}{\sigma^{\mu\nu}}{\omega} \cdot \contractAMB{\bar\psi}{\sigma_{\mu\nu}}{\chi}
            \right\},
    \end{equation}
    where $g_\pm \equiv (g_{\text{S}} \pm g_{\text{P}})/2$.
    In fact, the vector and the axial vector terms have already been evaluated above, so that it remains only to evaluate the
    tensor one
    \begin{eqnarray}
        \frac12\contractAMB{\bar\varphi}{\sigma^{\mu\nu}}{\omega} \cdot \contractAMB{\bar\psi}{\sigma_{\mu\nu}}{\chi} &=&
        \contractAMB{\bar\varphi}{\bvec\Sigma}{\omega} \cdot \contractAMB{\bar\psi}{\bvec\Sigma}{\chi}
        +\contractAMB{\bar\varphi}{\bvec\alpha}{\omega} \cdot \contractAMB{\bar\psi}{\bvec\alpha}{\chi}, \\
        \contractAMB{\bar\psi}{\bvec\Sigma}{\chi} &=& \frac{\sqrt{2}}{V} \sum_{\bvec{q}, \gamma} \bvec\zeta_\gamma(\bvec{q})
                                                     \wp_{c\gamma, d\, -\gamma}(\bvec{q}), \\
        \contractAMB{\bar\psi}{\bvec\alpha}{\chi} &=& \frac{\sqrt{2}}{V} \sum_{\bvec{q}, \gamma} \gamma\bvec\zeta_\gamma(\bvec{q})
                                                     \wp_{c\gamma, d\, -\gamma}(\bvec{q}), \\
        \contractAMB{\bar\varphi}{\bvec\Sigma}{\omega} &=& \sqrt{\frac{2}{V}}
                                                  \big\{\delta_{bd} \bvec\zeta_\beta(\bvec{p}) \rho_{A,c\,-\beta}(\bvec{p})
                                                   + \delta_{ad} \bvec\zeta_{-\alpha}(\bvec{p}) \rho_{c\, -\alpha, B}(\bvec{p}) \big\}, \\
        \contractAMB{\bar\varphi}{\bvec\alpha}{\omega} &=& \sqrt{\frac{2}{V}}
                                                  \big\{\delta_{bd} \beta \bvec\zeta_\beta(\bvec{p}) \rho_{A,c\,-\beta}(\bvec{p})
                                                   - \delta_{ad} \alpha\bvec\zeta_{-\alpha}(\bvec{p}) \rho_{c\, -\alpha, B}(\bvec{p})
                                                   \big\}.
    \end{eqnarray}
    Again, working in the same way as we did in the case of V--A interaction, we transform Eq.~\eqref{rhoDeriv_SP_2} into
    \begin{eqnarray}
        \ii\left(\frac{\pd \rho(\bvec{p})}{\pd t}\right)_{\text{SP}} &=& \Bigl[h_{\text{self}}^{\text{SP}}(\bvec{p}),
        \rho(\bvec{p})\Bigr], \\
        h_{\text{self}}^{\text{SP}}(\bvec{p}) &=& \frac{G_{\text{F}}\sqrt2}{V} \sum_{\bvec{q}}
        (1 - \hat{\bvec{p}} \cdot \hat{\bvec{q}})
        \Bigl\{
        g_- \bigl(\wp^\diagPart(\bvec{q}))^{\text{T}} + g_+ e^{\ii \Gamma(\hv{p},\hv{q}) \mathcal{G}} \bigl(\wp^\offdiagPart(\bvec{q}))^{\text{T}}
        \Bigl\},
        \label{h_SP}
    \end{eqnarray}
    where we have expressed the scalar products $\bvec\zeta_\pm(\bvec{p})\cdot \bvec\zeta_\pm(\bvec{p}) = e^{\pm \ii \Gamma(\hv{p},\hv{q})} (1 - \hv{p} \cdot \hv{q}) / 2$
    in terms of a single complex phase $\Gamma(\hv{p},\hv{q})$ (see Appendix~\ref{app:MajoranaIdentities}). The block-off-diagonal part of the matrix above is defined analogously to Eq.~\eqref{diagPart}
    \begin{equation}\label{offdiagPart}
        \wp^\offdiagPart \equiv \frac12(\wp - \mathcal{G} \wp \mathcal{G})
         = \twoMatrix{0}{\wp_{\nu\bar\nu}}{\wp_{\bar\nu\nu}}{0},\qquad
        \wp = \wp^\diagPart + \wp^\offdiagPart.
    \end{equation}
    In fact, this block-off-diagonal part anticommutes with $\mathcal{G}$, which underpins hermiticity of the $g_+$ part of the Hamiltonian.
    Further, the complex phase $\Gamma(\hv{p},\hv{q})$ comes from the phases included in the helicity eigenstates $\chi_\pm(\bvec{p}),
    \chi_\pm(\bvec{q})$; one would therefore like to rephase helicity eigenstates, thus getting rid of the $e^{\ii\Gamma \mathcal{G}}$ term in
    the NSSI Hamiltonian. However, this is impossible to achieve on the whole Bloch sphere $\hv{p}, \hv{q} \in S^2$,
    since the parallel transport of the helicity eigenstates across this sphere is characterized with a nonzero Berry curvature
    \cite{Urbantke}.

    As we observe, both in the electroweak and the nonstandard scalar-pseudoscalar case, the time derivative of the neutrino flavor
    density matrix $\rho(\bvec{p})$ can be cast into a von Neumann form \eqref{Heisenberg_rho}, i.e., a form of a commutator with a hermitian
    matrix $h(\bvec{p})$. Writing down all the contributions to this time derivative together
    (see Eqs.~\eqref{lambda_vac}, \eqref{lambda_mat}, \eqref{lambda_AMM}, \eqref{h_VA}, \eqref{h_SP}), we finally arrive at
    the desired evolution equation on $\rho$
    \begin{eqnarray}\label{rho_evolution}
        \ii\frac{\pd \rho(\bvec{p})}{\pd t} &=& \Bigl[h(\bvec{p}), \rho(\bvec{p})\Bigr],\\
        h(\bvec{p}) &=& h_{\text{vac}}(\bvec{p}) + h_{\text{mat}} + h_{\text{AMM}}(\bvec{p}) + h_{\text{self}}(\bvec{p})  \nonumber\\
                    &\equiv& \frac{1}{2|\bvec{p}|}\twoMatrix{M^2}{0}{0}{M^2} +
                    G_{\text{F}}\sqrt2 \twoMatrix{n_e \mathds{P}_e - \frac{n_n}{2} \idMatrix}{0}{0}{-n_e \mathds{P}_e^{\text{T}} + \frac{n_n}{2} \idMatrix}
                    -\ii\twoMatrix{0}{\mathfrak{m}\cdot B_\perp(\bvec{p})}{\mathfrak{m}\cdot B_\perp\cc(\bvec{p})}{0} \nonumber\\
                    &+&
                    \frac{G_{\text{F}}\sqrt2}{V} \sum_{\bvec{q}}
                    (1 - \hat{\bvec{p}} \cdot \hat{\bvec{q}})
                    \Bigl\{
                        \tr\bigl(\rho(\bvec{q})\mathcal{G}\bigr) \mathcal{G} + \wp^\diagPart(\bvec{q})
                        + g_- \bigl(\wp^\diagPart(\bvec{q}))^{\text{T}}
                        + g_+ e^{\ii\Gamma(\hv{p},\hv{q}) \mathcal{G}} \bigl(\wp^\offdiagPart(\bvec{q}))^{\text{T}}
                    \Bigr\},
        \label{rho_evolution_Hamiltonian}
    \end{eqnarray}
    where one can show that $B_\perp(\bvec{p}) \equiv \sqrt2 \bvec\zeta_+(\bvec{p})\cdot \bvec{B}$ is indeed a complex number with the
    absolute value equal to the strength of the magnetic field across the neutrino momentum (in accordance with the notation we
    use). The equations of motion possess a gauge freedom with respect to rephasing the helicity eigenstates
    \begin{gather}
        \rho(\bvec{p}) \to e^{\ii \gamma(\hv{p})\mathcal{G}/2} \rho(\bvec{p}) e^{-\ii \gamma(\hv{p})\mathcal{G}/2}, \\
        B_\perp(\bvec{p}) \to B_\perp(\bvec{p}) e^{\ii\gamma(\hv{p})}, \quad
        \Gamma(\hv{p}, \hv{q}) \to \Gamma(\hv{p}, \hv{q}) + \gamma(\hv{p}) + \gamma(\hv{q}),
        \label{gauge_transformation}
    \end{gather}
    where $\gamma(\hv{p})$ is a real function on a sphere describing the gauge transformation. This transformation virtually
    tunes the relative phases of neutrino and antineutrino states and becomes trivial for a block-diagonal density matrix, i.e.,
    if neutrinos do not coherently mix with antineutrinos.

    It is also instructive here to emphasize that the off-diagonal blocks of $\rho(\bvec{p})$ do not behave as scalars under rotations.
    Indeed, the spinor representation of
    a rotation $R(\bvec{n}\vartheta) \in SO(3)$ around the $\bvec{n}$ axis on $\vartheta$ radians has the form
    \begin{equation}
        \hat{U}(R(\bvec{n}\vartheta)) \nu_a(\bvec{x},t) \hat{U}\hc(R(\bvec{n}\vartheta)) =
        e^{\ii\bvec{n} \cdot \bvec\Sigma \vartheta / 2} \nu_a( R(\bvec{n}\vartheta) \bvec{x}, t),
    \end{equation}
    from which one readily obtains the transformation of the neutrino annihilation operators
    \begin{equation}
        \hat{U}(R(\bvec{n}\vartheta)) \hat{a}_{A\bvec{p}} \hat{U}\hc(R(\bvec{n}\vartheta))
        =
        e^{\ii \alpha\Xi(\bvec{p},\bvec{n}\vartheta)} \hat{a}_{A,R(\bvec{n}\vartheta)\bvec{p}},
    \end{equation}
    where the phase $\Xi$ is connected with the complex phases chosen in the helicity eigenstates,
    $e^{\ii \bvec{n} \cdot \bvec\sigma \vartheta / 2} \chi_\pm(R(\bvec{n}\vartheta)\bvec{p}) \equiv
    e^{\pm\ii\Xi(\bvec{p},\bvec{n}\vartheta)}\chi_\pm(\bvec{p})$. Now, if one rotates the neutrino state,
    $\cket\Phi \to \cket{\Phi'} = \hat{U}\hc(R(\bvec{n}\vartheta)) \cket\Phi$, the density matrix
    $\rho_{AB}(\bvec{p}) = \brac{\Phi} \hat{a}\hc_{B\bvec{p}} \hat{a}_{A\bvec{p}} \cket\Phi$ will transform as
    \begin{equation}
        \rho(\bvec{p}) \to \rho'(\bvec{p}) = e^{-\ii \Xi(\bvec{p},\bvec{n}\vartheta) \mathcal{G}}
        \rho(R(\bvec{n}\vartheta)\bvec{p}) e^{\ii \Xi(\bvec{p},\bvec{n}\vartheta) \mathcal{G}}.
    \end{equation}
    Just like in the case of a gauge transformation, the diagonal blocks are left intact, while the off-diagonal ones get
    rephased. In particular, one can demonstrate that for the neutrino-antineutrino block,
    \begin{equation}
        \rho'_{\nu\bar\nu}(\bvec{p}) \bvec\zeta_-(\bvec{p}) = \rho_{\nu\bar\nu}(R\bvec{p})\;R^{-1}\bvec\zeta_-(R\bvec{p}),
    \end{equation}
    i.e., its product with the $\bvec\zeta_-$ vector transforms as a vector field without additional phases.
    As a result, for an isotropic neutrino gas ($\rho'(\bvec{p}) = \rho(\bvec{p})$), the above equation tells us that
    $\rho_{\nu\bar\nu}(\bvec{p})\bvec\zeta_-(\bvec{p})$ is a spherically-symmetric vector field. However, because
    $\bvec{p}\cdot\bvec\zeta_-(\bvec{p}) = 0$ (see Appendix~\ref{app:MajoranaIdentities}),
    this requires $\rho_{\nu\bar\nu}(\bvec{p}) = 0$. That is, nontrivial neutrino-antineutrino mixing violates isotropy,
    so the NSSI Hamiltonian~\eqref{rho_evolution_Hamiltonian} loses its $g_+$ term when applied to an isotropic neutrino gas.

    \vsp

    In the next two sections, we will study the effect of the nontrivial neutrino NSSIs on the flavor evolution in
    the neutrino bulb model using the so-called single-angle scheme \cite{Duan2010_review}. To obtain the corresponding evolution equations,
    we focus on stationary processes with propagating neutrinos and make a replacement transforming the
    von Neumann equation \eqref{Heisenberg_rho} into a quantum Liouville equation
    \begin{equation}
        \frac{\pd\rho(\bvec{p};t)}{\pd t} \to \left(\frac{\pd }{\pd t} + \hv{p} \cdot \frac{\pd }{\pd \bvec{x}}\right)
        \rho(\bvec{p},\bvec{x}) = (\hv{p} \cdot \bvec\nabla) \rho(\bvec{p},\bvec{x}).
    \end{equation}
    Next, construction of the single-angle scheme should include introduction of the so-called geometric factor $\mathcal{D}(r / R_\nu)$,
    (partially) accounting for the non-spherically symmetric distribution of the neutrino numbers far from the neutrino sphere;
    even in the Standard-Model case, there exist several versions of it \cite{Duan2010_review, Duan2006_swaps, Dasgupta2008_GeomFactor}.
    In our case, since the same factor $1-\hv{p}\cdot\hv{q}$ appears in front of both the
    electroweak and NSSI $g_-$ terms in the collective part of the Hamiltonian \eqref{rho_evolution_Hamiltonian}, it is natural to equip
    their single-angle counterparts with the same geometric factor. Regarding the $g_+$ term having an extra $e^{\ii\Gamma(\hv{p},\hv{q}) \mathcal{G}}$
    factor, the same may not be the case, however, we will keep the same geometric factor, qualitatively relying upon the
    argument that far from the neutrino sphere the solid angle spanned by the neutrino momenta $\bvec{q}$ is small and one can
    rephase the helicity eigenstates within it to approximately eliminate the phase factor $e^{\ii\Gamma \mathcal{G}}$. A
    more rigorous discussion of the geometric factor for the $g_+$ term of the Hamiltonian is to be given elsewhere.

    With these points in mind, we can now write down the evolution equations of the single-angle scheme. Namely,
    in the flavor basis and the two-flavor approximation, a neutrino with energy $E$, escaping the protoneutron star radially from
    its neutrino sphere $r = R_{\nu}$ outwards, is described by equations
    \begin{eqnarray}\label{rho_evolution_singleAngle}
        \ii\frac{\pd \rho_E(r)}{\pd r} &=& \bigl[h_E(r), \rho_E(r)\bigr],\qquad
        h_E(r) = h_{\text{vac},E} + h_{\text{mat}}(r) + h_{\text{AMM}}(r) + h_{\text{self}}(r),\\
        h_{\text{vac},E} &=& \frac{\eta \Delta m^2}{4E}\twoMatrix{\mathbb{M}}{0}{0}{\mathbb{M}}, \quad
                    \mathbb{M} \equiv \twoMatrix{-\cos2\theta}{\sin2\theta}{\sin2\theta}{\cos2\theta}, \\
        h_{\text{mat}}(r) &=& G_{\text{F}}\sqrt2 \diag\Bigl(n_e(r) - \frac{n_n(r)}{2}, \; -\frac{n_n(r)}{2}, \; -n_e(r) + \frac{n_n(r)}{2}, \; \frac{n_n(r)}{2} \Bigr), \\
        h_{\text{AMM}}(r) &=& \twoMatrix{0}{\mu_{12} B_\perp(r)\sigma_2}{\mu_{12}B_\perp\cc(r)\sigma_2}{0}, \label{h_AMM_2flavors}\\
        h_{\text{self}}(r) &=&
                    G_{\text{F}}\sqrt2 \; \mathcal{D}\Bigl(\frac{r}{R_\nu}\Bigr) n_\nu(r) \int_0^\infty \diff{E'}
                    \; \Bigl\{ \tr\bigl(\rho_{E'}(r)\mathcal{G}\bigr) \mathcal{G} + \wp^\diagPart_{E'}(r)
                            +g_- \bigl(\wp^\diagPart_{E'}(r))^{\text{T}} + g_+ \bigl(\wp^\offdiagPart_{E'}(r))^{\text{T}}
                    \Bigr\}, \label{hSelf_our}
    \end{eqnarray}
    where $\Delta m^2$ is the mass squared difference, $\eta = \pm 1$ marks the normal/inverted mass hierarchy, $\theta$ is the vacuum mixing
    angle, and $\mu_{12}$ is the transition magnetic moment (note that the diagonal entries $\mu_{11} = \mu_{22} = 0$ for Majorana neutrinos \cite{Giunti2009_nuEMP});
    for brevity, we use the same notation $\rho$ for the density matrix in the flavor basis.
    To get rid of additional factors, we have renormalized the density matrix so that
    \begin{equation}\label{rho_normalization}
        \int_0^\infty \tr\rho_{E}(r) \diff{E} = 1;
    \end{equation}
    after such a renormalization, the total neutrino number density $n_\nu(r)$ appears in the nonlinear collective term. Note
    also that in the two-flavor case, a nontrivial Majorana phase can be eliminated from all terms but the $g_-$ NSSI term in
    the self-interaction~\eqref{hSelf_our} by a gauge transformation~\eqref{gauge_transformation} \; (for details, see
    Appendix~\ref{app:rho_flavor}). Thus, technically, the $g_-$ term in~\eqref{hSelf_our} should also contain a Majorana phase,
    however, we are not focusing on this term further and thus limit ourselves to a simplified expression ignoring it.

    The initial condition for the above system of equations is usually placed at the neutrino sphere, where different
    neutrino flavors/helicities are assumed to be thermalized with well-defined energy spectra
    \begin{equation}\label{rho_initial_condition}
        \rho_E(R_\nu) = \diag\Bigl( s_{\nu_e}(E), s_{\nu_x}(E), s_{\bar{\nu}_e}(E), s_{\bar{\nu}_x}(E)\Bigr),
    \end{equation}
    with the normalization convention \eqref{rho_normalization} requiring that
    \begin{equation}
        \int_0^\infty \bigl( s_{\nu_e}(E) + s_{\nu_x}(E) + s_{\bar{\nu}_e}(E) + s_{\bar{\nu}_x}(E) \bigr) \diff{E} = 1.
    \end{equation}

    \vsp

    Equations similar to the above ones for collective oscillations of Majorana neutrinos have been derived earlier in a number of
    contexts. First of all, V--A neutrino-neutrino interactions were included into the effective Hamiltonian in
    Ref.~\cite{Dvornikov2012_Heff}, allowing for a nontrivial structure of this interaction in the flavor space. The result we obtained here
    agrees with Ref.~\cite{Dvornikov2012_Heff} in the absence of nonstandard interactions ($g_\pm = 0$).
    In a recent paper \cite{NSSI_Yang2018_SPint}, the collective NSSI Hamiltonian was derived in the absence of the neutrino magnetic
    moment. One also observes agreement of the matrix structure of our effective Hamiltonian with this paper for $\mu_{12} = 0$,
    however, our derivation goes qualitatively beyond this special case. Namely, interaction with the external magnetic field
    via the magnetic moment introduces mixing between neutrinos and antineutrinos, so that their states cannot be described by two $N_{\text{f}}\times N_{\text{f}}$
    density matrices anymore, but a single $2N_{\text{f}}\times 2N_{\text{f}}$ matrix is necessary to refer to both neutrino and antineutrino
    flavors and their coherent mixtures. In fact, in the $\mu_{12} = 0$ (or $\bvec{B} = 0$) case, our density matrix becomes block-diagonal and
    its $\rho_{\nu\nu}$ and $\rho_{\bar\nu \bar\nu}$ blocks representing neutrinos and antineutrinos,
    respectively, do obey the evolution equations of the form~\cite{NSSI_Yang2018_SPint}.

    In fact, careful comparison of the coefficient in front of the $g_-$ NSSI term in our
    Hamiltonian~\eqref{rho_evolution_Hamiltonian} with Ref.~\cite{NSSI_Yang2018_SPint} reveals a two-fold discrepancy: our
    result is two times greater. We have analyzed the two derivations and reckon that the discrepancy comes from the very
    method, rather than from a mistake in the calculations. The authors of Ref.~\cite{NSSI_Yang2018_SPint} introduce a
    mean-field Hamiltonian $\mathcal{H}_{\text{MF}}$ by reducing quartic neutrino field products to partially
    contracted quadratic ones, analogously to our Wick's theorem~\eqref{Wick2}, and then use these quadratic operators
    in the effective evolution equation for the density matrix. In particular, being written in the notation we are using here,
    their equation~(9) with $\alpha = \beta = a, \; \xi = \eta = b$ virtually claims that
    $\NProd{\bar\nu_a \nu_a\bar\nu_b \nu_b} \approx 2\contractAMB{\bar\nu}{{}_a\gamma^\mu_{\text{L}}}{\nu}{}_b \; \NProd{\bar\nu_a\gamma_{\mu\text{L}}\nu_b}$,
    which can be transformed into
    $\contractAMB{\bar\nu}{{}_a}{\nu}{}_a \NProd{\bar\nu_b \nu_b} + \NProd{\bar\nu_a \contractAMB{\nu}{{}_a}{\bar\nu}{}_b \nu_b} + \NProd{\contractAMB{\bar\nu}{{}_a\nu_a}{\bar\nu}{}_b \nu_b}$
    using the Fierz identity. However, even though the expectation of this quadratic operator coincides with $\brac\Phi \NProd{\bar\nu_a \nu_a \bar\nu_b \nu_b}
    \cket\Phi$, this does not apply to its commutator with $\hat{a}\hc_{B\bvec{p}} \hat{a}_{A\bvec{p}}$ one needs in the
    evolution equation~\eqref{rho_Ehrenfest}: contracted operators are $c$-numbers and do not contribute to the
    commutator, while in Eq.~\eqref{rho_Ehrenfest} all the four neutrino operators in $\bar\nu_a \nu_a \bar\nu_b \nu_b$ do contribute.
    Roughly speaking, a mean-field expression for a quartic Hamiltonian $\mathcal{H}(\nu) = g \nu^4$ should come from
    its quadratic part around the mean value $\langle\nu\rangle$, i.e., $\mathcal{H}_{\text{MF}} = \mathcal{H}''(\langle\nu\rangle) (\delta\nu)^2 / 2
    = 6 g \langle\nu\rangle^2 (\delta\nu)^2$, where $\delta\nu \equiv \nu - \langle\nu\rangle$; this expression features six
    terms, in contrast to the three terms assumed in Ref.~\cite{NSSI_Yang2018_SPint}. In view of the above, we are inclined
    to treat our $g_-$ coefficient in the Hamiltonian \eqref{rho_evolution_Hamiltonian} as correct.

    Next, we should also mention two papers \cite{deGouvea2012_2013} by A.~de Gouv\^ea and S.~Shalgar, in
    which the single-angle scheme for collective oscillations of Majorana neutrinos was analyzed in the $\mu_{12} \ne 0$ case.
    The papers analyzed the electroweak (V--A) four-fermion neutrino interaction, arriving at the neutrino self-interaction
    Hamiltonian of the form
    \begin{equation}\label{hSelf_dGS}
        h_{\text{self}}^{\text{(dGS)}}(r) =
                    G_{\text{F}}\sqrt2 \mathcal{D}\Bigl(\frac{r}{R_\nu}\Bigr) n_\nu(r) \int_0^\infty \diff{E'}
                    \; \Bigl\{ \tr\bigl(\rho_{E'}(r)\mathcal{G}\bigr) \mathcal{G} + \mathcal{G}
                                \wp_{E'}(r) \mathcal{G}
                    \Bigr\}.
    \end{equation}
    Note that such an interaction, in principle, includes off-diagonal blocks mixing neutrino and antineutrino states. In
    contrast, our collective Hamiltonian \eqref{hSelf_our} does not have such blocks in the Standard-Model regime $g_\pm = 0$, even if the
    density matrix $\rho_E(r)$ contains them. As we will see below, this makes a considerable difference in the instability
    spectra of the two Hamiltonians and, as a result, in the evolution of the neutrino spectra in magnetic field. In a
    nutshell, collective oscillations governed by interaction Hamiltonian \eqref{hSelf_dGS} favor neutrino-antineutrino mixing
    due to the presence of nontrivial off-diagonal blocks, thus, a tiny mixing introduced into the system by the magnetic
    moment term is followed by its exponential growth due to instabilities; however, nothing like occurs in the Standard-Model
    regime of our evolution equations \eqref{rho_evolution_singleAngle}, since the only source of neutrino-antineutrino mixing is the
    (linear, noncollective) magnetic-moment term.

    In any case, despite the disagreement of the results of Ref.~\cite{deGouvea2012_2013} with ours,
    as well as with the earlier Ref.~\cite{Dvornikov2012_Heff}, the Hamiltonian presented in the former paper can be treated as
    yet another type of NSSI containing nontrivial off-diagonal blocks. Moreover, it is spectacular that both the equations
    from Ref.~\cite{deGouvea2012_2013} and our equations \eqref{rho_evolution_singleAngle} conserve
    the block-diagonality property of the density matrix $\rho_{E}(r)$ along the trajectory, provided that $\mu_{12} = 0$ and that
    the initial condition $\rho_E(R_\nu)$ is block-diagonal (e.g., the one we chose \eqref{rho_initial_condition}). In this case,
    $\rho = \rho^\diagPart = \diag(\rho_{\nu\nu}, \rho_{\bar\nu \bar\nu})$ and Hamiltonian \eqref{hSelf_dGS} can be replaced by
    \begin{equation}\label{hSelf_noAMM}
        h_{\text{self}}^{\text{(dGS)}}(r) \to
                    G_{\text{F}}\sqrt2 \mathcal{D}\Bigl(\frac{r}{R_\nu}\Bigr) n_\nu(r) \int_0^\infty \diff{E'}\;
                    \twoMatrix{\wp_{\nu\nu,E'}(r)}{0}{0}{\wp_{\bar\nu \bar\nu,E'}(r)},
    \end{equation}
    which is nothing but the well-known expression within the Standard Model, also valid for Dirac neutrinos \cite{Duan2010_review}.
    Our Hamiltonian \eqref{rho_evolution_singleAngle} also takes the above conventional form, when additionally, $g_- = 0$,
    i.e., the scalar and the pseudoscalar couplings are equal in the NSSI Lagrangian \eqref{L_NSSI}. If, in contrast, $g_- \ne 0$,
    then scalar-pseudoscalar NSSI terms are present in the diagonal blocks of the density matrix and survive in the $\mu_{12} =
    0$ regime. Their effect has recently been analyzed in Ref.~\cite{NSSI_Yang2018_SPint}, revealing considerable deviations of the
    neutrino spectra from the predictions of the electroweak theory for nonvanishing $g_-$. As a result, constraints
    can be placed on the $g_-$ coupling from the possible measurements of the neutrino spectra from a supernova, regardless of
    the magnitude of the neutrino magnetic moment.

    Our Hamiltonian, however, contains another, `hidden' sector with the coupling $g_+$, which produces no effect in the zero
    magnetic moment case. It is this coupling that is able to introduce block-off-diagonal terms into the self-interaction
    Hamiltonian \eqref{hSelf_our}, possibly leading to nontrivial evolution of neutrino-antineutrino mixing. In the special
    case $g_- = 0$, $g_+ = -1$, the corresponding interaction term takes a form resembling the Hamiltonian~\eqref{hSelf_dGS} from
    Ref.~\cite{deGouvea2012_2013}
    \begin{equation}\label{hSelf_SM_noAMM}
        h_{\text{self}}(r) =
                    G_{\text{F}}\sqrt2 \mathcal{D}\Bigl(\frac{r}{R_\nu}\Bigr) n_\nu(r) \int_0^\infty \diff{E'}
                    \; \Bigl\{ \tr\bigl(\rho_{E'}(r)\mathcal{G}\bigr) \mathcal{G} + \wp^\diagPart_{E'}(r)
                     - \bigl(\wp^\offdiagPart_{E'}(r))^{\text{T}}
                    \Bigr\}, \quad g_+ = -1
    \end{equation}
    (note that $\mathcal{G} \wp_{E'} \mathcal{G} = \wp^\diagPart_{E'} - \wp^\offdiagPart_{E'}$), yet, it differs
    in the transposition of the off-diagonal part of the density matrix. As we will see below, however, the transposition leads to qualitatively
    different effects, and we have to conclude that the interaction \eqref{hSelf_dGS} introduced in Ref.~\cite{deGouvea2012_2013}
    cannot originate from scalar/pseudoscalar four-fermion interactions of Majorana neutrinos.

    Finally, we would like to mention that we briefly analyzed the case of Dirac neutrinos in our recent paper \cite{Kharlanov2019},
    where coherent mixing between two (anti)neutrino helicity states was introduced by the magnetic moment. It turns out that
    the corresponding interaction Hamiltonian is also block-diagonal in the absence of NSSIs and, as a result, the effect of the
    magnetic moment on the neutrino spectra is suppressed, not being subject to instabilities. In the present paper, we are not
    focusing on Dirac neutrinos, and the analysis which follows will be restricted to the Majorana case.

    \section{Linear stability analysis}%
    \label{sec:Stability}%

    Before making numerical simulations of the single-angle scheme in various scenarios, it is instructive to take a
    look at the linear stability of the evolution equations. Our specific interest here regards the effect of a small
    neutrino magnetic moment; the crucial question is whether it can trigger new types of neutrino flavor instabilities
    that are hidden in the $\mu_{12} = 0$ regime. 

    As we mentioned above, when $\mu_{12} = 0$ and the initial condition is of the form \eqref{rho_initial_condition},
    the density matrix $\rho_E^{(0)}(r)$ retains its block-diagonal form during the whole evolution $r \ge R_\nu$. For such a
    matrix, the effect of a nonzero $g_+$ NSSI term is absent, while the effect of the block-diagonal $g_-$ term was studied in a
    recent paper \cite{NSSI_Yang2018_SPint}. Namely, in that paper, it was shown that the $g_-$ term is able to trigger instabilities even in
    the absence of neutrino magnetic moment. Now, on top of possibly nonzero $g_\pm$ couplings, let us switch on an infinitesimal magnetic moment
    $\mu_{12} \to 0$ and write down the equations of motion linearized in this small $\mu_{12}$ quantity
    \begin{gather}\label{rho_evolution_linearized}
        \frac{\pd \delta\rho_E}{\pd r} + \ii \bigl[h_{\text{vac},E} + h_{\text{mat}} + h_{\text{self}}^{(0)}, \delta\rho_E\bigr]
                                          + \ii \bigl[\delta h_{\text{self}}, \rho^{(0)}_E \bigr] = \mathcal{J}_E(r), \\
        \mathcal{J}_E(r) \equiv -\ii [h_{\text{AMM}}(r), \rho^{(0)}_E(r)] = \bigO(\mu_{12}),\\
        \rho_E(r) = \rho_E^{(0)}(r) + \delta\rho_E(r), \quad \delta\rho_E(R_\nu) = 0; \quad \delta\rho_E(r) = \bigO(\mu_{12}),
    \end{gather}
    where $h_{\text{self}}^{(0)}$ is the neutrino-neutrino Hamiltonian calculated for the density matrix $\rho_E^{(0)}$ and
    $\delta h_{\text{self}}$ is its variation resulting from a variation $\delta\rho_E$ of the density matrix. For both standard
    and nonstandard neutrino interaction terms, one can separate the block-diagonal and block-off-diagonal parts of the above
    equation
    \begin{eqnarray}\label{rho_evolution_linearized_split_eq1}
        \frac{\pd \delta\rho_E^\diagPart}{\pd r} + \ii \bigl[h_{\text{vac},E} + h_{\text{mat}} + h_{\text{self}}^{(0)}, \delta\rho_E^\diagPart \bigr]
                                          + \ii \bigl[\delta h_{\text{self}}^\diagPart, \rho^{(0)}_E \bigr] &=& 0, \\
        \frac{\pd \delta\rho_E^\offdiagPart}{\pd r} + \ii \bigl[h_{\text{vac},E} + h_{\text{mat}} + h_{\text{self}}^{(0)}, \delta\rho_E^\offdiagPart \bigr]
        + \ii \bigl[\delta h_{\text{self}}^\offdiagPart, \rho^{(0)}_E \bigr]
        &=& \mathcal{J}_E(r). \label{rho_evolution_linearized_split_eq2}
    \end{eqnarray}
    Indeed, $h_{\text{vac},E} + h_{\text{mat}} + h_{\text{self}}^{(0)}$ is a block-diagonal matrix and so commutation with it does not mix the
    diagonal with off-diagonal contributions; the same applies to commutation with $\rho^{(0)}_E$; finally, the `source' term
    $\mathcal{J}_E$ is purely block-off-diagonal because of the form of $h_{\text{AMM}}$. Note also that block-diagonal and
    block-off-diagonal parts of $\delta h_{\text{self}}$ are determined by the corresponding parts of $\delta\rho_E$, so that
    evolutions of $\delta\rho_E^\diagPart(r)$ and $\delta\rho_E^\offdiagPart(r)$, defined by the two above equations,
    do indeed decouple in the linear regime.
    Technically, this means, that one can search for block-diagonal and block-off-diagonal unstable modes separately.

    If one sets $g_+$ to zero, i.e., turns off the block-off-diagonal part of the interaction, then the second equation becomes
    \begin{equation}
        \frac{\pd \delta\rho_E^\offdiagPart}{\pd r} + \ii \bigl[h_{\text{vac},E} + h_{\text{mat}} + h_{\text{self}}^{(0)}, \delta\rho_E^\offdiagPart \bigr]
        = \mathcal{J}_E(r),
    \end{equation}
    which features a fixed Hamiltonian $h_{\text{vac},E} + h_{\text{mat}} + h_{\text{self}}^{(0)}$. In other words,
    this equation virtually describes a noncollective flavor evolution and it is straightforward to show that
    \begin{equation}\label{deltaRho_offdiag_bound}
        \|\delta\rho_E^\offdiagPart(r)\|_{\text{F}} \le \int_{R_\nu}^r \|\mathcal{J}_E(r')\| \diff{r'},
    \end{equation}
    where $\|\mathcal{M}\|_{\text{F}} = \sqrt{\tr(\mathcal{M}\hc\mathcal{M})}$ is the Frobenius norm of a matrix and $\|\mathcal{J}_E(r')\|$
    is the spectral norm. The latter being equal to $|\mu_{12} B_\perp(r')| \cdot \|\rho_E^{(0)}(r')\|$ up to a factor of order unity,
    we conclude that block-off-diagonal perturbations $\delta\rho^\offdiagPart$ grow linearly with distance, not exhibiting exponential
    growth. As for the block-diagonal part $\delta\rho^\diagPart$ of the perturbation, it obeys the same equation
    \eqref{rho_evolution_linearized_split_eq1} as in the $\mu_{12} = 0$ case, when the density matrix is exactly
    block-diagonal at all $r$. Instabilities are well-known to exist here, both in the absence of NSSIs, $g_+ = g_- = 0$ \cite{Duan2010_review},
    and in their presence, $g_+ = 0, g_- \ne 0$ \cite{NSSI_Yang2018_SPint}, but they contain no signatures of a nonzero magnetic moment in the linear regime.

    We are thus able to conclude that in the absence of a scalar-pseudoscalar NSSI, i.e., within the electroweak theory,
    a tiny magnetic moment of a Majorana neutrino does not trigger new unstable modes that could result in observable signatures
    in the neutrino spectra. In order to leave such a signature, the transition magnetic moment should be at least of the order of
    $1/|B_\perp| L$, where $L$ is the distance covered by the neutrino (see Eq.~\eqref{deltaRho_offdiag_bound}); for $L \sim R_\nu \sim 50~\text{km}$
    and $|B_\perp| \sim 10^{12}~\text{Gauss}$, this amounts to about $0.5\times 10^{-15}\mu_{\text{B}}$. In the next section, we will
    show numerically that this statement also holds true beyond the linear stability regime. Note also that the argument that has led to such a
    conclusion was based only on the block structure of the Hamiltonian matrix, thus, it can be repeated beyond the
    single-angle scheme, i.e., for the Hamiltonian \eqref{rho_evolution_Hamiltonian}.

\begin{figure}[tbh]
        \includegraphics[width=6.5cm]{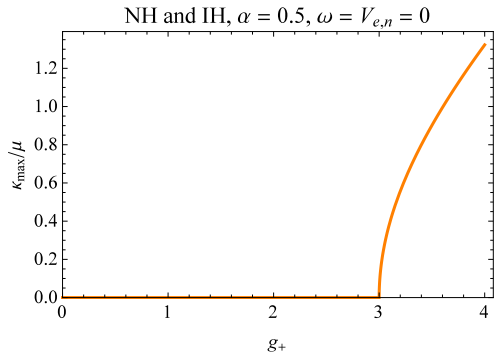}
        \caption{Growth rates corresponding to a fast unstable mode for a monochromatic neutrino flux, depending on the NSSI coupling $g_+$.
                 The plot demonstrates the asymptotic case with both the vacuum oscillation frequency $\omega$
                 and the matter potentials $V_{e,n}$ neglected}
        \label{fig:instabilityRates_mono_fast}
    \end{figure}

    Let us now turn on the block-off-diagonal NSSI, $g_+ \ne 0$. Even in this case, Eq.~\eqref{rho_evolution_linearized_split_eq1} tells us
    that the fate of block-diagonal perturbations is determined solely by $g_-$ and provides no signatures of a nonzero $\mu_{12}$.
    The block-off-diagonal modes follow Eq.~\eqref{rho_evolution_linearized_split_eq2}, which, in general, has to be treated numerically. However, to probe the
    possibility of unstable modes, it is usually instructive to carry out a Lyapunov stability analysis. For that, firstly,
    the source term in the evolution equation~\eqref{rho_evolution_linearized_split_eq2} is omitted but is replaced by a nontrivial
    initial perturbation $\delta\rho^\offdiagPart_E(R_\nu)$ on top of a diagonal reference matrix $\rho^{(0)}_E(R_\nu)$
    of the form~\eqref{rho_initial_condition}. Secondly, assuming that the mixing angle is virtually zero and (for
    our toy model) that neutrinos are monochromatic with energy $E_0$, we arrive at a constant nonperturbed solution
    \begin{equation}
        \rho^{(0)}_E(r) = \rho_E^{(0)}(R_\nu) \equiv \varrho^{(0)} \delta(E-E_0), \quad
        \varrho^{(0)} = \diag\Bigl( s_{\nu_e}, s_{\nu_x}, s_{\bar{\nu}_e}, s_{\bar{\nu}_x}\Bigr).
    \end{equation}
    Finally, let us assume that inhomogeneities of the matter and neutrino number densities are negligible at the typical growth scale of unstable perturbations.
    Under these assumptions, we arrive at a linear homogeneous equation with constant coefficients governing the growth of perturbation
    $\delta\rho^\offdiagPart_E(r) \equiv \delta\varrho(r) \delta(E-E_0)$
    \begin{gather}\label{EoM_varrho}
        \ii\frac{\pd \delta\varrho}{\pd r} = \mathbb{L}(\delta\varrho), \qquad
        \mathbb{L}(\delta\varrho) \equiv \bigl[h_{\text{vac},E_0} + h_{\text{mat}} + h_{\text{self}}^{(0)}, \delta\varrho \bigr]
        + \bigl[\delta h_{\text{self}}^\offdiagPart, \varrho^{(0)} \bigr],\\
        \delta h_{\text{self}}^\offdiagPart = \mu g_+ (\delta\varrho^{\text{T}} - \delta\varrho^{\text{c}}),
        \label{EoM_varrho_deltah}
    \end{gather}
    where $\mu = G_{\text{F}} \sqrt2 \mathcal{D}(r / R_\nu) n_\nu(r)$ is the neutrino-neutrino coupling strength. Now the
    question of linear stability reduces to existence of non-real eigenvalues of a linear map $\mathbb{L}$.
    Parameterizing the (block-off-diagonal) perturbation matrix as
    \begin{equation}
        \delta\varrho \equiv \begin{pmatrix}
                                0 & 0 & \xi_5 & \xi_1 \\
                                0 & 0 & \xi_2 & \xi_6 \\
                                \xi_7 & \xi_3 & 0 & 0 \\
                                \xi_4 & \xi_8 & 0 & 0
                             \end{pmatrix},
    \end{equation}
    one writes the eigenvalue equation for $\mathbb{L}(\delta\varrho) = \lambda \delta\varrho$ in a vectorized form
    \begin{gather}
        \mathbb{L}_{\text{vec}} \xi = \lambda \xi, \quad
        \mathbb{L}_{\text{vec}} \equiv \begin{pmatrix}
                                          \mathbb{L}_{\text{A}} & 0 \\
                                          0              & \mathbb{L}_{\text{B}}
                                      \end{pmatrix}, \\
        \mathbb{L}_{\text{A}} = \begin{pmatrix}
                                \Omega_- + 3\mu(\Delta s_e + \Delta s_{x}) & 0 & \mu g_+(s_{\nu_e}-s_{\bar\nu_x}) & -\mu g_+(s_{\nu_e}-s_{\bar\nu_x}) \\
                                0 & \Omega_+ + 3\mu(\Delta s_e + \Delta s_{x}) & -\mu g_+(s_{\nu_x}-s_{\bar\nu_e}) & \mu g_+(s_{\nu_x}-s_{\bar\nu_e})  \\
                                -\mu g_+(s_{\nu_x}-s_{\bar\nu_e}) & \mu g_+(s_{\nu_x}-s_{\bar\nu_e}) & -\Omega_+ - 3\mu(\Delta s_e + \Delta s_{x}) & 0 \\
                                \mu g_+(s_{\nu_e}-s_{\bar\nu_x}) & -\mu g_+(s_{\nu_e}-s_{\bar\nu_x}) & 0 & -\Omega_- - 3\mu(\Delta s_e + \Delta s_{x})
                             \end{pmatrix},
        \\
        \mathbb{L}_{\text{B}} = 2 G_{\text{F}}\sqrt{2}
                         \diag\Bigl(n_e - \frac{n_n}{2}, - \frac{n_n}{2}, -n_e + \frac{n_n}{2}, \frac{n_n}{2}\Bigr)
            \nonumber\\
            \qquad\qquad\qquad\qquad\quad
            + 2\mu
            \diag\Bigl(
                \Delta s_e + \Delta s_{x},
                \Delta s_e + 2\Delta s_{x},
                -2\Delta s_e + \Delta s_{x},
                -\Delta s_e - 2\Delta s_{x}
            \Bigr),
    \end{gather}
    where $\xi = (\xi_1,\xi_2,\ldots,\xi_8)^{\text{T}}$, $\Delta s_{e,x} \equiv s_{\nu_{e,x}} - s_{\bar\nu_{e,x}}$,
    $\Omega_\pm = G_{\text{F}}\sqrt2 (n_e - n_n) \pm \eta \omega$, $\omega = \Delta m^2 / 2 E_0$ is the vacuum oscillation frequency, and
    $\eta = \pm1$ is the mass hierarchy. The diagonal block $\mathbb{L}_{\text{B}}$ has real eigenvalues,
    while the eigenvalues of a non-hermitian $4\times 4$ matrix $\mathbb{L}_{\text{A}}$
    may be complex if the initial neutrinos possess a flavor imbalance, $s_{\nu_{e,x}} \ne s_{\bar\nu_{x,e}}$.
    Further we will plot the instability rates $\kappa_{\text{max}}$, i.e., the maximum imaginary parts of the eigenvalues of $\mathbb{L}_A$,
    for $s_{\nu_e} = 1$, $s_{\bar\nu_e} = \alpha \ge 0$, and $s_{\nu_x} = s_{\bar\nu_x} = 0$, changing the neutrino intensity
    parameter $\mu$ and the NSSI coupling $g_+$. In fact, for the matter potential $V_{e,n} \equiv G_{\text{F}}\sqrt2 n_{e,n}$
    set to zero, the eigenvalues of $\mathbb{L}_A$ can be found analytically,
    \begin{eqnarray*}
        \lambda_{1,2,3,4} &=& \pm \sqrt{\omega^2 + (1 - \alpha)^2 (9-g_+^2/2)\mu^2 \pm \mu \sqrt{D}}, \\
        D &=& g_+^4 (1-\alpha)^4 \mu^2/4 + 4[9(1-\alpha)^2 + \alpha g_+^2] \omega^2 + 6 g_+^2 (1+\alpha)(1-\alpha)^2 \eta \omega \mu.
    \end{eqnarray*}
    From this equation, one observes that for a strong NSSI coupling $|g_+| > 3$, instabilities survive even in the ultradense
    neutrino gas regime $\omega = V_{e,n} = 0$, their growth rate being $\kappa_{\text{max}} = \sqrt{g_+^2 - 9} \; \mu |1-\alpha|$. This
    corresponds to a so-called \emph{fast} unstable mode (see, e.g., Refs.~\cite{Glas2020_FastInstabilities, Johns2020_FastInstabilities, Capozzi2020_FastInstabilities}),
    whose growth rates for $\omega, V_{e,n} \ll \mu$ are of the order of $\mu$.
    It is well known that there are no such modes in the single-angle scheme with purely electroweak (V--A)
    neutrino-neutrino interactions: all instabilities disappear as $\omega \to 0$ \cite{Duan2010_review}. Moreover,
    interestingly, the unstable mode we have obtained survives in the absence of antineutrinos ($\alpha = 0$),
    whereas Standard-Model interactions do not generate instabilities in this case, within the monochromatic setup being discussed.
    In contrast, scalar-pseudoscalar NSSI, as we see, does generate a fast unstable mode even for monochromatic neutrinos,
    and its growth rate is plotted in Fig.~\ref{fig:instabilityRates_mono_fast}. However, we will not discuss this mode here, since we are more interested in unstable
    modes present for a small NSSI coupling. These come from an eigenvalue branch with $\lim\limits_{\omega \to 0} \IIm\lambda(\omega) = 0$,
    thus, these instabilities are \emph{slow}, i.e., suppressed in the $\omega \ll \mu$ regime.

\begin{figure}[tbh]
        $\begin{array}{ccc}
            \includegraphics[width=5.5cm]{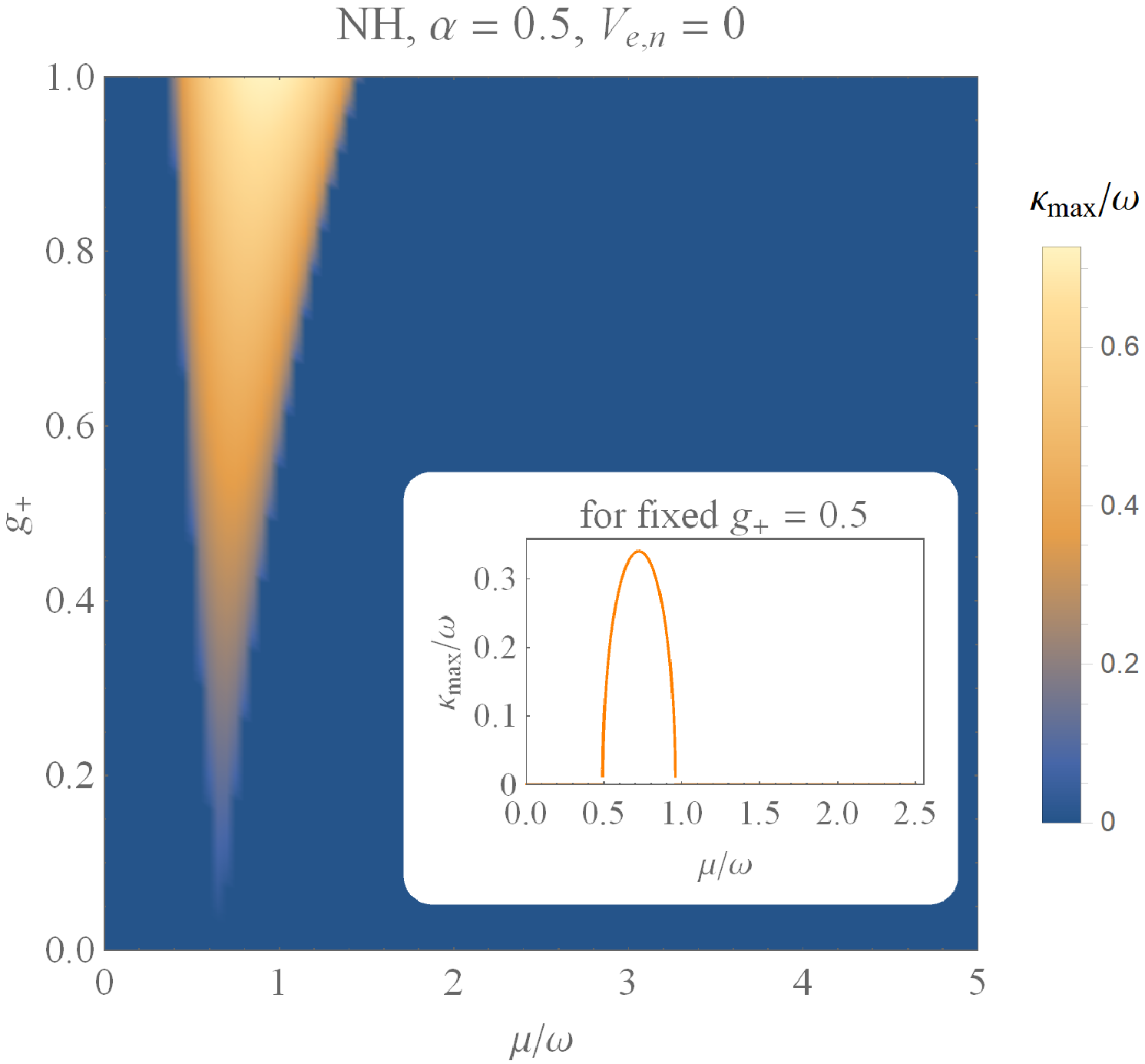} &
            \includegraphics[width=5.5cm]{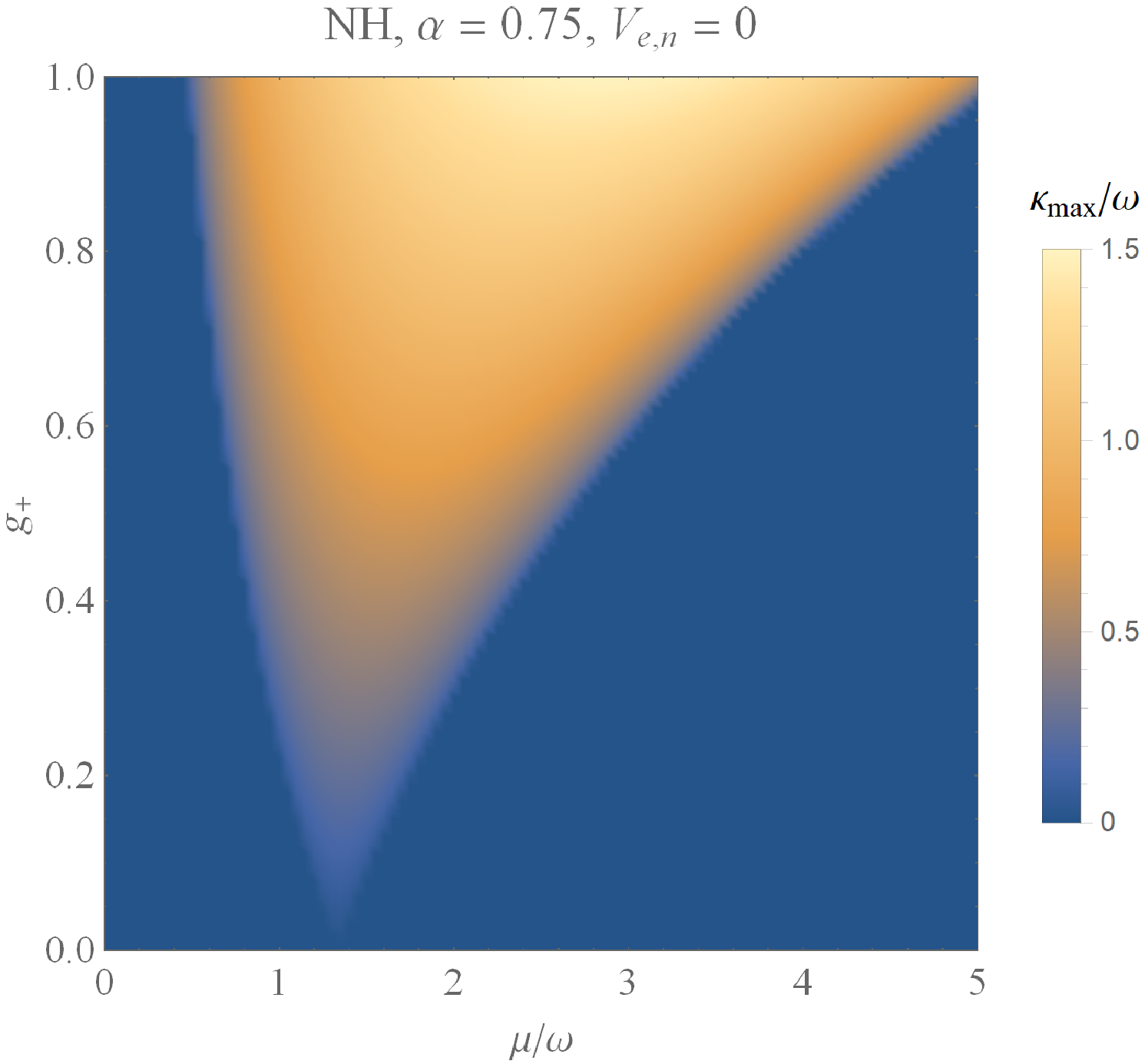} &
            \includegraphics[width=5.5cm]{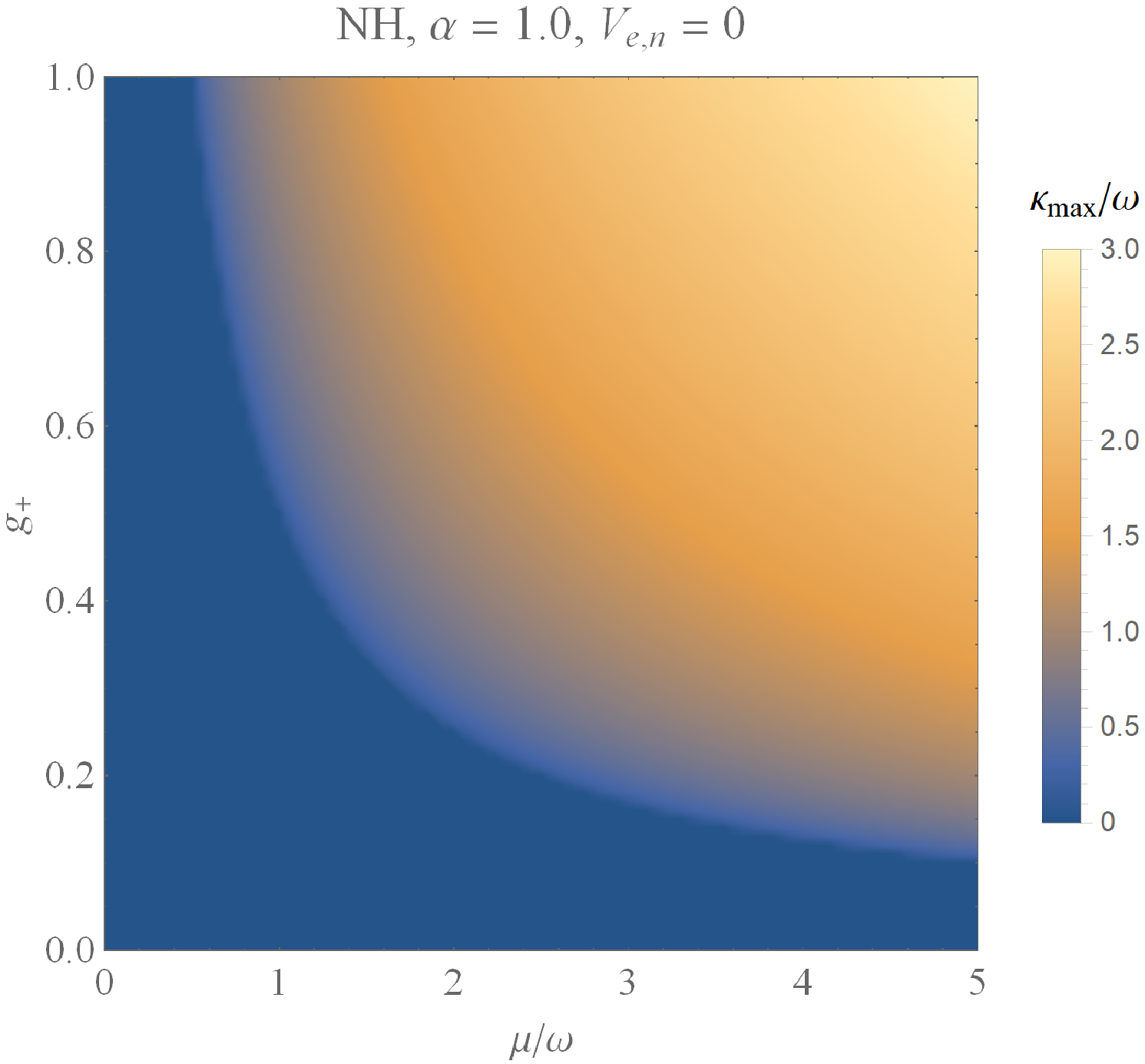} \\
            \includegraphics[width=5.5cm]{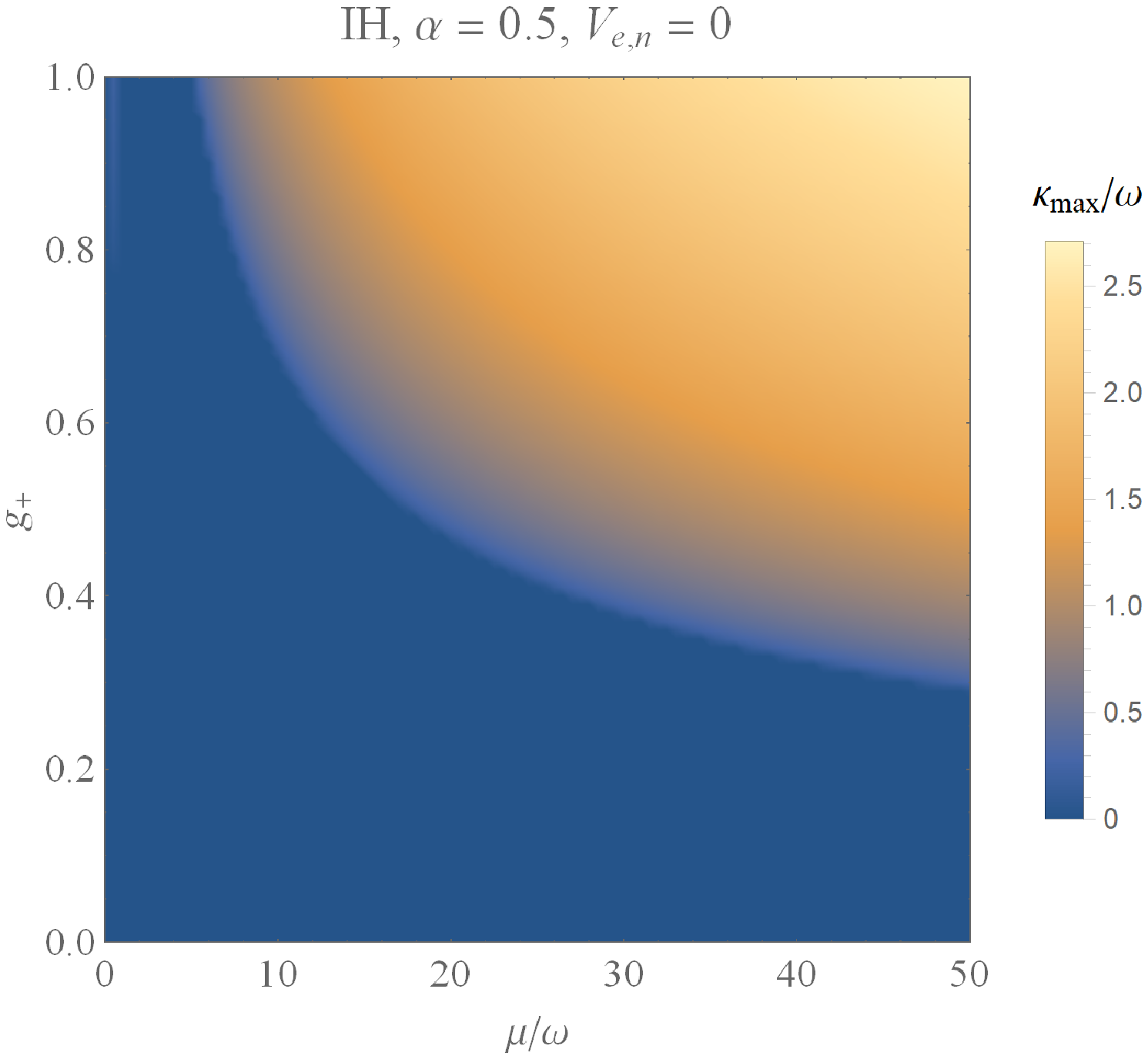} &
            \includegraphics[width=5.5cm]{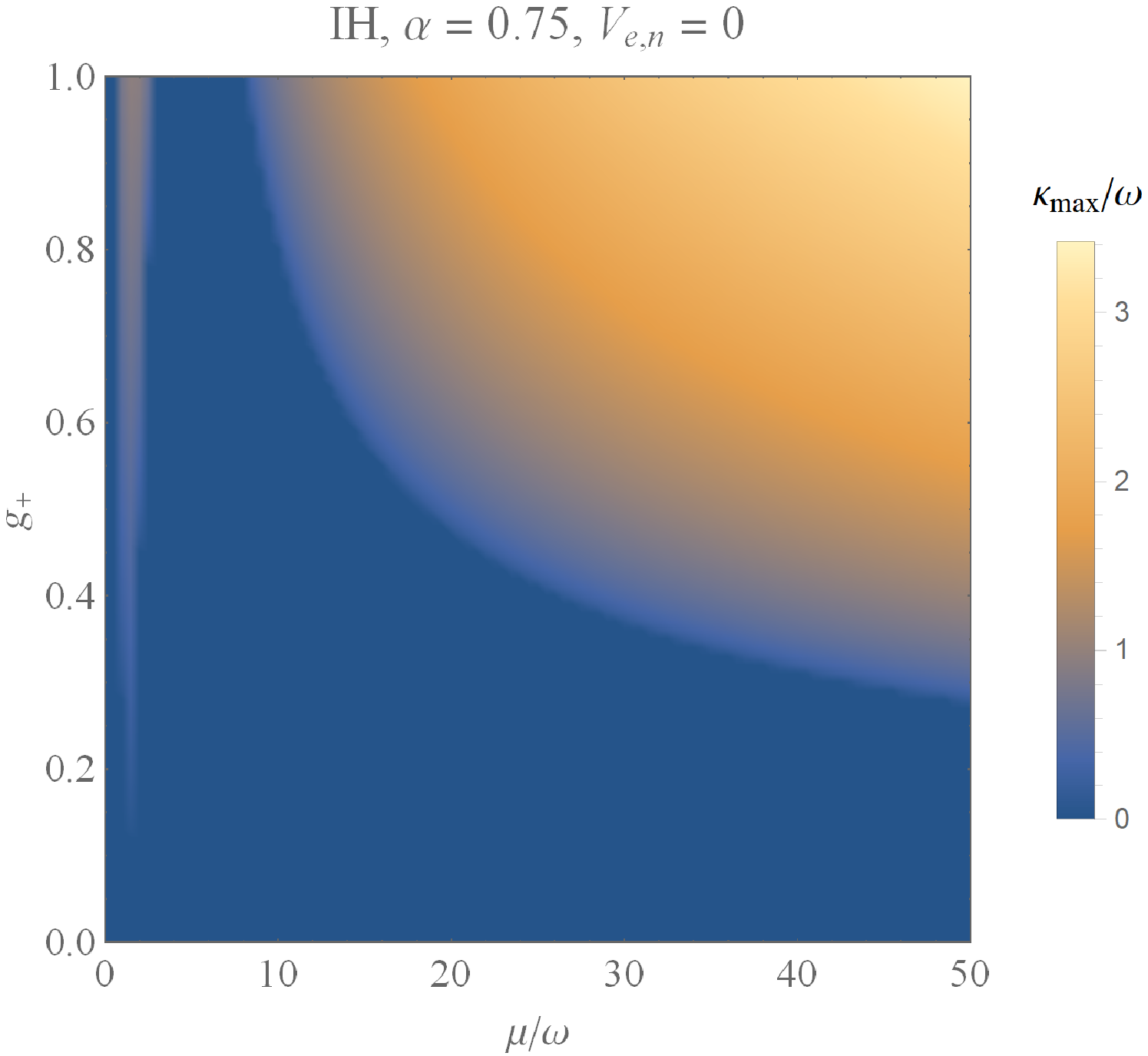} &
            \includegraphics[width=5.5cm]{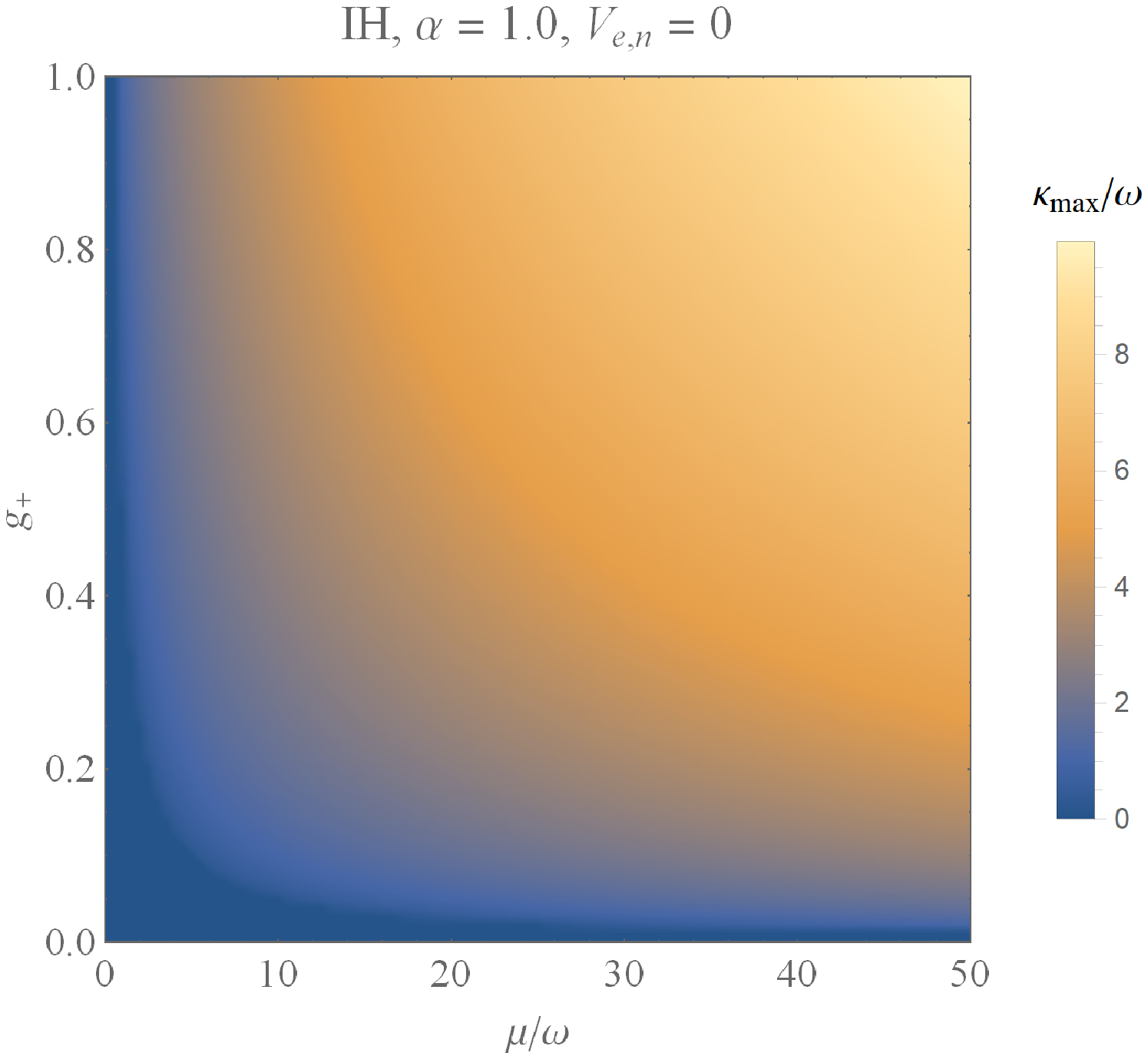}
        \end{array}$
        \caption{Instability growth rates for a monochromatic neutrino flux due to presence of a block-off-diagonal
                NSSI with parameter $g_+$. Inset: instability growth rate as a function of the neutrino number density
                for a fixed NSSI coupling $g_+ = 0.5$. The background matter density is set to zero in all panels}
        \label{fig:instabilityRates_mono}
    \end{figure}

\begin{figure}[tbh]
        \begin{center}
            \includegraphics[width=6cm]{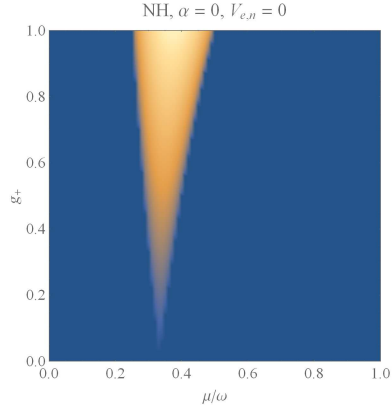}
        \end{center}
        \caption{Instability growth rates for a monochromatic purely electron neutrino flux due to presence of a block-off-diagonal
                NSSI with parameter $g_+$. The background matter density is set to zero}
        \label{fig:instabilityRates_mono_nu_e}
    \end{figure}

    The instability rates for the slow modes are plotted in Figs.~\ref{fig:instabilityRates_mono}, \ref{fig:instabilityRates_mono_nu_e}
    for different neutrino/antineutrino ratios $\alpha$ and both hierarchies. One observes that presence of antineutrinos considerably
    enhances the instability region; for normal hierarchy ($\eta = +1$), in fact, the widest instability `sector' in the $(\mu/\omega, g_+)$
    plane corresponds to $\alpha = 1$. This sector touches the $g_+ = 0$ line at a resonant neutrino density $\mu = \omega / 3\vert 1-\alpha \vert$,
    i.e., for such a density, the studied type of instabilities arises for arbitrarily small $g_+$. The $\alpha = 0$ case, corresponding to
    flavor instabilities of a monochromatic purely $\nu_e$ flux, is worth special attention: as we mentioned above, instabilities are absent
    here in the conventional V--A case, while NSSI triggers them for the normal hierarchy \emph{only} near the resonance $\mu \sim \omega / 3$ (Fig.~\ref{fig:instabilityRates_mono_nu_e}).
    This effect may have implications for the flavor evolution of predominantly electron
    neutrinos produced during the early neutronization phase of a supernova explosion \cite{HandbookOfSN}. In the presence of antineutrinos,
    instabilities do arise for both mass hierarchies, but the patterns for the inverted one are quite different
    (see the lower row in Fig.~\ref{fig:instabilityRates_mono}), also presenting another, high-neutrino-density instability region.
    In both cases, as one observes, these \emph{slow} instabilities should grow at typical scales smaller than the vacuum oscillation length, i.e., at
    several kilometers or even smaller.

\begin{figure}[tbh]
        \includegraphics[width=6cm]{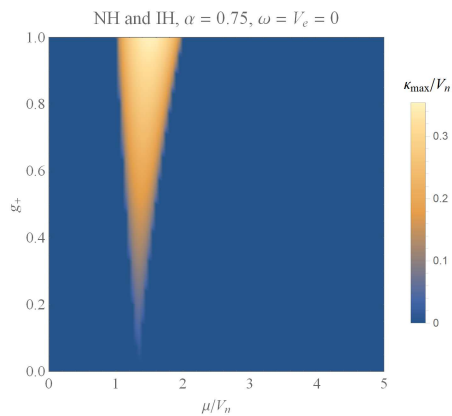}
        \caption{Instability growth rates for a monochromatic neutrino flux due to presence of a block-off-diagonal
                NSSI interaction with parameter $g_+$. In contrast to Fig.~\ref{fig:instabilityRates_mono}, a neutron
                background is added with Wolfenstein potential $V_n$, while the contribution of
                background electrons $V_e$ and the vacuum oscillation frequency $\omega$ are neglected}
        \label{fig:instabilityRates_mono_MSW}
    \end{figure}

    Let us now discuss the effect of background matter, i.e., nonzero $V_{e,n}$ potentials, on the stability properties of our system.
    Interestingly, in the presence of neutrino-antineutrino NSSI coupling, a transformation to a corotating frame in the
    flavor space does not let one eliminate the MSW term, as it does for $g_\pm = 0$ \cite{Duan2010_review}. Namely, a substitution
    \begin{equation}
        \delta\varrho(r) = \exp\left\{-\ii \int_{R_\nu}^r h_{\text{mat}}(r') \diff{r'} \right\} \;
                           \delta\tilde\varrho(r) \;
                           \exp\left\{\ii \int_{R_\nu}^r h_{\text{mat}}(r') \diff{r'} \right\}
    \end{equation}
    into the equation of motion~\eqref{EoM_varrho} for the instability does not totally `hide' the MSW term into the unitary
    transformation because of the matrix structure of the block-off-diagonal NSSI interaction \eqref{EoM_varrho_deltah}. Note that
    both the electroweak ($g_\pm = 0$) interaction Hamiltonian and the Hamiltonian~\eqref{hSelf_dGS} from
    Ref.~\cite{deGouvea2012_2013} feature a combination of the form $\rho - \rho^{\text{cT}}$ instead of a
    scalar-pseudoscalar $\rho^{\text{T}} - \rho^{\text{c}}$ in Eq.~\eqref{EoM_varrho_deltah}, which lets one eliminate the MSW
    term by a unitary transformation to a corotating frame. Physically, this results in the fact that in the presence of NSSI,
    background matter modifies the instability rates; in particular, one observes that even when $\omega = 0$ and $V_{e,n} \ne 0$,
    there are `intermediate' (neither fast, nor slow) unstable modes whose rates $\kappa$ are proportional to $V_{e,n}$.
    Indeed, in a typical supernova situation, $\mu \gtrsim V_{e,n} \gg \omega$ near the neutrino sphere, thus, these
    should be considerably faster than the \emph{slow} ones. Correspondingly, the two possibly non-real eigenvalues in the $\omega = 0$
    case read
    \begin{equation}
        \lambda_{1,2} = \pm \sqrt{(V_e - V_n + 3 (1-\alpha)\mu)^2 - g_+^2 (1-\alpha)^2 \mu^2}.
    \end{equation}
    In particular, from this expression, it follows that for a small NSSI coupling, instabilities take place in a narrow
    resonance region around $\mu = (V_n - V_e) / 3 (1 - \alpha)$. In Fig.~\ref{fig:instabilityRates_mono_MSW}, we demonstrate
    their growth rates for an idealized situation $\omega = V_e = 0$; the pattern looks quite similar to
    Fig.~\ref{fig:instabilityRates_mono}, but now the rates are measured in $V_n$ instead of $\omega$, i.e., they are higher.

    As we see, the block-off-diagonal NSSI coming from the scalar and pseudoscalar terms in the Lagrangian may lead to fast,
    slow, and intermediate instabilities that remain hidden in the absence of neutrino magnetic moment (or external magnetic field).
    Within the purely electroweak interaction model, these modes are also absent.

    \section{Numerical simulation}\label{sec:Simulation}

    Lyapunov stability analysis carried out in the previous section provides an insight into the effect of a small magnetic
    moment of a Majorana neutrino on collective oscillations, however, it is based on a number of artificial assumptions
    limiting the status of its conclusions. Moreover, in reality, one is interested in non-negligible NSSI/magnetic moment signatures in
    the observable neutrino energy spectra, which is beyond the regime of linear perturbations on top of a stationary
    nonperturbed solution. To explore the issue and estimate the sensitivities of the neutrino spectra to
    the magnetic moment, in the present section we carry out a numerical analysis of collective oscillations with
    the block-off-diagonal NSSI within the single-angle scheme.

\begin{figure}[tbh]
        $\begin{array}{cc}
            \includegraphics[height=5cm]{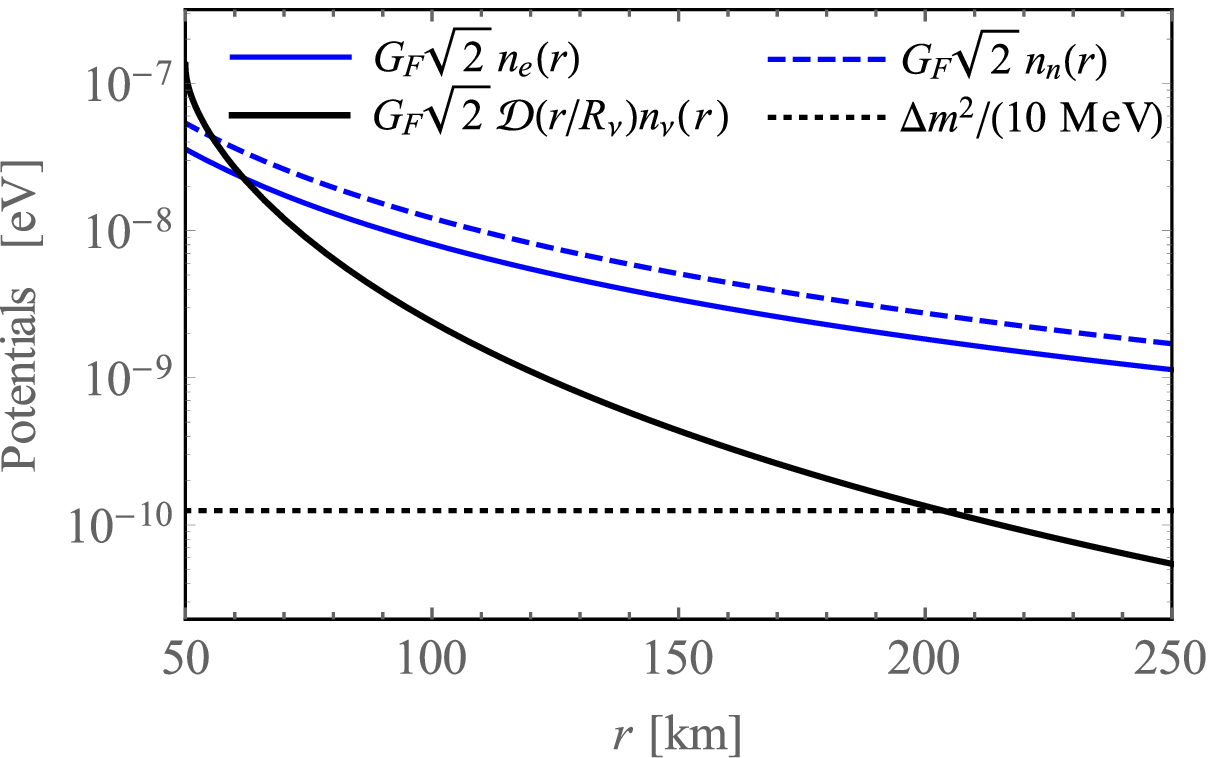} & \includegraphics[height=5cm]{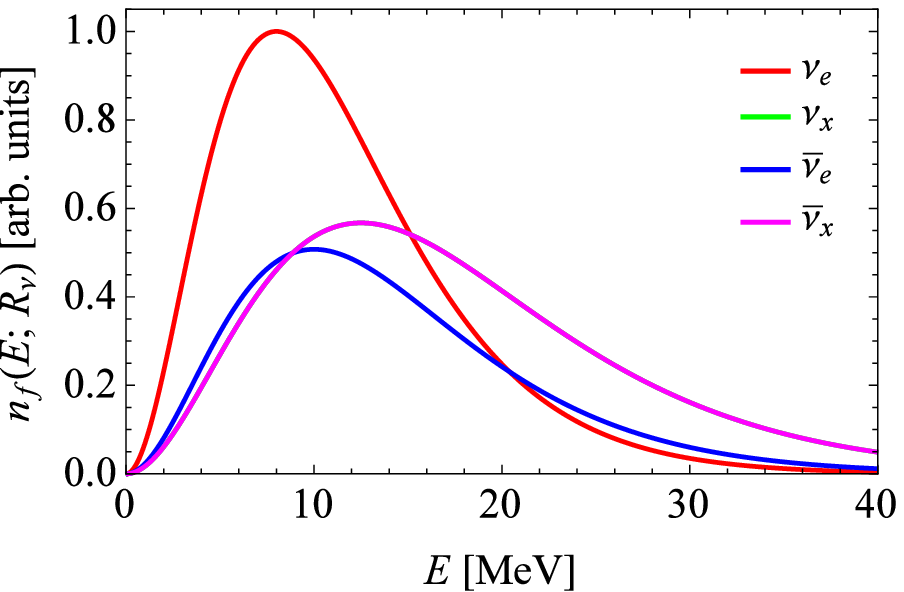} \\
            \text{(a)} & \text{(b)}
        \end{array}$
        \caption{(a) Comparison of matter (neutron, electron) potentials, the neutrino self-coupling, and the energy scale of
                 vacuum oscillations for a star with luminosity $L = 10^{55} \text{ sec}^{-1}$;
                 (b) the initial flavor/energy spectra at the neutrino sphere $r = R_\nu$ used in the simulations }
        \label{fig:potentials_initSpectra}
    \end{figure}

    For our simulation, we work with two neutrino flavors with $\Delta m^2 = 2.4\times 10^{-3}\text{ eV}^2$ and $\theta = 9\degree$ \cite{PDG}.
    The supernova setup is analogous to the one used, e.g., in \cite{deGouvea2012_2013}. Namely, a neutrino is leaving the neutrino sphere
    with the radius $R_\nu = 50\text{ km}$ radially in the equatorial plane, so that the transversal component of the magnetic field
    decays as an inverse power law with $r$,
    \begin{equation}
        B_\perp(r) = B_{\text{surf}} \left(\frac{R_\nu}{r}\right)^2, \quad B_{\text{surf}} = 10^{12}\text{ Gauss}.
    \end{equation}
    The neutrino density profile and the geometric factor together form an effective neutrino-neutrino coupling
    \begin{equation}
        \mu(r) \equiv G_{\text{F}} \sqrt{2} \; n_\nu(r) \; \mathcal{D}(r / R_\nu) = G_{\text{F}} \sqrt{2} \times \frac{L}{2\pi R_\nu^2} \times
        \frac12\Bigl( 1 - \sqrt{1 - (R_\nu / r)^2} \Bigr)^2,
    \end{equation}
    where $L$ is the neutrino luminosity parameter, namely, the number of emitted neutrinos per second (we follow the convention
    of Ref.~\cite{Duan2010_review} dividing it by $2\pi R_\nu^2$). By default we take $L = 10^{55}\text{ sec}^{-1}$ corresponding to the total radiation power
    $L \times 10\text{ MeV} \sim 1.5\times 10^{50}\text{ erg/sec}$.
    The electron density is chosen following the single-angle simulations in Ref.~\cite{deGouvea2012_2013}, while we also add
    neutrons with the density $n_n = 1.5 n_e$. The potentials $V_{e,n}(r) = G_{\text{F}} \sqrt2 n_{e,n}(r)$ and the above neutrino-neutrino coupling
    are shown in Fig.~\ref{fig:potentials_initSpectra}a, together with the energy scale $\omega = \Delta m^2 / (2\times 10\text{ MeV})$ of vacuum oscillations.
    At the neutrino sphere, the flavor density matrix is diagonal (see Eq.~\eqref{rho_initial_condition}),
    with the energy spectra taken from a simulation \cite{Keil2003_InitialSpectra} (see Fig.~\ref{fig:potentials_initSpectra}b);
    these spectra are nothing but Fermi distributions for the three different decoupling temperatures
    of $\nu_e$, $\bar\nu_e$, and $\nu_x/\bar\nu_x$. Upon evolution \eqref{rho_evolution_singleAngle}, the flavor/energy spectra
    are extracted from the density matrix with the help of a flavor/helicity projector $\mathcal{P}_{f}$
    \begin{equation}\label{flavorProbabilities}
        n_f(E; r) = \tr(\rho_E(r)\mathcal{P}_f) = \bigl(\rho_E(r)\bigr)_{f,f}, \quad f = e, x, \bar{e}, \bar{x} \; (e-, x-, e+, x+).
    \end{equation}

    We study the region $R_\nu \le r \lesssim 250 \text{ km}$ far from the MSW resonance ($V_{e,n} \gg \Delta m^2 \cos2\theta / 2E$),
    in which the oscillations are mainly driven by collective effects. Moreover, in reality, the neutrino self-coupling falls off quite rapidly,
    so that the oscillations virtually freeze around $r = 200-250\text{ km}$ producing well-defined `final' neutrino spectra on exit
    from the region. It is these spectra produced by nonlinear, collective effects that we are interested in within our
    analysis of NSSI-induced flavor instabilities.

\begin{figure}[tbh]
        $\begin{array}{ccc}
            \includegraphics[height=4cm]{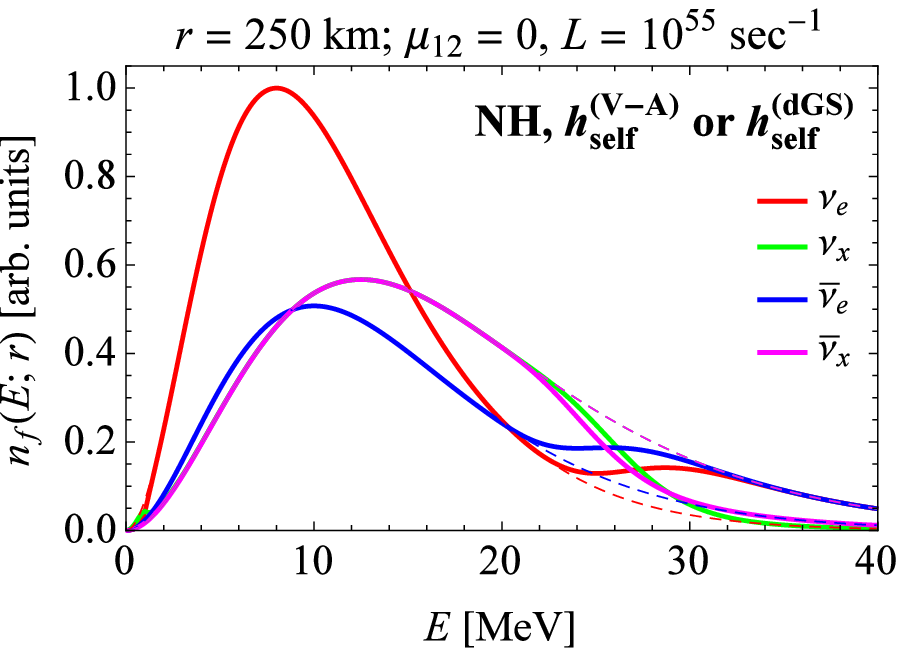} &
            \includegraphics[height=4cm]{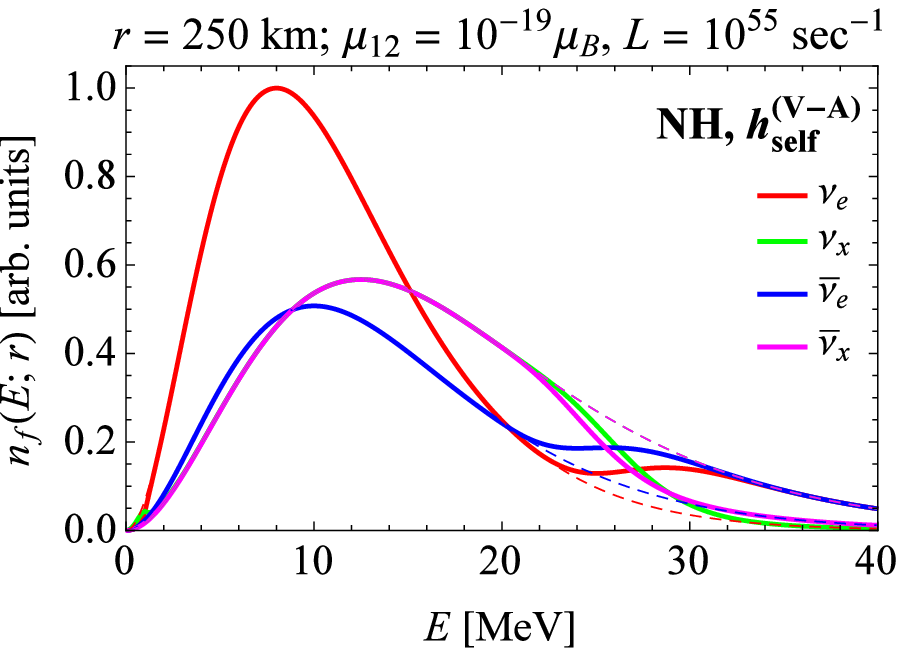} &
            \includegraphics[height=4cm]{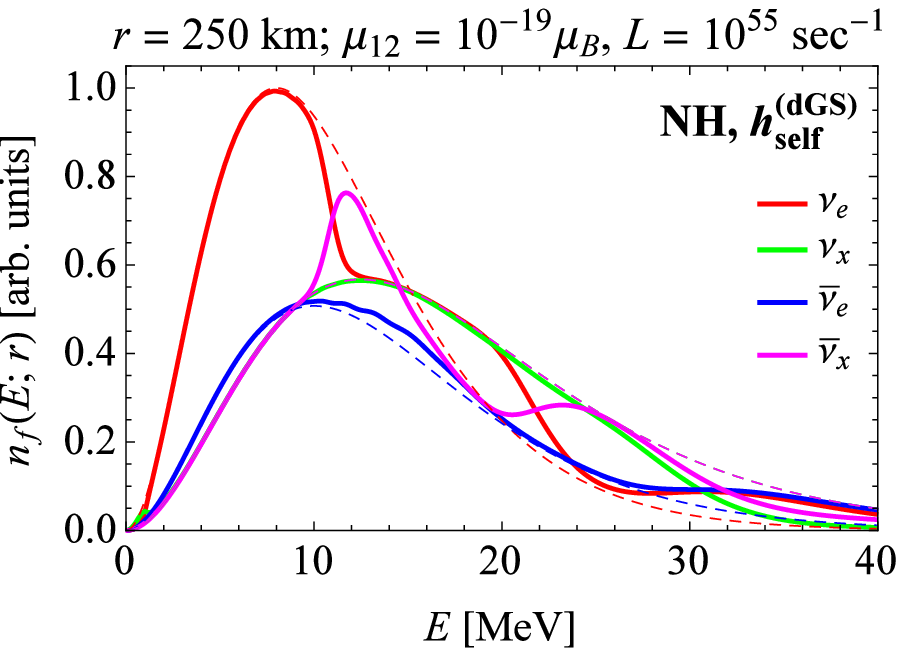} \\[0.5em]
            \text{(a)} & \text{(b)} & \text{(c)}            \\[1em]
            \includegraphics[height=4cm]{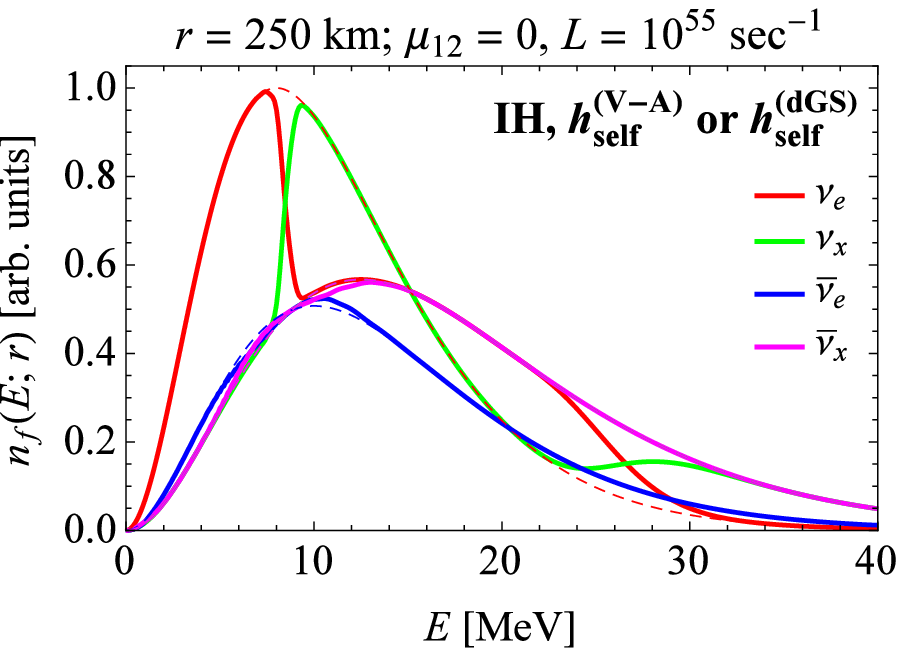} &
            \includegraphics[height=4cm]{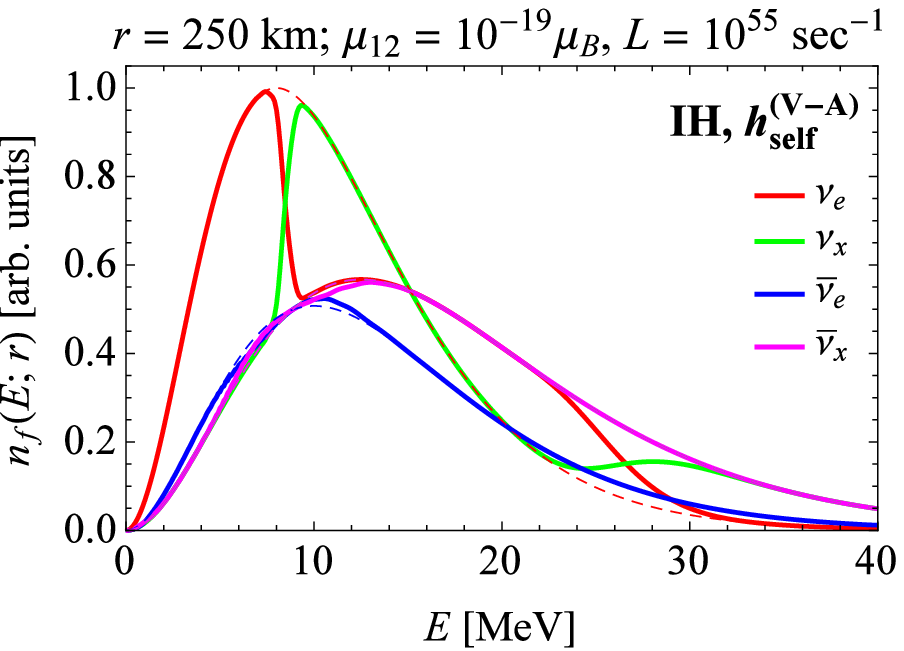} &
            \includegraphics[height=4cm]{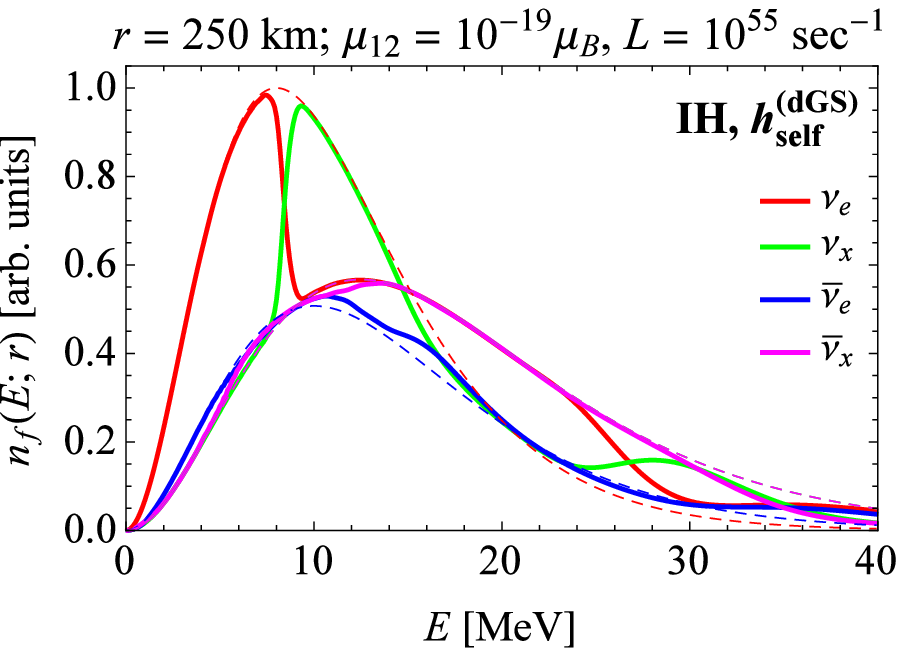} \\[0.5em]
            \text{(d)} & \text{(e)} & \text{(f)}
        \end{array}$
        \caption{The effect of the neutrino magnetic moment on the neutrino spectra without (pseudo)scalar NSSI for $L = 10^{55}\text{ sec}^{-1}$.
        The upper and the lower rows correspond to the normal and the inverted hierarchies, respectively. The first column (a, d) shows the
        no-AMM case $\mu_{12} = 0$, the second (b, e) the $\mu_{12} \ne 0$ case with self-interaction Hamiltonian
        \eqref{hSelf_our}, and the last one (c, f) also to $\mu_{12} \ne 0$, but based on the Hamiltonian \eqref{hSelf_dGS} from Ref.~\cite{deGouvea2012_2013}.
        Dashed lines sketch the initial spectra at the neutrino sphere (Fig.~\ref{fig:potentials_initSpectra}b).
        In the $\mu_{12} = 0$ case, the two Hamiltonians produce the same result. Visual identity of (a,c) and (b,d)
        plots, respectively, holds up to at least $|\mu_{12}| \sim 10^{-15}\,\mu_{\text{B}}$}
        \label{fig:OK_dGS}
    \end{figure}

    Let us first discuss the flavor evolution without (pseudo)scalar NSSIs, i.e., within the (minimally extended) Standard Model, comparing the effect of the self-interaction
    Hamiltonian \eqref{hSelf_our} derived by us and that of the Hamiltonian \eqref{hSelf_dGS} claimed in Ref.~\cite{deGouvea2012_2013}.
    As mentioned in Sec.~\ref{sec:EvolutionEquation}, these two Hamiltonians lead to identical flavor evolutions in the $\mu_{12} =
    0$ case. Fig.~\ref{fig:OK_dGS} shows the results of the simulation for $\mu_{12} \ne 0$, in the cases where the effect is visually noticeable.
    It turns out that with our Hamiltonian the final spectra are virtually insensitive to the neutrino magnetic moment up to
    at least $|\mu_{12}| \sim 10^{-15}\,\mu_{\text{B}}$, while the evolution with self-interaction \eqref{hSelf_dGS} contains
    considerable magnetic moment signatures already for $\mu_{12} = 10^{-19}\,\mu_{\text{B}}$, especially in the normal
    hierarchy (the latter was, in fact, stated in \cite{deGouvea2012_2013}). For a luminosity $L = 10^{54}\text{ sec}^{-1}$ one order of
    magnitude lower than that in Fig.~\ref{fig:OK_dGS}, the magnetic moment signatures virtually disappear, obviously because the
    instabilities causing them get suppressed. Note that the low sensitivity of the evolution equations \eqref{rho_evolution_singleAngle}
    to the magnetic moment of a Majorana neutrino agrees with the linear stability analysis in Sec.~\ref{sec:Stability}, so
    that we have to conclude that within the Standard Model without NSSIs, the effect of a small
    magnetic moment should be quite hard to observe. This has also been noted in our previous analysis
    \cite{Kharlanov2019}; the case of a large AMM has been recently studied in \cite{Abbar2020},
    including the angular neutrino distributions, and the analysis demonstrates that for a smaller AMM/weaker magnetic field,
    their signatures are suppressed. Regarding the Hamiltonian \eqref{hSelf_dGS} that was claimed to hold within the Standard Model in Ref.~\cite{deGouvea2012_2013},
    we have to consider it a nonstandard neutrino self-interaction instead, which leads to a pronounced effect of the
    neutrino transition magnetic moment.

\begin{figure}[tbh]
        $\begin{array}{ccc}
            \includegraphics[height=4cm]{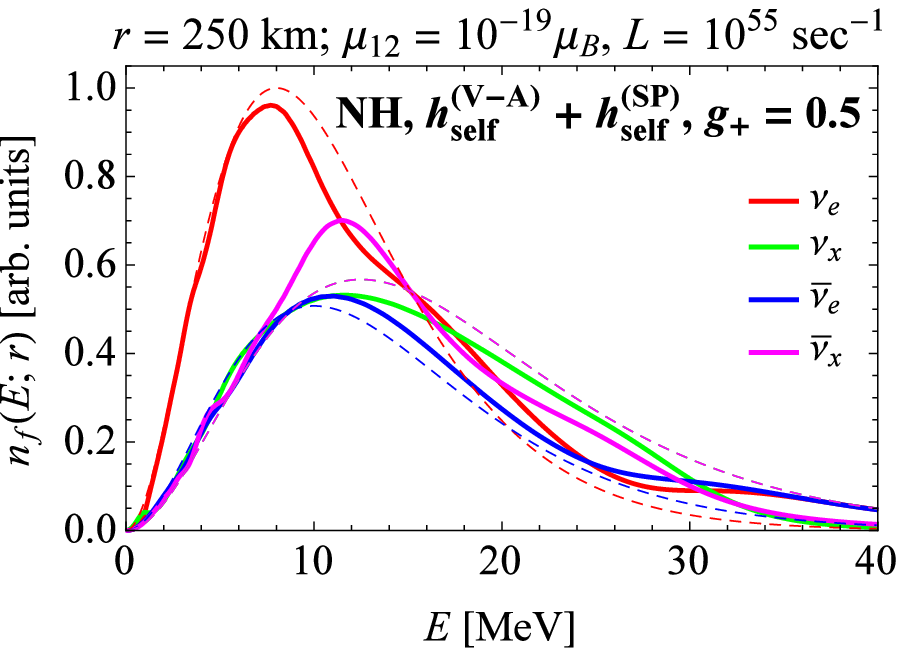} &
            \includegraphics[height=4cm]{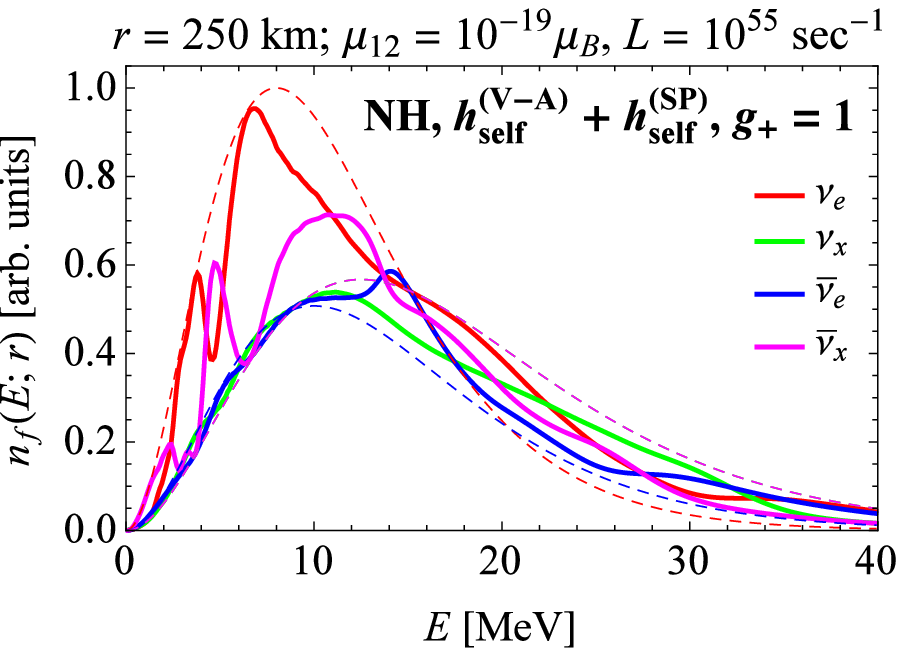} &
            \includegraphics[height=4cm]{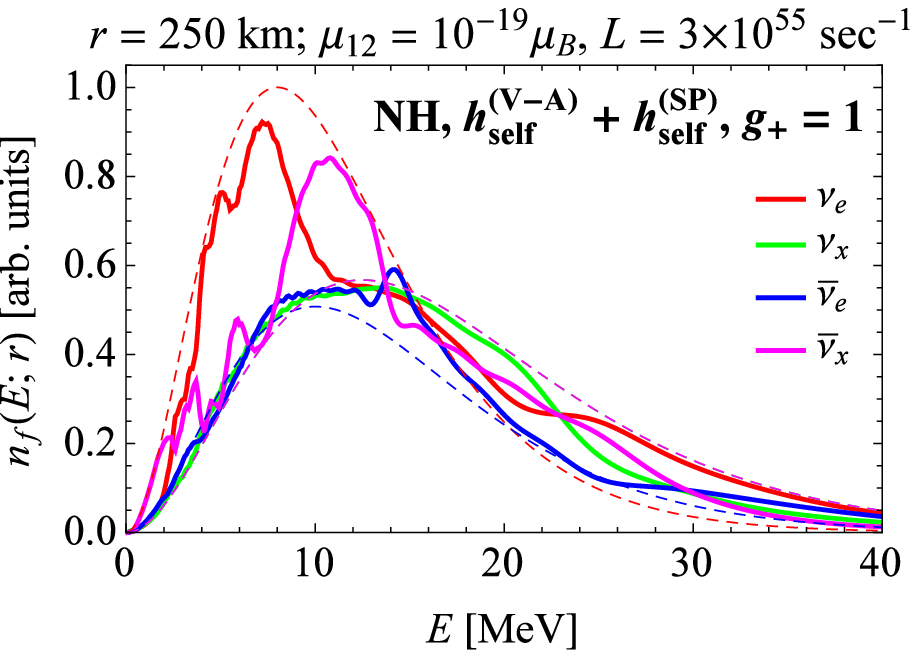} \\[0.5em]
            \text{(a)} & \text{(b)} & \text{(c)}            \\[1em]
            \includegraphics[height=4cm]{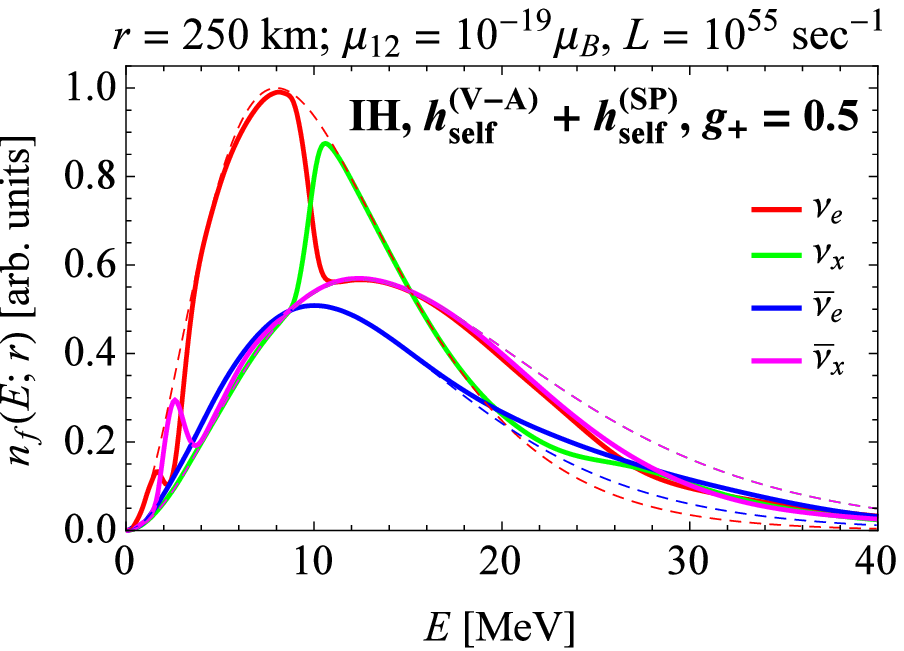} &
            \includegraphics[height=4cm]{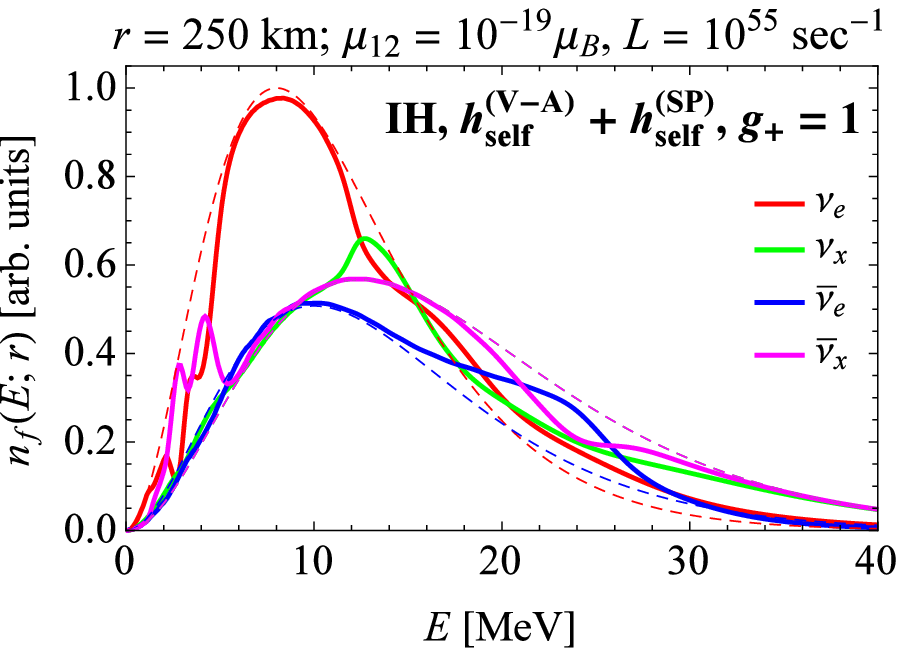} &
            \includegraphics[height=4cm]{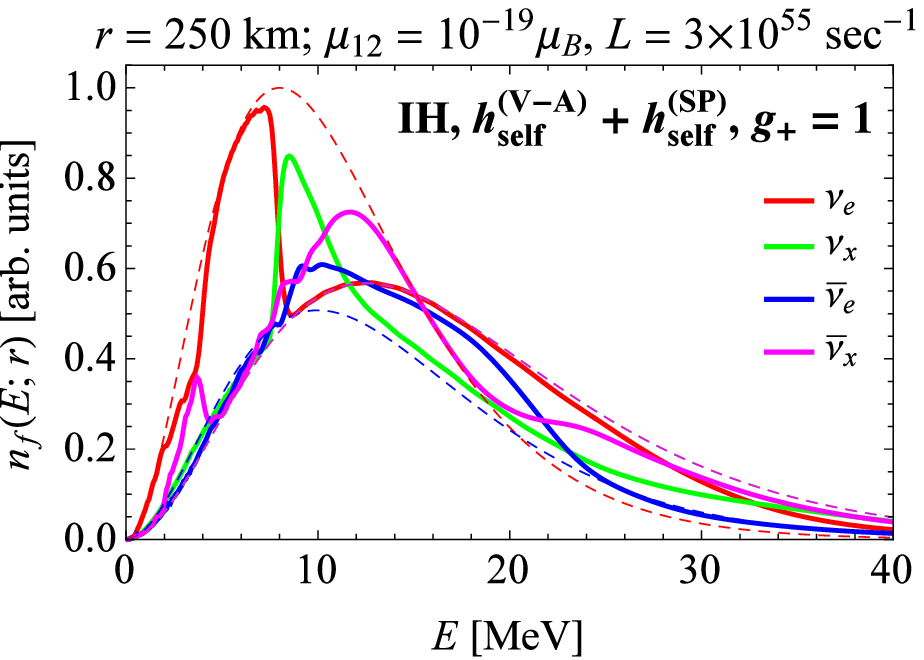} \\[0.5em]
            \text{(d)} & \text{(e)} & \text{(f)}
        \end{array}$
        \caption{The effect of the block-off-diagonal NSSI on the neutrino spectra for the neutrino magnetic moment $\mu_{12} = 10^{-19}~\mu_{\text{B}}$
        and different NSSI couplings $g_+$ and luminosities $L$. The upper and the lower rows correspond to the normal and the inverted hierarchies, respectively.
        The first (a, d) and the second (b, e) columns demonstrate the cases with luminosity $L = 10^{55}~\text{sec}^{-1}$ and
        different NSSI couplings; the last column (c, f) shows the results for $L = 3\times10^{55}~\text{sec}^{-1}$.
        Dashed lines sketch the initial spectra at the neutrino sphere (Fig.~\ref{fig:potentials_initSpectra}b)}
        \label{fig:NSSI_spectra}
    \end{figure}

    \vsp
    Now let us switch to the main issue of the present paper, namely, to the effect of nontrivial scalar/pseudoscalar NSSIs on
    the collective neutrino flavor evolution. Namely, for the same luminosity $L = 10^{55} \text{ sec}^{-1}$, we set the
    block-diagonal NSSI coupling $g_-$ to zero (as mentioned earlier, this coupling has been analyzed in other papers
    \cite{NSSI_Yang2018_SPint}), keeping only the block-off-diagonal term with the $g_+$ coupling, and analyze the effect of this small
    NSSI for a small transition magnetic moment $\mu_{12} = 10^{-19}\,\mu_{\text{B}}$. Again, when the magnetic moment is zero,
    the density matrix becomes block-diagonal and the $g_+$ coupling produces zero effect on the oscillations. For a nonzero
    magnetic moment, in contrast, nonstandard interactions radically change the neutrino spectra, see
    Fig.~\ref{fig:NSSI_spectra}.

    Indeed, neutrino-antineutrino instability caused by the block-off-diagonal NSSI and triggered by the $\mu B$ interaction
    shows up in both hierarchies provided that the NSSI coupling is not too small, $|g_+| \gtrsim 0.3$, and rapidly leads to a
    complicated, if not a chaotic pattern. This is clearly manifested in the wiggly patterns in Fig.~\ref{fig:NSSI_spectra}
    resulting from probability exchange between $\nu_e$ and $\bar\nu_x$ flavors. For large $g_+$ couplings, this exchange even
    results in a $\nu_e-\bar\nu_x$ spectral split, which replaces the $\nu_e-\nu_x$ split in the inverted hierarchy case.
    In fact, animations of the flavor evolution we made also reveal that $g_+ \ne 0$ flavor evolution
    starts from rapid synchronized oscillations of the spectra with typical lengths of $10-100\text{ m}$ (in contrast to the
    no-NSSI case, where the oscillation lengths are of the order of $10\text{ km}$), which, in principle, agrees with our
    conclusions on the fast/intermediate instabilities in Sec.~\ref{sec:Stability}.

    \begin{figure}[tbh]
        $\begin{array}{cc}
            \includegraphics[height=5.5cm]{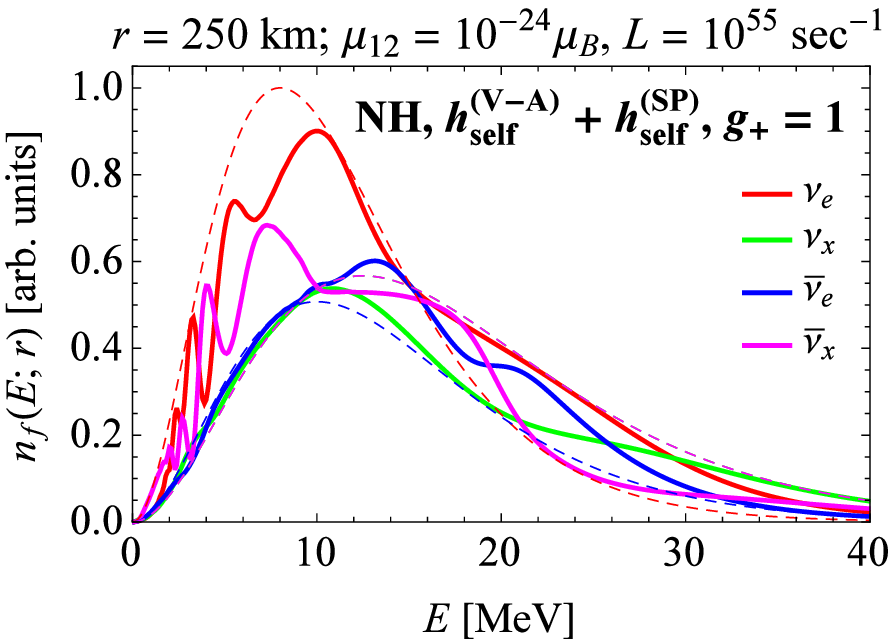} &
            \includegraphics[height=5.5cm]{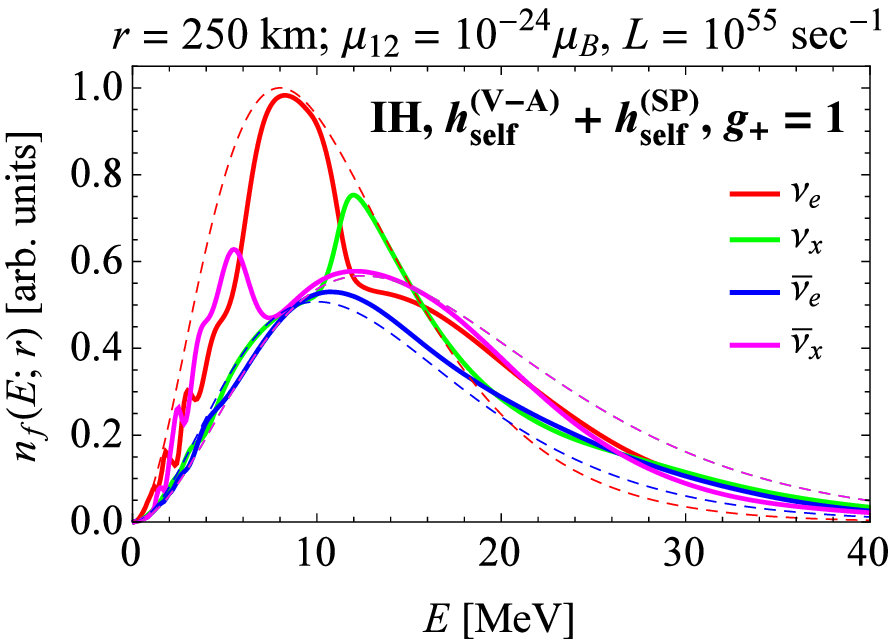}
            \\
            \text{(a)}& \text{(b)}
        \end{array}$
        \caption{Impact of scalar-pseudoscalar NSSI with $g_+ = 1$ on the neutrino flavor evolution for $L = 10^{55}\text{ sec}^{-1}$
                 and a very small magnetic moment $\mu_{12} = 10^{-24}\,\mu_{\text{B}}$ for the two mass hierarchies}
        \label{fig:NSSI_spectra_smallMu}
    \end{figure}

    The effects do not drastically depend on the magnetic moment, since it acts just as a seed for an exponentially
    growing instability; deformation of the spectra can be observed even for $\mu_{12} \sim 10^{-24}\,\mu_{\text{B}}$
    (see Fig.~\ref{fig:NSSI_spectra_smallMu})\footnote{Interestingly, such a small value of the magnetic moment does not lead
    to large roundoff errors in the numerical integration, roughly speaking, coming from addition of `large' and `small' matrices
    in expressions of the form $\rho(r + \Delta{r}) \approx \rho(r) -\ii\Delta{r} [h(r), \rho(r)]$. The reason is that during the early stages
    of the evolution $\rho$ is approximately block-diagonal, so adding a `small' but block-off-diagonal matrix $-\ii [h_{\text{AMM}}, \rho]$
    to it does not round the latter one down to zero: the `large' blocks are not added to `small' blocks of the density matrix.
    During the later stages, it is mainly the self-interaction and not the tiny magnetic moment that drives the evolution of the density matrix.}.
    On the other hand, the rates of instabilities resulting in nontrivial NSSI signatures strongly depend on
    the degree of nonlinearity of the equations, i.e., on the luminosity, and these signatures are virtually absent for $L = 10^{54}\text{ sec}^{-1}$.

\begin{figure}[tbh]
        $\begin{array}{cc}
            \includegraphics[height=5.5cm]{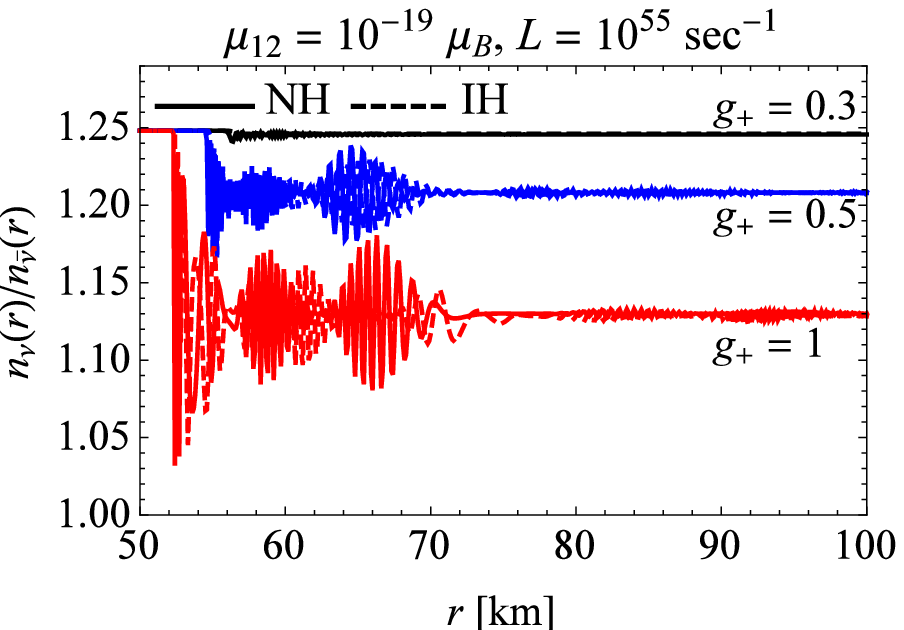} &
            \includegraphics[height=5.5cm]{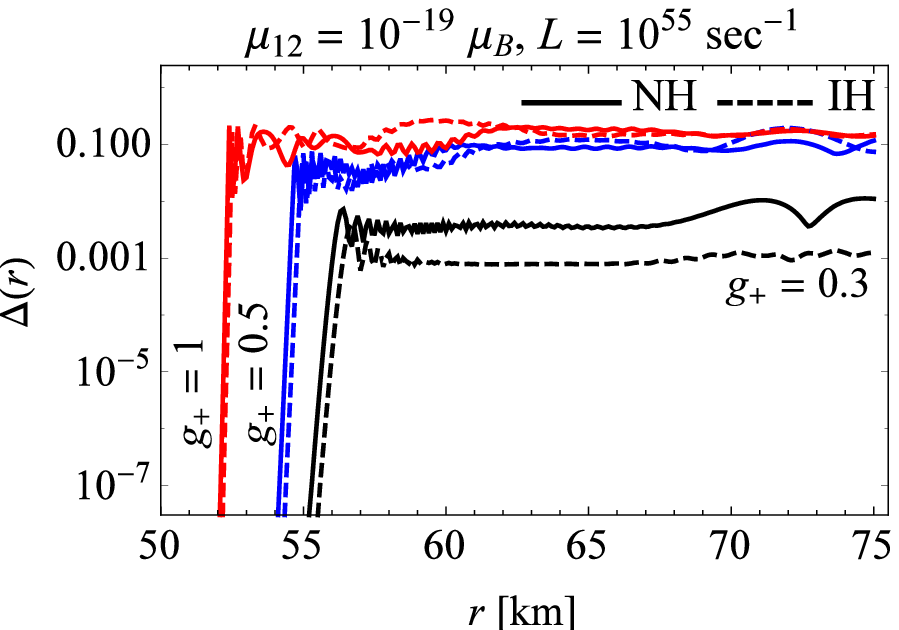}
            \\
            \text{(a)}& \text{(b)}
        \end{array}$
        \caption{Impact of scalar-pseudoscalar NSSI on the neutrino flavor evolution for $L = 10^{55}\text{ sec}^{-1}$ and
        $\mu_{12} = 10^{-19}\mu_{\text{B}}$ for different NSSI couplings and mass hierarchies:
        (a) evolution of neutrino-to-antineutrino number ratio, (b) evolution of spectral residuals \eqref{spectral_residual}}
        \label{fig:spectralImpact}
    \end{figure}

    Let us now take a closer look at the development of NSSI-induced instabilities for $L = 10^{55}\text{ sec}^{-1}$ and address
    the issue of their potential observability. First of all, a natural parameter controlling these neutrino-antineutrino
    instabilities is the ratio of the total neutrino and antineutrino numbers $n_\nu(r) / n_{\bar\nu}(r)$ obtained after integration
    over the whole energy spectrum. This ratio is conserved in the absence of magnetic moment $\mu_{12}$; however, as we saw above,
    it is also conserved to a high accuracy even when $\mu_{12}$ is nonzero, in both noncollective oscillations and collective oscillations
    without NSSIs, neither of which contain neutrino-antineutrino instabilities. Interestingly, Fig.~\ref{fig:spectralImpact}a demonstrates that
    when NSSIs come into play, the total neutrino and antineutrino numbers rapidly reach an equilibrium plateau. This phenomenon resembles
    a sort of equilibration discussed in Ref.~\cite{Abbar2020}, but for a value of $\mu_{12}B$ that is many orders of magnitude smaller.
    The equilibrium neutrino-antineutrino ratio is then conserved up to the neutrino detector, so that antineutrino excess could, in principle,
    serve as a signature of NSSIs mixing neutrinos with antineutrinos, such as scalar and pseudoscalar interactions.

\begin{figure}[tbh]
        $\begin{array}{cc}
            \includegraphics[height=5.5cm]{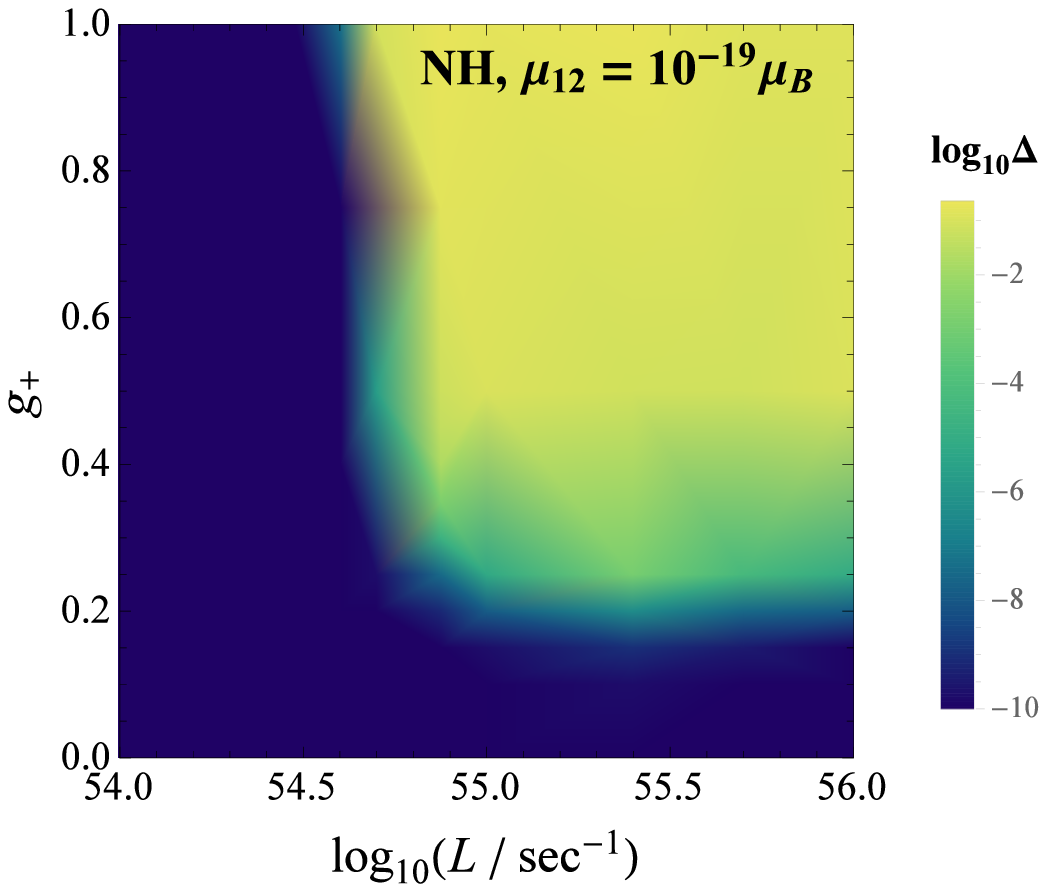} &
            \includegraphics[height=5.5cm]{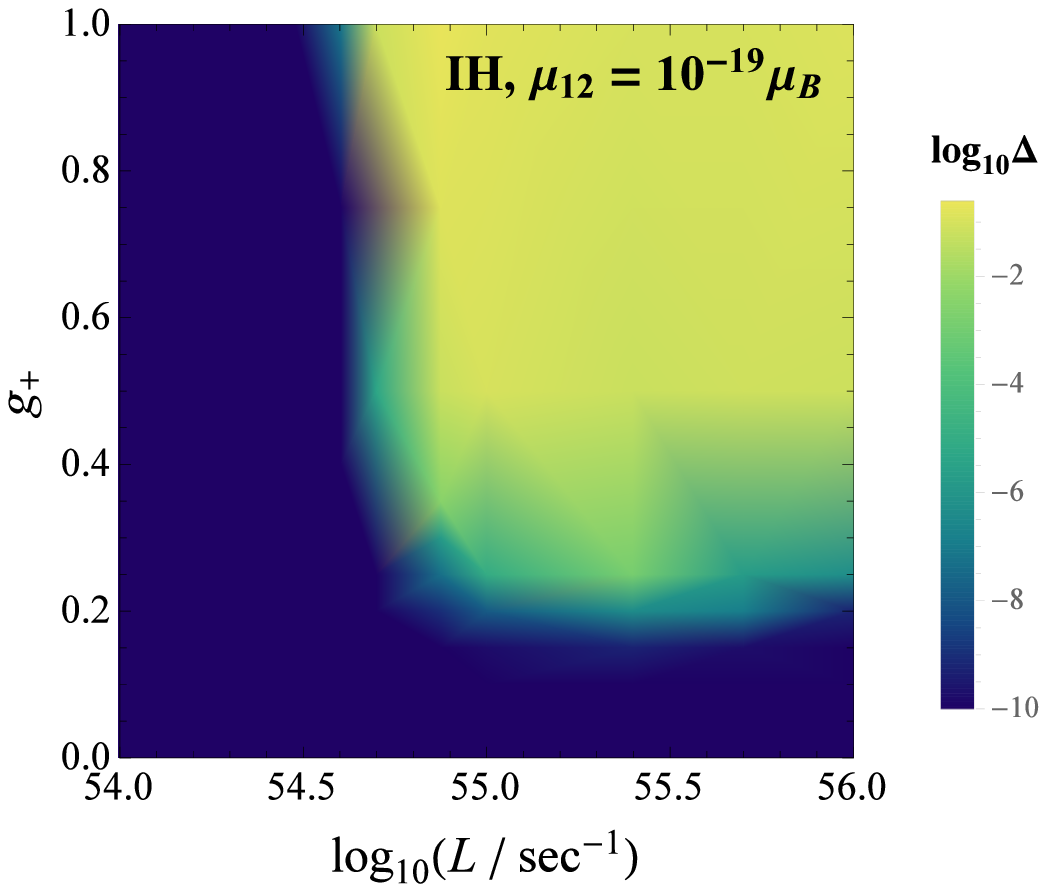}
            \\
            \text{(a)}& \text{(b)} \\
            \includegraphics[height=5.5cm]{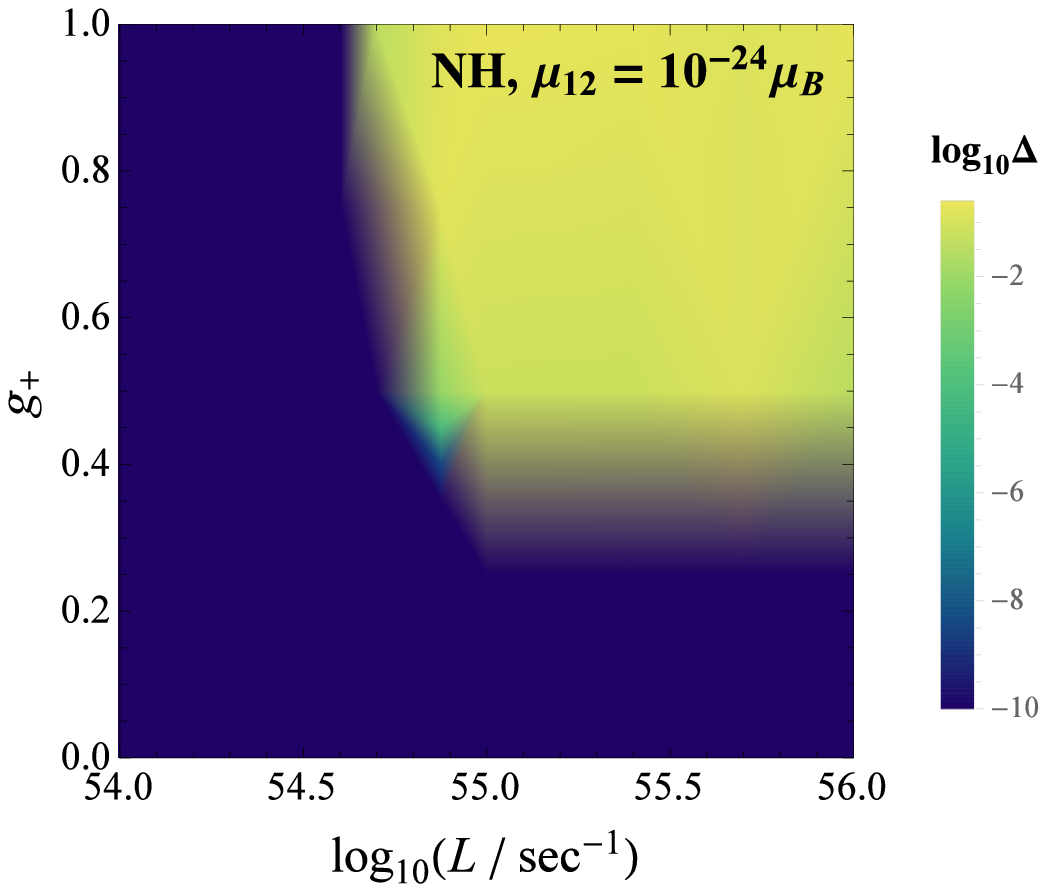} &
            \includegraphics[height=5.5cm]{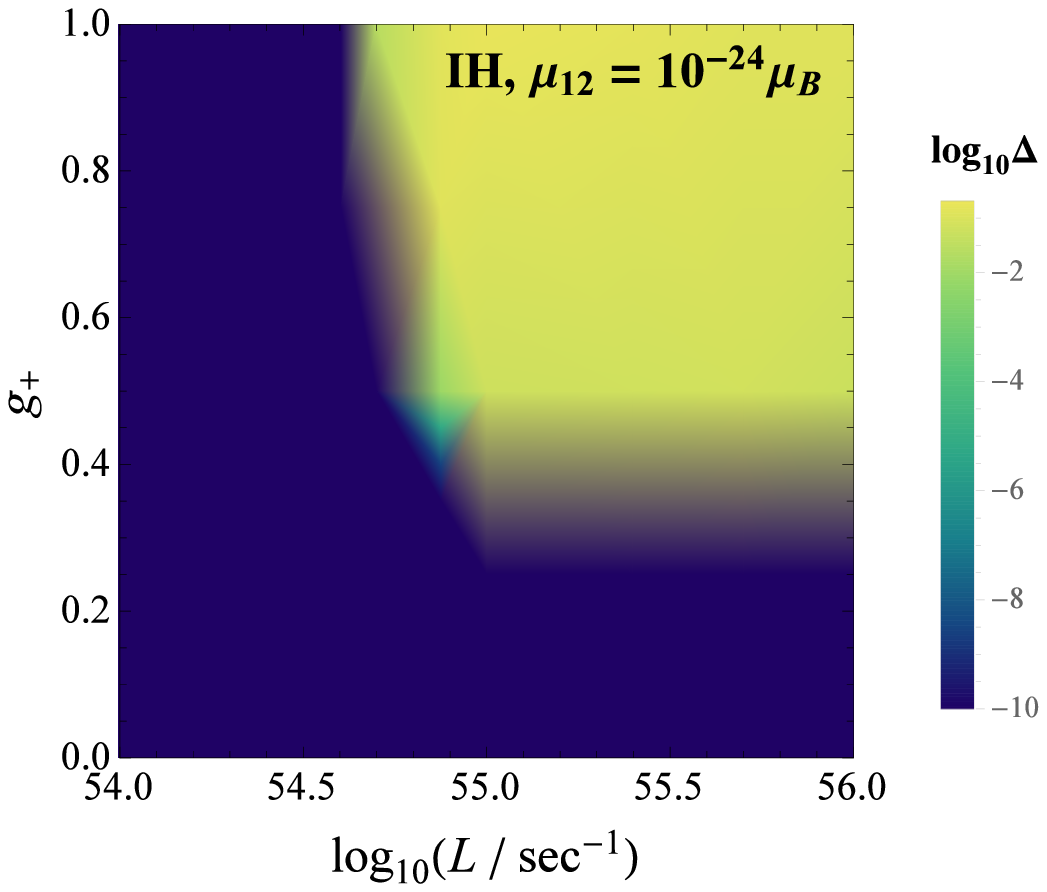}
            \\
            \text{(c)}& \text{(d)}
        \end{array}$
        \caption{Sensitivity of collective neutrino oscillations to the NSSI coupling $g_+$ in terms of the spectral residual
        $\Delta(250~\text{km})$, depending on the coupling $g_+$ and the luminosity $L$. The magnetic moment is
        set to $\mu_{12} = 10^{-19}\mu_{\text{B}}$ and $10^{-24}\mu_{\text{B}}$ in the two upper and lower panels, respectively}
        \label{fig:NSSI_sensitivity}
    \end{figure}

    Fig.~\ref{fig:spectralImpact}a also reveals a \textit{fast} character of the instability, as opposed to
    \textit{slow} instabilities with growth scales of the order of $10~\text{km}$. To quantify this
    instability and the resulting impact of the neutrino spectra, we introduce a spectral residual as the relative integral
    deviation of the neutrino flavor spectra from the no-AMM case, in which the effect of NSSIs is absent
    \begin{equation}\label{spectral_residual}
        \Delta(r) = \frac{\sum\limits_{f} \int{\bigl| n_f(E;r) - n_f^{(0)}(E;r)\bigr| \diff{E}}}
                                    {\sum\limits_{f} \int n_f^{(0)}(E;r) \diff{E}},
    \end{equation}
    where $n^{(0)}_f(E;r)$ correspond to the flavor-energy spectra for $\mu_{12} = 0$ and the summation is performed over both
    neutrino and antineutrino flavors. Now, Fig.~\ref{fig:spectralImpact}b clearly demonstrates a drastic difference between the
    evolutions of neutrino spectra for $g_+ \le 0.3$ and $g_+ \gtrsim 0.5$: in the latter case, the residuals $\Delta(r)$ rapidly saturate
    to quite observable values of the order of $10^{-1}$ and keep this level during further evolution. Note that the level $\Delta \sim 0.01-0.1$
    corresponds, in fact, to quite a large spectral distortion, such as those shown in Fig.~\ref{fig:NSSI_spectra}. Finally, the spectral residuals
    at the `final' point $r = 250~\text{km}$ are shown in Fig.~\ref{fig:NSSI_sensitivity} versus the luminosity $L$ and the NSSI coupling $g_+$.
    One observes that at least for $L \le 10^{56}~\text{sec}^{-1}$, one can probe the NSSI coupling with the sensitivity around $0.25-0.4$
    in both mass hierarchies. It is important, however, that this order of sensitivity can be achieved for extremely small
    values of the neutrino transition magnetic moments, comparable with the figures predicted within the minimally extended
    Standard Model with the standard, V--A structure of weak interactions \cite{Giunti2009_nuEMP, GIM}.

    \section{Discussion}
    \label{sec:Discussion}%

    The analysis given in the above sections has led us to a number of interesting conclusions regarding the effect of nonstandard, scalar
    and/or pseudoscalar four-fermion neutrino interactions on the collective oscillations taking place during a supernova
    explosion. Firstly, it turned out that such NSSIs include the terms that mix neutrinos and antineutrinos, and these are
    able to open up a new channel of fast neutrino-antineutrino instabilities considerably deforming the neutrino flavor-energy spectra.
    In fact, these spectral features, including nonstandard splits and chaotic-looking patterns, need a seed to develop;
    however, a minuscule transition neutrino magnetic moment $\mu_{12} \sim 10^{-24}~\mu_{\text{B}}$ proves to be enough to trigger
    the effect via neutrino-antineutrino (helicity flip) transitions in the stellar magnetic field. Note that such a transition
    magnetic moment could be generated even at the one-loop level of the Standard Model, in which a Glashow--Iliopoulos--Maiani (GIM)
    cancelation takes place \cite{Giunti2009_nuEMP, GIM}.

    Secondly, the discussed type of $\nu-\bar\nu$ instability is also quite nonstandard in that the electron/neutron background nontrivially
    affects the instability rates even for monochromatic neutrinos and in that these instabilities survive in the absence of antineutrinos
    (see Sec.~\ref{sec:Stability} devoted to linear stability analysis). From this analysis we also observed that, quite as usual,
    a neutrino-antineutrino ratio close to unity favors the development of instabilities.

    Thirdly, the sensitivity of (anti)neutrino spectra to the NSSI coupling stays at the level of $g_+ \sim 0.3$
    for neutrino luminosities $L \gtrsim 10^{55}~\text{sec}^{-1}$, notably, virtually regardless of the (nonzero) value of the
    magnetic moment. Note that the spectral residual $\Delta(250~\text{km})$ chosen by us to quantify the NSSI
    signatures in the neutrino spectra (see Eq.~\eqref{spectral_residual}) can change outside the $r = 250~\text{km}$ sphere,
    where collective oscillations are not important, while other effects, such as the MSW effect in a turbulent medium, may come into
    play \cite{Kneller2010_Turbulence}. Importantly, however, there is another natural measure of scalar/pseudoscalar NSSI signatures in question, that does not change under
    (conventional) noncollective oscillations: the neutrino-antineutrino ratio $n_\nu / n_{\bar\nu}$.
    Both this ratio for neutrinos of a given energy $E$ and the one integrated over the whole energy spectrum are almost conserved
    in noncollective oscillations (the magnetic moment has a negligible effect once the collective oscillations are over),
    thus, the neutrino-antineutrino ratio (see Fig.~\ref{fig:spectralImpact}) should bring the information on the NSSI in the innermost
    supernova layers to the neutrino detector. Moreover, even though in the present paper we have discussed the NSSI effect on collective
    oscillations within a simplified picture not including angular degrees of freedom, it is highly likely that the neutrino-antineutrino ratio
    should depend on the `latitude', i.e., on the angle between the directions to the stellar magnetic pole and the observer. Therefore, helicity-flipping
    scalar/pseudoscalar NSSIs could manifest themselves in a periodic variation of the neutrino-antineutrino flux ratio from
    an explosion of a (rotating) supernova.

    Finally, from our analysis, both numerical and analytical, it inevitably turns out that within the framework with a pure
    V--A interaction, the effective neutrino oscillation Hamiltonian takes the form derived in Ref.~\cite{Dvornikov2012_Heff},
    which is unable to exponentially catalyze the growth of coherent neutrino-antineutrino mixing initially introduced via the
    transition magnetic moment. Note that the linear stability analysis in Sec.~\ref{sec:Stability} that led to this conclusion
    was not a Lyapunov stability analysis of a necessarily stationary solution, so it does not rely upon a `qualitatively right'
    assumption of the vanishing vacuum mixing angle. As a result, within the Standard Model extended with nonzero Majorana neutrino masses,
    the effect of a small nonzero magnetic moment is very unlikely to be observable.

    A couple of words should also be said on what was left beyond the present analysis. First of all, Dirac neutrinos should
    probably have very similar block-off-diagonal terms in the interaction Hamiltonian, if one introduces scalar and/or pseudoscalar
    terms into the Lagrangian, while in this case these blocks will describe superpositions of active, left-handed neutrinos
    with `sterile', right-handed neutrinos rather than neutrino-antineutrino mixing. In the absence of NSSIs, these blocks
    virtually vanish, so nothing is able to catalyze the helicity-mixing effect introduced by a small magnetic moment, analogously to the
    Majorana case \cite{Kharlanov2019}. Next, we have tested Majorana neutrinos with scalar/pseudoscalar
    interactions for instabilities initiated by the so-called induced magnetic moment \cite{Grigoriev2019}, which physically
    is a helicity-dependent Wolfenstein potential in a magnetized medium, and found no new instabilities. Among other possible reasons,
    this is definitely related to the fact that at least for a nonmoving background medium,
    the induced magnetic moment term in the Hamiltonian does not flip the neutrino helicity. However, this, as well as other
    issues regarding the effect of scalar and pseudoscalar nonstandard neutrino self-interactions on the evolution of
    dense neutrino flows require further study beyond the present paper.

    \section*{Acknowledgments}
    The authors are grateful to Maxim Dvornikov and Alexander Grigoriev for fruitful discussions. O.~K. would like to give
    special thanks to Sergei Gladchenko for his attention to the manuscript and for checking a lot of expressions in it after
    it was published. The research has been carried out using the equipment of the shared research facilities of HPC computing resources
    at Lomonosov Moscow State University \cite{Lomonosov}. O.~K. would also like to acknowledge kind support with the computational
    resources from Skolkovo Institute of Science and Technology.

    \appendix

    \section{Bilinear expressions for Majorana neutrinos}
    \label{app:MajoranaIdentities}%
    We list here the notations and relations for Majorana spinors in the ultrarelativistic regime,
    which are used to derive the effective Hamiltonian for collective oscillations in Sec.~\ref{sec:EvolutionEquation}.
    We work in the spinor representation, with Dirac matrices
    \begin{gather}\label{DiracMatrices}
        \gamma^0 = \twoMatrix{0}{1}{1}{0}, \;
        \gamma_5 = \ii\gamma^0 \gamma^1 \gamma^2 \gamma^3 = \twoMatrix{-1}{0}{0}{1}, \;
        \bvec\Sigma = \twoMatrix{\bvec\sigma}{0}{0}{\bvec\sigma},\\
        \bvec{\alpha} = \gamma_5 \bvec\Sigma = \twoMatrix{-\bvec\sigma}{0}{0}{\bvec\sigma}, \;
        \bvec\gamma = \gamma^0 \bvec\alpha = \twoMatrix{0}{\bvec\sigma}{-\bvec\sigma}{0},
    \end{gather}
    $\bvec\sigma$ being the Pauli matrices; the commutator of the gamma matrices $\sigma^{\mu\nu} \equiv \frac{\ii}{2} [\gamma^\mu, \gamma^\nu]$
    has components $\sigma^{0i} = \ii \alpha^i$ and $\sigma^{ij} = \epsilon_{ijk} \Sigma^k$. In this representation,
    the ultrarelativistic polarization bispinors $u_\alpha(\bvec{p})$ entering the neutrino field operator~\eqref{nu_operator} become chiral, taking
    the form
    \begin{equation}\label{u_spinors}
        u_+(\bvec{p}) = \twoCol{0}{\chi_+(\bvec{p})},\; u_-(\bvec{p}) = \twoCol{\chi_-(\bvec{p})}{0}, \quad
        u_\alpha^{\text{c}}(\bvec{p}) = u_{-\alpha}(\bvec{p}),
    \end{equation}
    where the two-component helicity eigenvectors are defined in the standard way
    \begin{gather}
        (\bvec\sigma \cdot \bvec{p}) \chi_\alpha(\bvec{p}) = \alpha \, |\bvec{p}|\, \chi_\alpha(\bvec{p}), \quad \alpha = \pm1,\\
        \chi_\alpha\hc(\bvec{p}) \chi_\beta(\bvec{p}) = \delta_{\alpha\beta}, \quad
        -\ii\alpha\sigma_2\chi\cc_\alpha(\bvec{p}) = \chi_{-\alpha}(\bvec{p}).
    \end{gather}
    Now a straightforward matrix-vector multiplication, together with the chirality properties of the polarization bispinors,
    yield a set of bilinear expressions
    \begin{eqnarray}
      \bar{u}_\alpha(\bvec{p}) u_\beta(\bvec{p}) &=& \bar{u}_\alpha(\bvec{p}) \gamma_5 u_\beta(\bvec{p}) = 0, \label{u_bilinears_SP} \label{u_bilinears_first}\\
      \bar{u}_\alpha(\bvec{p}) \gamma^\mu u_\beta(\bvec{p}) &=& \delta_{\alpha\beta} (1, \hat{\bvec{p}}),\\
      \bar{u}_\alpha(\bvec{p}) \gamma^\mu \gamma_5 u_\beta(\bvec{p}) &=& \alpha \delta_{\alpha\beta} (1, \hat{\bvec{p}}),\\
      \bar{u}_\alpha(\bvec{p}) \gamma^\mu_{\text{L}} u_\beta(\bvec{p}) &=& \delta_{\alpha,-}\delta_{\beta,-} (1, \hat{\bvec{p}}),  \label{u_bilinears_VA}\\
      \bar{u}_\alpha(\bvec{p}) \bvec\Sigma u_\beta(\bvec{p}) &=& \sqrt{2} \delta_{\alpha, -\beta} \bvec\zeta_\beta(\bvec{p}), \\
      \bar{u}_\alpha(\bvec{p}) \bvec\alpha u_\beta(\bvec{p}) &=& \sqrt{2} \beta \delta_{\alpha, -\beta}
      \bvec\zeta_\beta(\bvec{p}), \label{u_bilinears_last}
    \end{eqnarray}
    where $\bar{u}_\alpha(\bvec{p}) \equiv u\hc_\alpha(\bvec{p}) \gamma^0$ is a Dirac conjugate,
    $\hat{\bvec{p}} \equiv \bvec{p} / |\bvec{p}|$, and a complex 3-vector $\bvec\zeta_\beta(\bvec{p})$ is defined as
    \begin{equation}
        \bvec\zeta_\beta(\bvec{p}) = \frac{1}{\sqrt2} \chi\hc_{-\beta}(\bvec{p}) \bvec\sigma \chi_\beta(\bvec{p}).
    \end{equation}
    While a couple of properties of this vector follow directly from the above definitions,
    \begin{gather}
        \bvec{p}\cdot \bvec\zeta_\beta(\bvec{p}) = 0,
        \\
        |\bvec{B}\cdot \bvec\zeta_\beta(\bvec{p})| = \frac{1}{\sqrt{2}} |\bvec{B}\times \hv{p}|,
        \\
        \bvec\zeta_\beta\cc(\bvec{p}) = \bvec\zeta_{-\beta}(\bvec{p}),
        \\
        \bvec\zeta_\beta\cc(\bvec{p}) \cdot \bvec\zeta_\beta(\bvec{p}) = 1,
    \end{gather}
    others depend on the $U(1)$ phase $\phi(\bvec{p})$ in the definition of the two-component helicity eigenstates.
    Fixing it as
    \begin{equation}\label{chi_plusMinus}
        \chi_+(\bvec{p}) = e^{\ii\phi(\bvec{p})/2} \twoCol{\cos\frac{\vartheta}{2} e^{-\ii\varphi/2}}{\sin\frac{\vartheta}{2} e^{\ii\varphi/2}},
        \quad
        \chi_-(\bvec{p}) = e^{-\ii\phi(\bvec{p})/2} \twoCol{-\sin\frac{\vartheta}{2} e^{-\ii\varphi/2}}{\cos\frac{\vartheta}{2} e^{\ii\varphi/2}}
    \end{equation}
    for the momentum vector $\bvec{p} = |\bvec{p}| \; (\sin\vartheta \cos\varphi, \sin\vartheta \sin\varphi, \cos\vartheta)$ leads to
    an explicit expression for the $\bvec\zeta$ vectors
    \begin{equation}\label{zeta_explicit}
        \bvec\zeta_\pm(\bvec{p}) = \frac{e^{\pm\ii\phi(\bvec{p})}}{\sqrt2}
        \bigl(
            \cos\vartheta \cos\varphi \mp \ii \sin\varphi,
            \cos\vartheta \sin\varphi \pm \ii \cos\varphi,
            -\sin\vartheta
        \bigr).
    \end{equation}
    Now one can explicitly evaluate a scalar product entering the effective Hamiltonian (see Sec.~\ref{sec:EvolutionEquation})
    \begin{equation}\label{zetaProduct}
        \bvec\zeta_\pm(\bvec{p}) \cdot \bvec\zeta_\pm(\bvec{q}) = e^{\pm\ii \Gamma(\hv{p},\hv{q})} \frac{1 - \hv{p}\cdot\hv{q}}{2},
    \end{equation}
    where $\Gamma(\hv{p},\hv{q})$ is another $U(1)$ phase. Even though this phase obviously contains an additive term
    $\phi(\bvec{p}) + \phi(\bvec{q})$, it is easy to see that the $\Gamma$ phase cannot be eliminated completely by a gauge
    transformation of $\phi(\bvec{p})$.

    \section{An analogue of the Wick's theorem}%
    \label{app:WickTheorem}%
    While the original Wick's theorem relates the vacuum expectation value of a product of field operators to a set of pairwise contractions,
    the \emph{approximate} theorem we have used in Eqs.~\eqref{Wick1}, \eqref{Wick2} is related to a state $\cket\Phi$
    with definite neutrino numbers $N_{\bvec{p}}$. First of all, note that in a correlator of four neutrino fields
    \begin{equation}
        \brac\Phi \NProd{\bar\varphi^i \chi^j \bar\psi^k \omega^l} \cket\Phi,
    \end{equation}
    where $i, j, k, l = 1, 2, 3, 4$ are spinor indices, every field operator is a series of the form \eqref{nu_operator}
    over the neutrino momentum, i.e., a sum of creation/annihilation operators $\hat{a}^{(\dagger)}_{A\bvec{p}}$ with certain
    coefficients. Therefore, the correlator itself is a quadruple series of quartic expectation values of the form
    \begin{equation}\label{quartic_a_correlator}
        \brac\Phi \NProd{\hat{a}^{(\dagger)}_{A\bvec{p}} \hat{a}^{(\dagger)}_{B\bvec{q}} \hat{a}^{(\dagger)}_{C\bvec{r}}
                  \hat{a}^{(\dagger)}_{D\bvec{s}}}
        \cket\Phi;
    \end{equation}
    to prove the theorem, it is thus enough to relate the latter expectation to pairwise contractions of $\hat{a}/\hat{a}\hc$ operators.
    Now, since $\cket\Phi$ is characterized by fixed neutrino numbers, the above expectation vanishes unless there are exactly two creation
    and two annihilation operators amongst the four $\hat{a}^{(\dagger)}$'s and the four momenta match each other, namely,
    $\bvec{p} = \bvec{q}$ and $\bvec{r} = \bvec{s}$, or $\bvec{p} = \bvec{r}$ and $\bvec{q} = \bvec{s}$, or $\bvec{p} = \bvec{s}$ and $\bvec{q} = \bvec{r}$.
    If we neglect very `rare' terms with $\bvec{p} = \bvec{q} = \bvec{r} = \bvec{s}$, then \eqref{quartic_a_correlator}, if it does not vanish,
    represents an expectation value of a product of two pairs of operators acting in different momentum subspaces,
    which is a product of two quadratic expectation values
    \begin{equation}
        \avg{\NProd{\hat{a}^{(\dagger)}_{A\bvec{p}} \hat{a}^{(\dagger)}_{B\bvec{q}} \hat{a}^{(\dagger)}_{C\bvec{r}}
                  \hat{a}^{(\dagger)}_{D\bvec{s}}}
        }
        \approx
        \avg{\NProd{\hat{a}^{(\dagger)}_{A\bvec{p}} \hat{a}^{(\dagger)}_{B\bvec{q}}}}
        \avg{\NProd{\hat{a}^{(\dagger)}_{C\bvec{r}}\hat{a}^{(\dagger)}_{D\bvec{s}}}}
        +
        \avg{\NProd{\hat{a}^{(\dagger)}_{A\bvec{p}} \hat{a}^{(\dagger)}_{C\bvec{r}}}}
        \avg{\NProd{\hat{a}^{(\dagger)}_{D\bvec{s}}\hat{a}^{(\dagger)}_{B\bvec{q}}}}
        -
        \avg{\NProd{\hat{a}^{(\dagger)}_{A\bvec{p}} \hat{a}^{(\dagger)}_{D\bvec{s}}}}
        \avg{\NProd{\hat{a}^{(\dagger)}_{C\bvec{r}}\hat{a}^{(\dagger)}_{B\bvec{q}}}},
    \end{equation}
    where the approximate equality sign refers to the special case with four equal momenta, in which the equality is wrong.
    Now that we have established such an equality for $\hat{a}/\hat{a}\hc$ operators, we can write an analogous expression for
    the fields
    \begin{equation}
        \avg{\NProd{\bar\varphi^i \chi^j \bar\psi^k \omega^l}}
        \approx
        \contractWithIndices{\bar\varphi}{i}{\chi}{j} \contractWithIndices{\bar\psi}{k}{\omega}{l} +
        \contractWithIndices{\bar\varphi}{i}{\bar\psi}{k} \contractWithIndices[1.5ex]{\omega}{l}{\chi}{j} -
        \contractWithIndices{\bar\varphi}{i}{\omega}{l} \contractWithIndices{\bar\psi}{k}{\chi}{j},
    \end{equation}
    with a contraction defined as $\contractWithIndices{\bar\varphi}{i}{\chi}{j} \equiv \avg{\NProd{\bar\varphi^i \chi^j}}$.
    Note that for Majorana neutrinos allowing for coherent neutrino-antineutrino mixing, the second term on the r.h.s. may not
    vanish. Usually, the above expression is written in the form keeping the original order of the four operators, but
    with an explicitly specified contraction pattern
    \begin{equation}\label{WickTheorem}
        \avg{\NProd{\bar\varphi^i \chi^j \bar\psi^k \omega^l}}
        \approx
        \contraction{}{\bar\varphi}{{}^i}{\chi}
        \contraction{\bar\varphi^i\chi^j}{\bar\psi}{{}^k}{\omega}
        \bar\varphi^i \chi^j \bar\psi^k \omega^l
        +
        \contraction[1.5ex]{}{\bar\chi}{{}^i\chi^j}{\bar\psi}
        \contraction[1.2ex]{\bar\varphi^i}{\chi}{{}^j \bar\psi^k}{\omega}
        \bar\varphi^i \chi^j \bar\psi^k \omega^l
        +
        \contraction[1.5ex]{}{\bar\varphi}{{}^i\chi^j \bar\psi^k}{\omega}
        \contraction[0.8ex]{\bar\varphi^i}{\bar\chi}{{}^j}{\bar\psi}
        \bar\varphi^i \chi^j \bar\psi^k \omega^l.
    \end{equation}

    \section{Flavor density matrix in the flavor basis}%
    \label{app:rho_flavor}%
    In the derivation of the effective Hamiltonian~\eqref{rho_evolution_Hamiltonian}, the density matrix $\rho_{AB}(\bvec{p})$
    was defined in terms of operators related to neutrino mass states $a, b$ (see Eq.~\eqref{rho_def}), however, it is often handy
    to work in the flavor basis instead. Indeed, neutrino flavor currents can be written in terms of a $U(2N_{\text{f}})$
    transformed matrix $\rho_{ff'}^{\text{(fl)}}(\bvec{p})$
    \begin{gather}
        \brac\Phi \NProd{\bar\nu_f(\bvec{x}) \gamma^\mu_{\text{L}} \nu_{f'}(\bvec{x})} \cket\Phi =
         \sum_{\bvec{p}} \frac{(1, \hv{p})}{V} U\cc_{fa} U_{f'b} ( \rho_{b-,a-}(\bvec{p}) - \rho_{a+,b+}(\bvec{p}) )
        = \sum_{\bvec{p}} \frac{(1, \hv{p})}{V} ( \rho_{f'-,f-}(\bvec{p}) - \rho_{f+,f'+}(\bvec{p}) ),
        \\
        \rho^{\text{(fl)}}(\bvec{p}) = \twoMatrix{U}{0}{0}{U\cc} \rho(\bvec{p}) \twoMatrix{U\hc}{0}{0}{U^{\mathrm{T}}}
        \equiv \mathcal{U} \rho(\bvec{p}) \mathcal{U}\hc.
    \end{gather}
    Since this transformation is unitary, it induces a unitary transformation of the Hamiltonian
    $h^{\text{(fl)}}(\bvec{p}) = \mathcal{U} h(\bvec{p}) \mathcal{U}\hc$, and its straightforward application to
    Eq.~\eqref{rho_evolution_Hamiltonian} yields
    \begin{eqnarray}
        h^{\text{(fl)}}(\bvec{p}) &=&
        \twoMatrix{{M^{\text{(fl)}}}^2 / 2 |\bvec{p}| + G_{\text{F}}\sqrt2 (n_e \mathds{P}_e^{\text{(fl)}} - n_n \idMatrix / 2)}
                  {-\ii \mathfrak{m}^{\text{(fl)}} B_\perp(\bvec{p}) }
                  {\ii \mathfrak{m}^{\text{(fl)}\dagger} B_\perp\cc(\bvec{p})}
                  {{M^{\text{(fl)}\mathrm{T}}}^2 / 2 |\bvec{p}| - G_{\text{F}}\sqrt2 (n_e \mathds{P}_e^{\text{(fl)}} - n_n \idMatrix / 2)}
        \nonumber\\
        &+& \frac{G_{\text{F}}\sqrt2}{V} \sum_{\bvec{q}} (1 - \hv{p}\cdot \hv{q}) \mathcal{K}(\rho^{\text{(fl)}}(\bvec{q})),
        \\
        \mathcal{K}(\rho^{\text{(fl)}}) &\equiv&
                        \tr\bigl(\rho^{\text{(fl)}}\mathcal{G}\bigr) \mathcal{G} + \wp^{\text{(fl)}\diagPart}
                        + g_- \bigl(\mathcal{U}\cc\mathcal{U}\hc \wp^{\text{(fl)}} \mathcal{U}\mathcal{U}^{\mathrm{T}}\bigr)^{\diagPart\,\text{T}}
                        + g_+ e^{\ii\Gamma(\hv{p},\hv{q}) \mathcal{G}}
                        \bigl(\mathcal{U}\cc\mathcal{U}\hc \wp^{\text{(fl)}} \mathcal{U}\mathcal{U}^{\mathrm{T}}\bigr)^{\offdiagPart \, \text{T}},
    \end{eqnarray}
    where ${M^{\text{(fl)}}}^2 = U M^2 U\hc$, $\bigl(\mathds{P}_e^{\text{(fl)}}\bigr)_{ff'} = \delta_{f,e}\delta_{f',e}$,
    $\mathfrak{m}^{\text{(fl)}} = U \mathfrak{m} U^{\mathrm{T}} = -\mathfrak{m}^{\text{(fl)}\mathrm{T}}$, and the last term
    $\mathcal{K}(\rho^{\text{(fl)}})$ describes the self-interaction in the flavor basis. One observes immediately that if the
    PMNS matrix $U$ is real, i.e., no nontrivial CP or Majorana phases are present, $\mathcal{U}\cc\mathcal{U}\hc = \idMatrix$
    and the interaction Hamiltonian retains its original, mass-basis form in the flavor basis. In general, however, this is not
    the case. At the same time, it is interesting to note that in the two-flavor case,
    $\mathfrak{m} = \ii \mu_{12} \sigma_2$ and the off-diagonal blocks
    $\wp_{\mp \pm}^{\text{(fl)}} = \rho_{\mp \pm}^{\text{(fl)}} - \rho_{\mp \pm}^{\text{(fl)}\,\mathrm{T}}$ are also proportional to $\sigma_2$,
    so that simplifications take place when multiplicating by the mixing matrix $U = e^{\ii \theta \sigma_2} \diag(e^{\ii\alpha/2},1)$
    \begin{equation}
        \mathfrak{m}^{\text{(fl)}} = e^{\ii\alpha/2} \mathfrak{m}, \quad
        \bigl(\mathcal{U}\cc\mathcal{U}\hc \wp^{\text{(fl)}} \mathcal{U}\mathcal{U}^{\mathrm{T}}\bigr)^{\offdiagPart \, \text{T}}
        = e^{\ii \alpha \mathcal{G}} \bigl( \wp^{\text{(fl)}\, \offdiagPart} \bigr)^{\mathrm{T}},
    \end{equation}
    and the remaining Majorana phase $\alpha$ can be eliminated by a gauge transformation \eqref{gauge_transformation}, i.e.,
    by a rephasing of the two helicity eigenstates. This is how one arrives at a real antisymmetric magnetic
    moment matrix in Eq.~\eqref{h_AMM_2flavors} and at a conventional $g_+$ NSSI term in the self-interaction Hamiltonian~\eqref{hSelf_our}.
    Quite naturally, in the three-flavor case, this rephasing is not enough to absorb the two Majorana and one CP phase.
\end{document}